\documentclass[11pt, a4paper, oneside]{Thesis} 
\usepackage{units}
\usepackage{amsmath,bm}
\usepackage{flafter}
\usepackage{subfigure}
\usepackage{subfigmat}
\usepackage{transparent}
\usepackage{graphicx}
\usepackage{epstopdf}
\usepackage{import}
\usepackage{caption}
\hypersetup{urlcolor=blue, colorlinks=true} 
\title{\ttitle} 

\begin{document}
\frontmatter 

\setstretch{1.3} 

\fancyhead{} 
\rhead{\thepage} 
\lhead{} 

\pagestyle{fancy} 

\newcommand{\HRule}{\rule{\linewidth}{0.5mm}} 

\hypersetup{pdftitle={\ttitle}}
\hypersetup{pdfsubject=\subjectname}
\hypersetup{pdfauthor=\authornames}
\hypersetup{pdfkeywords=\keywordnames}


\begin{titlepage}
\begin{center}
\begin{figure}[hbt]
\centering
\includegraphics[width=0.55\textwidth]{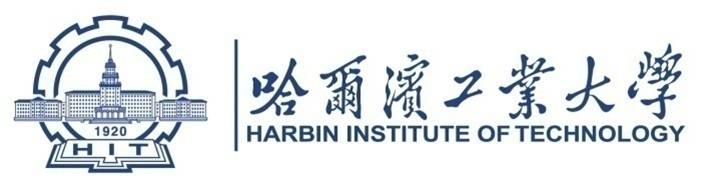}
\end{figure}
\textsc{\LARGE \univname}\\[1.5cm] 
\textsc{\Large Bachelor Thesis}\\[0.5cm] 

\HRule \\[0.4cm] 
{\huge \bfseries \ttitle}\\[0.4cm] 
\HRule \\[1.8cm] 
 
\begin{table}[h]
\centering
\large
\begin{tabular}{l l}
{\textbf{Author:}} & Hantian Zhang \\ \\ \\
{\textbf{Supervisor:}} & Prof. Wenhu Huang \\ [6pt]
{\textbf{Associate Supervisor:}} & Prof. Hutao Cui \\ [6pt]
\end{tabular}
\end{table}
\vspace{2.2cm}
 
\large \textit{A thesis submitted in fulfilment of the requirements\\ for the degree of \degreename}\\[0.3cm] 
\textit{in the}\\[0.4cm]
\deptname \\ \facname \\ [1.5cm] 
 
{\large June 2014}\\[4cm] 

\vfill
\end{center}

\end{titlepage}

%
%
%
%
%
%


\addtotoc{Abstract} 

\abstract{\addtocontents{toc}{\vspace{1em}} 

In this thesis, a non-penetrated and physically consistent non-smooth numerical approach has been proposed, by employing the Prox formulation and Moreau's mid-point time-stepping rule, for the contact dynamics with coupled and decoupled constraints. Under this circumstance, the robust impulse-based control has been successfully implemented and validated on the motion system of controlled frictional oscillator. Further improvement has been achieved by utilizing shooting method in the impulse estimating process instead of robust estimation. 
This non-smooth numerical technique has been applied to the under-actuated friction-coupled mulit-body system, by means of an implementation on the controlled frictional Furuta pendulum. The specifically designed impulse-based controller has successfully solved the problem of stabilization of the inverted frictional Furuta pendulum, which is suffered from the stiction effect of friction.\\ \\

Supervisor: Wenhu Huang\\
Title: Member of Chinese Academy of Engineering, Professor of Astronautics in Harbin Institute of Technology\\ \vspace{-0.3cm}\\
Associate Supervisor: Hutao Cui\\
Title: Professor of Astronautics in Harbin Institute of Technology\\ \vspace{-0.3cm} \\
}

\clearpage 


\begin{center}{\huge\bf Declaration\par}\end{center}

This Bachelor thesis, submitted to Harbin Institute of Technology in June 2014, is primarily based on the research project in the Institute of Applied Mechanics, Technische Universit\"{a}t M\"{u}nchen. During the period in TUM, I was a student assistant under supervisions from Dr. Thorsten Schindler and Johannes Mayet. The project follows David. M. Y. Wang's Master thesis titled "Impusle-based control of an inverted pendulum" \cite{wang13}, while in this thesis the numerical techniques for motion simulation have been reformulated correctly and tested, the mechanical modeling has been improved, and control laws have been adapted.
\\ \\
In particular, the Chapter 3 (except Section 3.3, and Figure \ref{fig:oscillator_control} - \ref{fig:oscillator_control_diagram}) was copied with minor revision from our completed report titled "Impulse-based control of a frictional oscillator" \cite{zhang14_1},  and Section 4.1 was copied from the first section of our incompleted report titled "Impulse-based control of a Furuta pendulum" \cite{zhang14_2}. The simulation techniques for non-smooth dynamics in this thesis was inspired by Dr. Thorsten Schindler, who also contributed greatly to the report \cite{zhang14_1}. 
\\ \\ \\
Hantian Zhang \\
Zurich, Switzerland, Feb. 2015.
%
%
%

\clearpage 


\setstretch{1.3} 

\acknowledgements{\addtocontents{toc}{\vspace{1em}} 

This research was carried out in the Department of Astronautics Engineering and Mechanics, School of Astronautics, Harbin Institute of Technology, and cooperated with the Institute of Applied Mechanics, Technische Universit\"{a}t M\"{u}nchen. 
\par
During my time in HIT and TUM, I am indebted to many people who provided helps and contributed to my thesis. The most significant contribution came from Dr. Thorsten Schindler and Johannes Mayet, who are my supervisors in TUM. Thanks to the student assistant position in the Institute of Applied Mechanics offered by them, I am able to be engaged in the leading-edge research in non-smooth mechanics field. Without the supervision from Dr. Schindler, this research cannot have been accomplished. His rigorous attitude on the scientific works has impressed and benefited me very much.
\par
I also thank my supervisors Prof. Wenhu Huang and Prof. Hutao Cui in HIT for their long-term cultivation. Prof. Huang is a prestigious professor in China, who had served as the former president of HIT, and the member of Chinese Academy of Engineering. 89 years old as he is, Prof. Huang always would like to discuss the scientific topics with me, along with his valuable comments on my current works, and prospective advices for the future research. I really appreciate his kind considerations on me.
\par
Furthermore, I would like to thank Prof. Qingjie Cao, who firstly introduced me into the non-smooth dynamics field, when we discussed the stability of bicycle motion in the presence of dry friction about two years ago. With his consecutive supervision, beyond the knowledge and ability on research, I fortunately learned how to critically consider my works and contributions. I also would like to thank Prof. Jie Zhao for her tailored tutorials and continuous helps for me on the advanced mathematical techniques in mechanics and control since I was a junior student.
\par
Moreover, I would like to convey my acknowledgements to Prof. Stephen John. Hogan in the University of Bristol. Prof. Hogan provide me the opportunity to investigate the piecewise smooth dynamical system in the Bristol Centre for Applied Nonlinear Mathematics. His supervision deepened my understanding of the non-smooth system on a mathematical view, and motivated me to set my goal to bridge the divides between the engineering and mathematics. 
\par
Especially, I would like to sincerely acknowledge Prof. Dong Eui Chang in the Department of Applied Mathematics, University of Waterloo. Prof. Chang is the one who influenced and encouraged me the most. His genius in applied mathematics and engineering broadened my horizon to the most cutting-edge research in control theory. Our cooperation provided me an excellent course on what is research and how to do it; his supervision, for the first time, made me realize how important the mathematics and computation are in academic works. It is the very reason that why I tend to become a mathematician, at the meanwhile, in addition to being an engineer.
\par
Last but not least, many friends have helped me over the years. Among them are Zhan Wang in the TUM, Chao Zhai, Benjamin Oscar and Victor F. Brena-Medina in the University of Bristol, Arman Tavakoli, Spideh Afshar and Amir Issaei in the University of Waterloo, Zhouhan Lin, Fan Fei, Rui Wang and Liang Zhang in the HIT, and so many others.
\par
In case I forgot anybody else, thank you to everyone who helped me.
\\ \\ \\
Hantian Zhang\\
Harbin, China, June 2014.
}
\clearpage 


\pagestyle{fancy} 

\lhead{\emph{Contents}} 
\tableofcontents 

\setstretch{1.3} 

\pagestyle{empty} 

\dedicatory{Dedicated to my parents, who encouraged me during my overseas and domestic times with their love and supports.} 

\addtocontents{toc}{\vspace{2em}} 


\mainmatter 

\pagestyle{fancy} 



\chapter{Introduction} 

\label{Chapter1} 

\lhead{Chapter 1. \emph{Introduction}} 


The non-smooth numerical approaches of the multi-body and multi-contact dynamics and control is a state-of-art field of investigation \cite{Aca08, brogliato2000} in recent decades, especially in space engineering and robotics. It is actively involved in various applications, such as control of space manipulators \cite{VanWoerkom95}, landing gear, dynamics of sand piles and planetary rings \cite{nikolai96, farhang96}, and walking robots \cite{Hurmuzlu2004}, where the frictions and impacts resulted from the contact constraints cannot be neglected.
\par
The problems of contacts and dry friction, along with the impacts, regarded as the inequality problems, can be treated with the tools from convex analysis described by the convex cones instead of linear spaces. The friction, contact force and impact are not expressed by classical force laws any more, but by a set-valued force law, which has been discussed in detail by Glocker in his book \cite{Glo01}. Glocker extend the set-valued interactions for non-smooth dynamics to derive the forces from scalar potential functions, of which the practical application on spatial Coulomb friction is presented by Leine and him \cite{Lei03}. The differential equations of non-smooth systems is therefore transformed into the differential inclusions with discontinuous right-hand side. This particular discontinuous dynamical system called Filippov-type has been studied by Leine in his Ph.D. thesis \cite{Leine00} with the emphasis on bifurcations,  and the book \cite{Lei06} co-authored with Nijmeijer. Leine and Wouw \cite{Lei08} also investigate the stability and convergence trying to extend the Lyapunov stability framework to non-smooth mechanical systems.
\par
On the aspect of numerical simulation of non-smooth dynamics,  Brogliato et al. \cite{Bro02} wrote an excellent review article introducing the main techniques, mathematical tools, and existing algorithms up to then. The powerful augmented Lagrangian method from convex analysis called Prox function has been firstly proposed by Alart and Curnier \cite{Alart1991} to formulate the friction and contact problems instead of the well-known linear and nonlinear complementarity formulations. Schindler et al. \cite{Sch11a} provided a detailed discussion and comparison on these two formulations along with their algorithm schemes for solving contact impacts and forces. For the simulation algorithms of motions, Moreau \cite{Mor99} classified them into three types, the Event-driven schemes, the Time-stepping schemes, and the Penalized-constraint schemes. In his paper \cite{Mor99}, Moreau introduced his time-stepping schemes, the sweeping process which is known as the Moreau's time-stepping mid-point rule by now, using convex analysis with prox function. The resulting contact problems are solved cyclically independent of each other by iteration. Schoeder, Ulbrich and Schindler \cite{Sch12d} proposed a time-stepping scheme based on Moreau's mid-point rule, prox formulation, and the Gear-Gupta-Leimkuhler method. It maintains the physical consistency of impulsive discretization and satisfies the non-penetration constraints on both velocity and position levels.
\par
With the existence of uncertain friction, the motion controlled system is suffered from the undesired effects such as stick-slip limit cycling and non-zero steady-state errors. Armstrong-H\'{e}louvry et al.\cite{Armstrong1994} reviewed the control strategies for frictional systems, such as PID feedback control, frictional compensation method, and adaptive control. However, they all suffered the drawbacks stated before. Wouw and Leine \cite{Wou12} proposed a robust impulsive feedback control strategy, which guarantees the robust stability of the set-point system in facing of frictional uncertainties. The main idea of the impulsive control is to apply the estimated control impulse to enforce the controlled target to escape from the sticking status to the desired status directly, when the system gets stuck at a non-zero steady-state error.
\par
Following the idea of impulsive control, Wang \cite{wang13} attempted to stabilize the pre-existing frictional Furuta pendulum to the unstable equilibrium (inverted) position, but failed. The control problem of Furuta pendulum, as a benchmark of under-actuated mechanical system, is first proposed by Furuta et al. \cite{Furuta92} with a swing-up control strategy. It is a two-degree freedom pendulum with an actuated driving arm and an under-actuated pendulum arm. Cazzolato and Prime \cite{caz2011pendulum} presented a clear and completed formulation for the whole mechanical system of Furuta pendulum without friction. Acosta \cite{Acosta10} concluded the nonlinear control methods for the Furuta pendulum, including energy shaping method, forwarding method, input-output feedback linearization method, and singular perturbation method, but merely considered the friction as an approximate smoothing function or regardless of friction. Chang mathematically developed the energy shaping method \cite{chang10lagrangians} with dissipation, which provide Lyapunov stabilizability via matching conditions of PDEs, and extend it into the fields of one degree of underactuation control \cite{chang10one_under} and two degrees of underactuation control \cite{chang13two_under}, along with the specific application on the Furuta pendulum \cite{chang13application}.
\par
However, taking the frictions within the joints into account, the pendulum will stick on the undesired positions outside the inverted position, where the impulsive control is necessary to be applied. Therefore, in this thesis, we develop the reliable non-smooth numerical schemes for the contacted motion system and impulse-based control to the frictional Furuta pendulum.

\section{Thesis Overview}
In Chapter \ref{Chapter2}, we present the main mathematical tools in non-smooth mechanics to illustrate how the contact dynamics problems can be solved numerically in a reliable way. The mathematical descriptions of contact constraints and Coulomb's friction law, along with the complementarity and Prox formulations are presented. The impact laws and numerical integration schemes are introduced. Particularly, the proposed time-stepping scheme based on Moreau's mid-point rule and prox function is briefly reviewed.
\par
In Chapter \ref{Chapter3}, we apply the impulse-based control law to the frictional oscillator employing the simulation methods presented in Chapter \ref{Chapter2}. The complementarity conditions of the motion of the oscillator are analyzed, and the numerical method for the simulation of oscillator free motion on velocity-impulse level is implemented. The impulsive feedback control law, impulse estimating methods, and their simulating results are provided.
\par
In Chapter \ref{Chapter4}, we derive the simplified equation of motion of Furuta pendulum from Lagrangian equations with frictional moment taken into account. The normal contact forces and frictional moments are formulated under the assumption of rigid body and uniform pressure provided by the contact surface of the joints. The proposed numerical method is applied and validated on the frictional coupled system on its free motion.
\par
In Chapter \ref{Chapter5}, we provide the general form of impulse-based control law for the frictional Furuta pendulum, and reduced it into the specific form with shooting method employed for the impulse estimating process. The simulation results and comparison are presented under the circumstances that the feedback and impulsive control torques are switched on or off.



\chapter{Mathematical Theory of Non-Smooth Mechanics} 

\label{Chapter2} 

\lhead{Chapter 2. \emph{Prox and Complementarity Formulations of Non-Smooth Mechanics}} 


\section{Description of Contact Systems}
The Prox and complementarity formulations are two powerful tools in describing the contact system both through set-based arguments. Basically, these two formulations are equivalent, but can be solved by different algorithm schemes. Here we provide a brief review on the two methods for the description of contact system, the detailed discussion has been presented by Schindler \cite{Sch11a}.
\subsection{Contact and friction law}
\begin{figure}[hbt]
\centering
 \begin{subfigmatrix}{2}
  \subfigure	\centering
	\def\svgwidth{0.4\textwidth}
	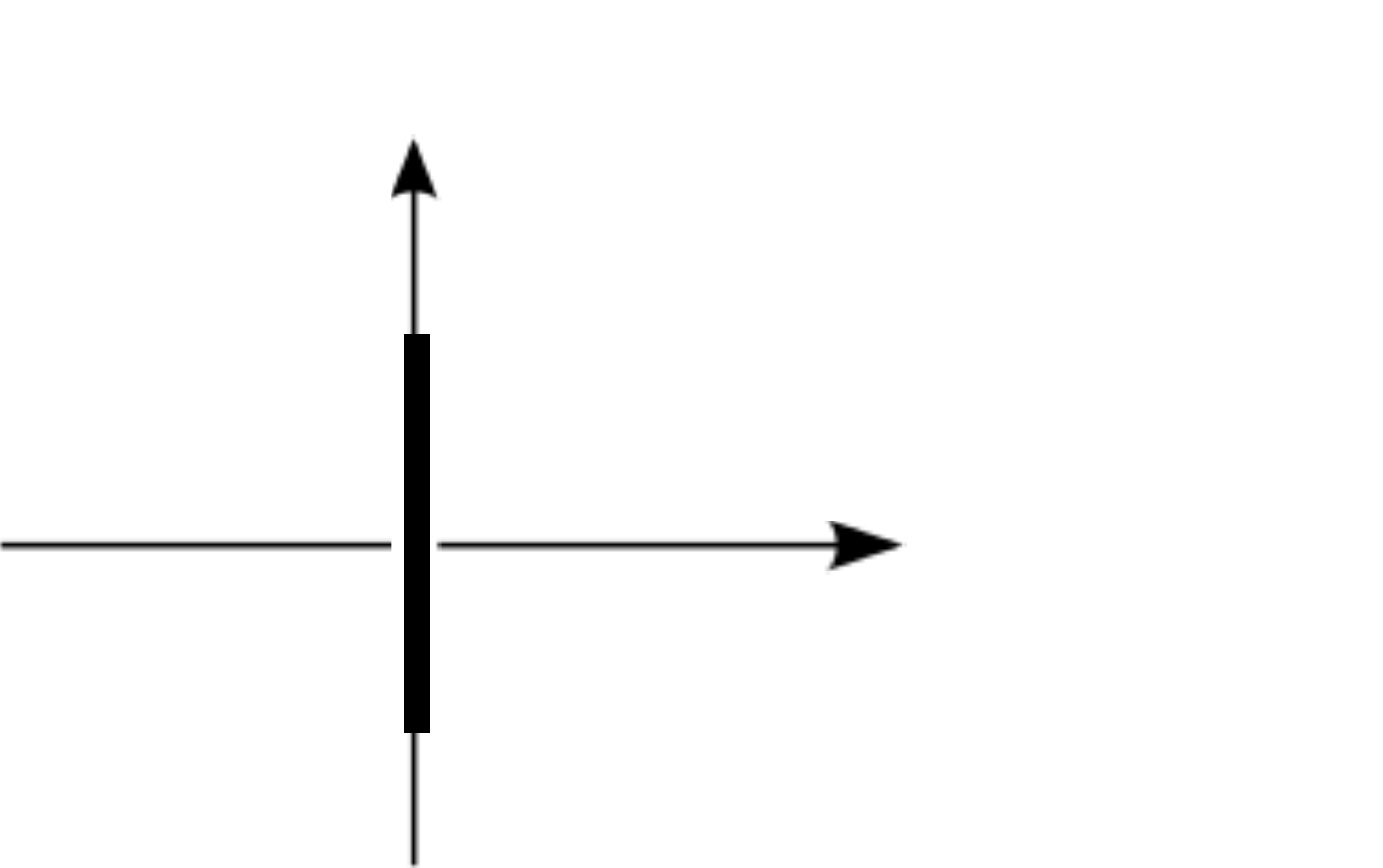
  \subfigure	\centering
	\def\svgwidth{0.4\textwidth}
	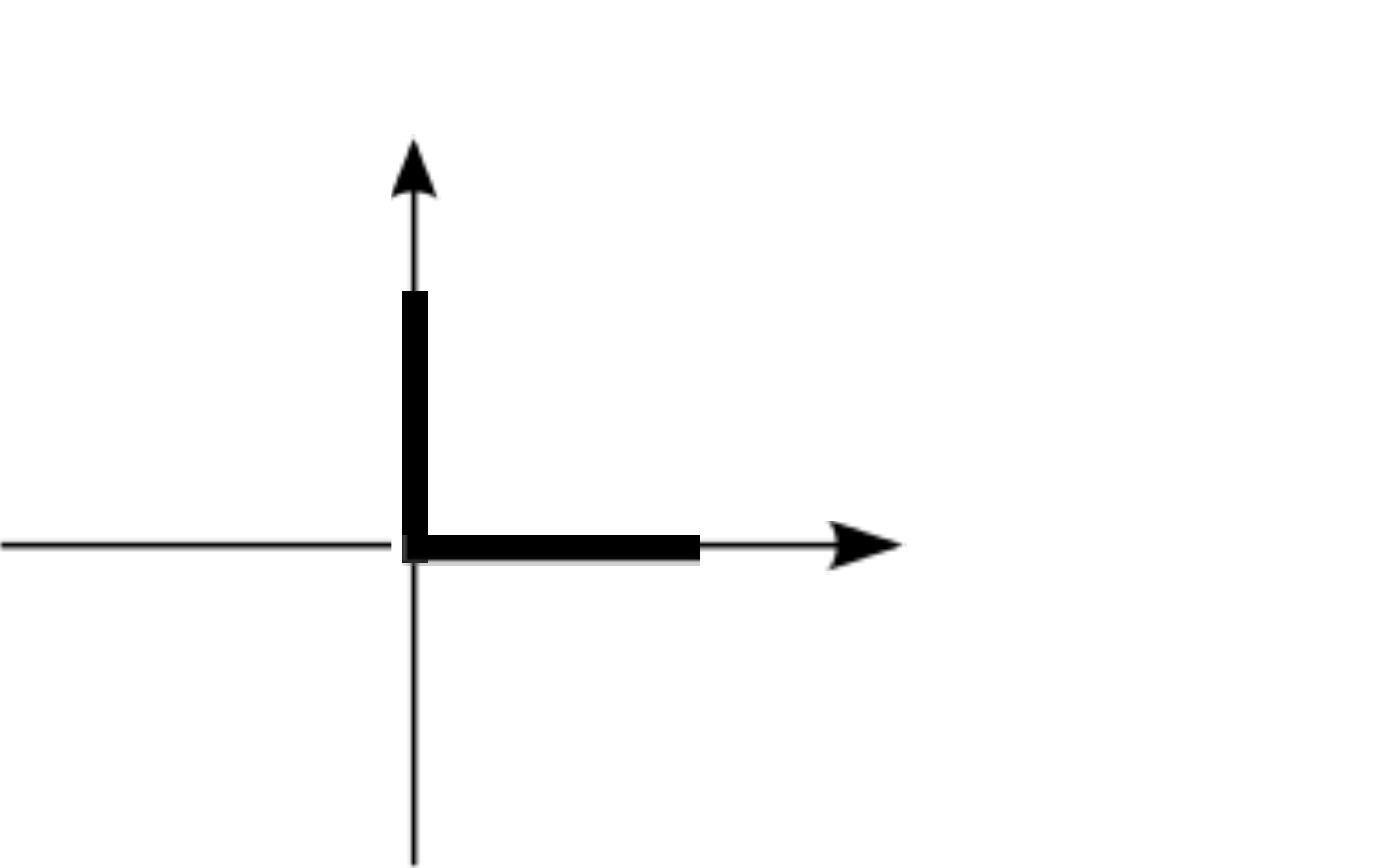
 \end{subfigmatrix}
 \caption{Normal force laws for contacts \\ (Left: Bilateral constraint. Right: Unilateral constraint)}
 \label{fig:contact}
\end{figure}

The contacts can be classified into two kinds, the bilateral contact and the unilateral contact. The normal force laws of these two kinds of constraints are shown in Fig. \ref{fig:contact} respectively.
In this figure, $g_B, g_U$ are the gap functions that denote the distance on the normal direction between two contact surface, and $\lambda_B, \lambda_U$ denote the normal contact forces.
\par
\begin{figure}[hbt] 
	\centering
	\def\svgwidth{0.5\textwidth}
	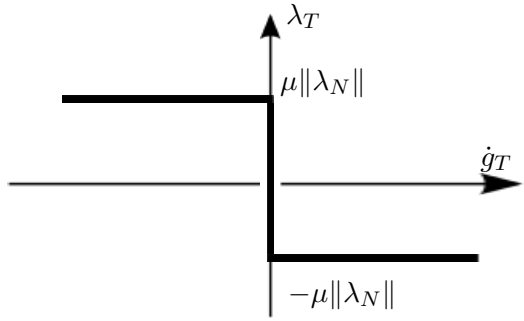
  \caption{Coulomb friction law.}
  \label{fig:friction}
\end{figure}

For $\lambda_N \in \{ \lambda_B, \lambda_U \}$, the tangential friction force can be represented by Coulomb friction law
\begin{align*}
\lambda_{T}\left\{ \begin{array}{lll}
= -\mu \| \lambda_{N} \| & \text{if} \; \dot g_T > 0 \\ 
\in [-\mu \| \lambda_{N} \|, \mu \| \lambda_{N} \|] & \text{if} \; \dot g_T = 0 \\
=\mu \| \lambda_{N} \| & \text{if} \; \dot g_T < 0
\end{array} \right. \;,
\end{align*}
where $\dot g_T$ indicates the relative velocity which is tangential to the contact surface. The friction law is depicted in Fig. \ref{fig:friction}.
The vertical line at the point of $\dot{g}_T=0$ represents the real physics called set-valuedness, which reveals that the force within this set is uncertain. Any available friction force may occur to compensate the motion system but without any changes in the relative velocity.

\subsection{Complementarity formulation}
In complementarity formulation, the bilateral constraint can be represented by the set
\begin{align}
\label{comple_bi}
B_C:=\{ (g_B, \lambda_B) \in \mathbb R \times \mathbb R \mid g_B=0 \}\;,
\end{align}
the unilateral constraint is represented by the set
\begin{align}
\label{comple_uni}
U_C:=\{ (g_U, \lambda_U) \in \mathbb R \times \mathbb R \mid 0 \leq g_U \perp \lambda_U \geq 0 \}\;,
\end{align}
where the symbol $\perp$ indicates the orthogonality, i.e. $g_U \lambda_U=0$. The conditions of the contact are expressed at configuration (position) level in \eqref{comple_bi}-\eqref{comple_uni}. Furthermore, when the contact on the configuration level is closed, the hidden constraints on velocity and acceleration levels can be formulated by replacing $(g_B=0, g_U=0)$ with $(\dot{g}_B, \dot g_U)$, and replacing $(\dot{g}_B=0, \dot g_U=0)$ with $(\ddot{g}_B, \ddot g_U)$. For example, the formulation of unilateral constraint on the velocity level  is given by
\[
U_{Cv}:=\{ (\dot g_U, \lambda_U) \in \mathbb R \times \mathbb R \mid 0 \leq \dot{g}_U \perp \lambda_U \geq 0, g_U=0 \}\;.
\]
\par
Indicating $\lambda_U \in \{ \lambda_B, \lambda_U \}$, the tangential Coulomb friction is formulated by
\begin{align}
\begin{split}
T_{C\sigma}(\lambda_N):=\{ & (\dot{\bm g}_T, \bm \lambda_T) \in \mathbb R^2 \times \mathbb R^2, \sigma \in \mathbb R \mid \\
 & \bm 0 = \mu \lvert \lambda_N \rvert \dot{\bm g}_T + \sigma \bm \lambda_T, \\
 & 0 \leq \sigma \perp \mu \lvert \lambda_N \rvert - \| \bm \lambda_T \| \geq 0\} \; ,
\end{split}
\end{align}
which is equivalent to the set
\begin{align}
\begin{split}
T_C(\lambda_N):=\{ & (\dot{\bm g}_T, \bm \lambda_T) \in \mathbb R^2 \times \mathbb R^2 \mid \\
 & \dot{\bm g}_T=\bm 0 \Rightarrow \| \bm \lambda_T \| \leq \mu \lvert \lambda_N \rvert , \\
 & \dot{\bm g}_T \neq \bm 0 \Rightarrow \| \bm \lambda_T \| = - \frac{\dot{\bm g}_T}{\dot{\bm g}_T} \mu \lvert \lambda_N \rvert \}  \; .
\end{split}
\end{align}
If the contact experiences the sticking case $(\dot{\bm g}_T=0)$, then $\dot{\bm g}_T$ is replaced by $\ddot{\bm g}_T$ for the acceleration level to calculate forces.
The complementarity formulation is usually solved by the pivoting scheme.

\subsection{Prox formulation}
\begin{figure}[hbt] 
	\centering
	\def\svgwidth{0.35\textwidth}
	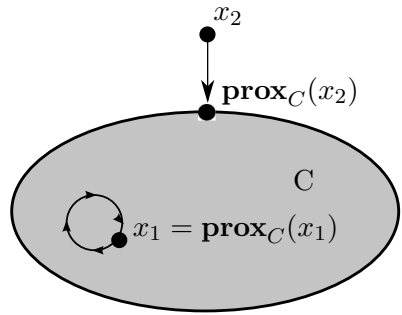
  \caption{Prox function in convex set.}
  \label{fig:prox_function}
\end{figure}

The Prox formulation is a more appropriate mathematical description for the contacts. Based on convex analysis, the Prox function can be obtained as a set-based argument. \\
For $ x\in \mathbb R^n$ and $n \in \mathbb N$, we define the proximal point to a given convex set $C \subset \mathbb R^n$:
\[
\mathbf{prox}_C (\mathbf x) = \underset{x^* \in C}{\operatorname{arg\;min\;}} \| \mathbf x - \mathbf x^* \|
\]
The proximal point is the point itself when it is in $C$, or the closet point on the boundary of $C$ otherwise. This projection process of Prox function is presented in Fig. \ref{fig:prox_function}.
\par
The contact conditions can be formulated as:\\
Bilateral Contact
\begin{align}
\begin{split}
B_P:=\{ & (g_B, \lambda_B) \in \mathbb R \times \mathbb R \mid \\
& f_B(\lambda_B, g_B) := \lambda_B - \text{prox}_{C_B}(\lambda_B - r \, g_B) =0 \} \; ,
\end{split}
\end{align}
where $r$ is the auxiliary parameter that $r \in \mathbb R^+$, and the corresponding convex set is
\[C_B=\mathbb R \; . \]
Unilateral Contact:
\begin{align}
\begin{split}
U_P:=\{ & (g_U, \lambda_U) \in \mathbb R \times \mathbb R \mid \\
& f_U(\lambda_U, g_U) := \lambda_U - \text{prox}_{C_U}(\lambda_U - r \, g_U) =0 \} \; ,
\end{split}
\end{align}
where the convex set is
\[C_U=\{ x \in \mathbb R \mid x \geq 0 \} \; . \]
Tangential Friction:
\begin{align}
\begin{split}
T_P(\lambda_N):= \{ & (\dot{\bm g}_T, \bm \lambda_T) \in \mathbb R^2 \times \mathbb R^2 \mid \\
& {f}_U(\bm \lambda_T, \dot{\bm g}_T) := \bm \lambda_T - \mathbf{prox}_{C_T(\lambda_N)}(\bm \lambda_T - r \, \dot{\bm g}_T) = \bm 0 \} \; ,
\end{split}
\end{align}
where the convex set is
\[C_{T} (y)=\{ \bm x \in \mathbb R^2 \mid \| \bm x \| \leq \mu \, | y | \, \} \; . \]
These formulation should be solved via a fixed point iteration scheme, which is influenced by the auxiliary parameter $r$ (See \cite{Sch11a}). The specific applications of Prox formulation, including numerical implementations and detailed algorithms, are presented in Chapter \ref{Chapter3} and Chapter \ref{Chapter4}.

\section{Non-smooth Numerical Treatment}
We propose a time stepping scheme based on Moreau's midpoint rule to satisfy the impact law and non-penetration constraints. With the Prox formulation, the nonsmooth equations are solved on velocity-impulse level to achieve the physical consistency on discretized integration interval.
\par
We consider the contact system with unilateral constraints as an illustration. The system is given by
\begin{align}
\label{system1}
&\dot{\bm q} = \bm v \; , \\ 
\label{system2}
&\mathbf M \dot{\bm v} = \bm h + \mathbf W \bm \lambda \; ,\\
\label{system3}
& \bm 0 \leq \bm g \perp \bm \lambda \geq \bm 0 \; ,
\end{align}
where $\bm q, \bm v$ are generalized positions and velocities, $\bm g$ is the gap function of the relative distance, $\mathbf M$ is the generalized mass matrix, $\bm h$ represents the generalized forces applied on the system, $\bm \lambda$ is the generalized contact force parameters, and $\mathbf W$ is the constraint matrix.

\subsection{Impact law and impulse force}
In the system \eqref{system1} - \eqref{system3}, the instantaneous impact will occur due to its rigidity and the existence of contact forces. In order to maintain the generalized positions $\bm q$ to be continuous functions of time to be physical consistent, the impact time has to be vanished to zero. However, in this case the generalized velocities $\bm v$ will jump and become discontinuous, thus the condition of \eqref{system2} will not hold any more, neither do the generalized accelerations $\dot{\bm v}$ and generalized contact force parameters $\bm \lambda$. Thus, the impact equations, instead of \eqref{system2} - \eqref{system3}, has to be proposed and solved
\begin{align}
& \mathbf M(\bm q (t_i)) \left( \bm v_i^+ - \bm v_i^- \right) = \mathbf W (\bm q (t_i)) \bm \Lambda_i \; , \\
& \text{if } \bm g \leq \bm 0, \quad \text{then } \bm 0 \leq \dot{\bm g}_i^+ + \varepsilon \dot{\bm g}_i^- \perp \bm \Lambda_i \geq 0 \; ,
\end{align}
where $\varepsilon$ is the kinematic coefficient of restitution, and the left-hand and right-hand limits of the generalized velocities are
\begin{align}
\label{impact1}
\bm v_i^+=\underset{t \downarrow t_i}{\operatorname{lim\;}} \bm v (t) \; , \\
\bm v_i^-=\underset{t \uparrow t_i}{\operatorname{lim\;}} \bm v (t) \; .
\label{impact2}
\end{align}
The impacts (or the finite impulses) are pseudo-instantaneous forces applied to the system, in the form of distribution we have
\begin{align}
\bm \Lambda_i = \underset{\delta \downarrow 0}{\operatorname{lim\;}} \int^{t_i}_{t_i - \delta} \bm \lambda \text{d} t \; .
\end{align}

\subsection{Event-driven Scheme}
The event-driven scheme integrate the non-impulsive system \eqref{system1}-\eqref{system3} by standard numerical integration schemes for continuous differential equations, at the meanwhile detecting the status of gap functions $\bm g$, closed contact or not. In the events that the slip-sticking transition captured (i.e. at least one contact closed, where the friction forces are compensate to the system, with $ h +  W  \lambda = 0,  \lvert v_i^- \rvert \leq \varepsilon$), the set-valued force law is evaluated and the impact equations are solved, which implies that the continuous solver will not activated until the next time instants when the sliding status are detected (i.e. $\lvert h +  W  \lambda \rvert > 0$).
\par
However, the event-driven scheme is not physical consistent resulting in the drawback that the existence of the zeno behavior, which implies that numerically it will provide solutions with infinite number of impacts or jumps in a finite time interval. With the occurrence of Zeno behavior, the numerical integration will never reach the end of the specific discretized time interval, but operating on the interior points within the interval. (i.e. it will never converge to the zero velocity, but merely oscillated around the zero.)

\subsection{Time-stepping Scheme based on Moreau's midpoint rule}
The time-stepping scheme discretize the equations of motion into the difference equations on velocity level in a physically consistent manner without the occurrence of Zeno behavior. We propose a time stepping scheme based on Moreau's midpoint rule and Prox function.
\par
We discretize the time interval $ I$ into intervals $ I_n = [t_n, t_{n+1}]$ by the fixed time step $\Delta t$. Then the numerical model of the system is given by
\begin{align}
& \bm q_{n+1}= \bm q_n + \frac{\bm v_{n+1} - \bm v_n}{2} \Delta t \; , \\
& \bm v_{n+1} = \bm v_n + \mathbf M^{-1}_{M_n} \left( \bm h_{M_n} \Delta t + \mathbf W_{M_n} \bm \Lambda_{n+1} \right) \; ,
\end{align}
with
\begin{align}
\mathbf M_{M_n} = \mathbf M \left( \bm q_n + \frac{\Delta t}{2} \bm v_n\right), \quad
\bm h_{M_n}= \bm h \left( \bm q_n + \frac{\Delta t}{2} \bm v_n, \bm v_n \right), \quad
\mathbf W_{M_n}= \mathbf W  \left( \bm q_n + \frac{\Delta t}{2} \bm v_n\right) \; .
\end{align}
It is noticed that the impulses $\bm \Lambda_{n+1}$ are evaluated at the end of the discretized time interval $I_n$, which is of the implicit form. Thus, the impulses have to be solved implicitly by Newton's impact law on velocity level. \\
The Prox function with convex set $C$ is formulated for this purpose by
\begin{align}
\label{prox_equation}
f(\bm \lambda_{n+1}, \dot{\bm g}_{n+1} ) = \bm \Lambda_{n+1} -\mathbf{prox}_C(\bm \Lambda_{n+1} - r \dot{\bm g}_{n+1}) = 0 \; ,
\end{align}
with
\begin{align}
& \dot{\bm g}_n =\mathbf W^T(\bm q_n) \bm v_n \; , \\
& \dot{\bm g}_{n+1}=\dot{\bm g}_n + \mathbf W^T_{M_n} \mathbf M^{-1}_{M_n} \left( \bm h_{M_n} \Delta t + \mathbf W_{M_n} \bm \Lambda_{n+1} \right) \; .
\end{align}
The Prox equation \eqref{prox_equation} is separated into two arguments within the iteration process by
\begin{align}
\label{prox_argument}
	f(\bm \Lambda_{n+1}, \dot{\bm g}_{n+1}) &= \begin{cases}r \dot{\bm g}_{n+1} = 0, & \text{if } \bm \Lambda_{n+1} - r \dot{\bm g}_{n+1} \in C \\ \bm \Lambda_{n+1} - \mathbf{b}_C=0,  & \text{if } \bm \Lambda_{n+1} - r \dot{\bm g}_{n+1} \notin C \end{cases} \; ,
\end{align}
where $\mathbf b_C$ is the proximal point on the boundary of the set $C$. Inserting the approximate relative velocity $\dot {\bm g}_{n+1}$ at the end of the time interval into the equation \eqref{prox_argument}, the resulting nonsmooth equations can be solved iteratively. Thus, the implicit impulses $\bm \Lambda$ at the end of the discretized time interval are obtained in a physically consistent sense.



\chapter{Impulse-based Control of Frictional Oscillator} 

\label{Chapter3} 

\lhead{Chapter 3. \emph{Impulse-based Control of Frictional Oscillator}} 


\section{Motion of a Planar Frictional Oscillator}
\begin{figure}[hbt] 
	\centering
	\def\svgwidth{0.6\textwidth}
	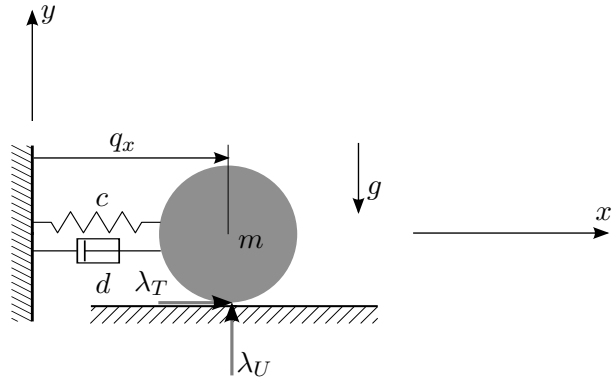
  \caption{Frictional oscillator from \cite{zhang14_1}.}
  \label{fig:oscillator}
\end{figure}

To mathematically describe the motion of a two-dimensional frictional oscillator $m$ on a plane with spring $c$ and damping $d$ effect~(Fig.~\ref{fig:oscillator}),
we use the following equations of motion with the gravitational acceleration $g>0$:
\begin{align}
  \dot{\mathbf q} &= \begin{pmatrix} \dot q_x \\ \dot q_y \end{pmatrix} = \begin{pmatrix} v_x \\ v_y \end{pmatrix} = \mathbf v \;, \label{eq:motion1}\\
  m \dot{\mathbf v} &= m\begin{pmatrix} \dot v_x \\ \dot v_y \end{pmatrix} = \mathbf{h} + \mathbf{W}\mathbf{\lambda} = \begin{pmatrix} -k_1 q_x -k_2 v_x \\ -mg \end{pmatrix} + \begin{pmatrix}0 & 1\\1 & 0\end{pmatrix}\begin{pmatrix} \lambda_U \\ \lambda_T \end{pmatrix} \;. \label{eq:motion2}
\end{align}
The additional constraint equations for the normal $\lambda_U$ and tangential $\lambda_T$ contact satisfy
\begin{align}
  & 0\leq q_y \perp \lambda_U \geq 0 \;, \label{eq:contact_normal}\\
  & \lambda_T = -\frac{v_x}{\| v_x \|}\mu |\lambda_U| \quad \text{for } v_x \neq 0 \;, \label{eq:contact_tangential_sliding}\\
  & \|\lambda_T\| \leq \mu |\lambda_U| \quad \text{for } v_x = 0 \label{eq:contact_tangential_sticking} 
\end{align}
whereby $q_y \perp \lambda_U$ means complementarity, i.e. $q_y\lambda_U=0$. The initial condition is given by $\begin{pmatrix}q_x & q_y\end{pmatrix}^T(t_0) = \begin{pmatrix}q_{x_0} & 0\end{pmatrix}^T$ and $\begin{pmatrix}v_x & v_y\end{pmatrix}^T(t_0) = \begin{pmatrix}v_{x_0} & 0\end{pmatrix}^T$.\par
For this example, normal and tangential direction are decoupled concerning the contact plane; focusing on the normal direction, we get
\begin{align}
  & m\dot{v}_y = -mg + \lambda_U \;,\\
  & 0\leq q_y \perp \lambda_U \geq 0 \;.
\end{align}
Hence, we can evaluate the initial and also the subsequent setting. The original constraint is satisfied ($q_y(t_0)=0$) but as it has to be valid for all times, we also have to check the so called hidden constraints on velocity and acceleration level:
\begin{align}
  & 0\leq v_y \perp \lambda_U \geq 0 \;,\\
  & 0\leq \dot{v}_y \perp \lambda_U \geq 0 \;.
\end{align}
Again, the first constraint is satisfied ($v_y(t_0)=0$) but we get an additional condition due to the second constraint by inserting the equation of motion:
\begin{align}
  0\leq -g + \frac{\lambda_U}{m} \perp \lambda_U \geq 0 \;.
\end{align}
If $\lambda_U$ was vanishing, we would get the contradiction $g\leq 0$. Hence $\lambda_U>0$ and by complementarity $\lambda_U=mg$. There will be no acceleration $\dot{v}_y$ and so also velocity $v_y=0$ and position $q_y=0$ will stay at their initial condition for all times.\par
We can reduce this example to a one-dimensional setting just focusing on the tangential direction but inserting $\lambda_U=mg$:
\begin{align}
\label{tan_motion1}
  \dot{q}_x &= v_x \;, \\
  m\dot{v}_x &= -k_1 q_x -k_2 v_x + \lambda_T
  \label{tan_motion2}
\end{align}
with
\begin{align}
  & \lambda_T = -\frac{v_x}{\| v_x \|}\mu mg \quad \text{for } v_x \neq 0 \;, \\
  & \|\lambda_T\| \leq \mu mg \quad \text{for } v_x = 0 \;.
\end{align}
Again the example simplifies in comparison to the general framework. Evaluating the current state $\begin{pmatrix}q_x & v_x\end{pmatrix}^T$ gives us an explicit idea of the state change $\begin{pmatrix}\dot{q}_x & \dot{v}_x\end{pmatrix}^T$ in case of $v_x \neq 0$.\par
In case of $v_x = 0$, we have to evaluate the still implicit formulation
\begin{align}
  \dot{q}_x &= 0 \;, \\
  m\dot{v}_x &= -k_1 q_x + \lambda_T  
\end{align}
with
\begin{align}
  & \|\lambda_T\| \leq \mu mg \quad \text{for } v_x = 0 \;.
\end{align}
For getting an explicit expression, we introduce the hidden constraint on acceleration level
\begin{align}
  & \lambda_T = -\frac{\dot{v}_x}{\| \dot{v}_x \|}\mu mg \quad \text{for } \dot{v}_x \neq 0 \;, \\
  & \|\lambda_T\| \leq \mu mg \quad \text{for } \dot{v}_x = 0 \;.
\end{align}
If $\dot{v}_x = 0$, we get $\lambda_T = k_1 q_x$ by the equation of motion and $\|\lambda_T\| \leq \mu mg$ by the constraint. So on the one hand necessarily $\|k_1 q_x\| \leq \mu mg$. On the other hand $v_x = 0$ and $\|k_1 q_x\| \leq \mu mg$ is also sufficient for $\dot{v}_x = 0$. If $\dot{v}_x \neq 0$, indeed we get
\begin{align}
  \lambda_T = m\dot{v}_x+k_1q_x \quad \wedge \quad \lambda_T = -\frac{\dot{v}_x}{\| \dot{v}_x \|}\mu mg
\end{align}
and so $\|k_1 q_x\| > \mu mg$ for the consistent case of $\text{sgn}(\dot{v}_x)$. Together with $v_x = 0$, the condition $\|k_1 q_x\| > \mu mg$ is also sufficient for $\dot{v}_x \neq 0$.\par
Summarizing the explicit formulation of the case $v_x = 0$, we get
\begin{align}
  &\text{if }\|k_1 q_x\| \leq \mu mg\text{ then }\dot{v}_x = 0 \;, \\
  &\text{if }\|k_1 q_x\| > \mu mg\text{ then }m\dot{v}_x = -k_1q_x-\frac{\dot{v}_x}{\| \dot{v}_x \|}\mu mg\;.
\end{align}
The last equation can be solved by trying the two possibilities for $\text{sgn}(\dot{v}_x)$, i.e. left or right sliding.

\section{Numerical Implementation on Velocity-Impulse Level}

Moreau's midpoint rule is employed to numerically solve the differential equation of the frictional oscillator. Although the tangential direction would be sufficient, we characterize the general framework:
\begin{align}
  & \mathbf v^{(i+1)} = \begin{pmatrix}  v_x^{(i)} \\  v_y^{(i)} \end{pmatrix}+\frac{1}{m}\left[ \begin{pmatrix} h_x(q^{(i)}_x+\frac{\Delta t}{2}v^{(i)}_x,v^{(i)}_x) \\  h_y(q^{(i)}_y+\frac{\Delta t}{2}v^{(i)}_y,v^{(i)}_y) \end{pmatrix} \Delta t + \begin{pmatrix} 0 & 1 \\ 1 & 0 \end{pmatrix} \begin{pmatrix} \Lambda_{U}^{(i+1)} \\ \Lambda_{T}^{(i+1)} \end{pmatrix} \right] \;, \\
  & \mathbf q^{(i+1)} = \mathbf q^{(i)}+\frac{\mathbf v^{(i+1)}+ \mathbf v^{(i)}}{2}{\Delta t}.
\end{align}
with the normal impulse $\Lambda_{U}^{(i+1)}$ 
\begin{align}
  \label{normal1}
  & \Lambda_U^{(i+1)} = mg \Delta t  \quad \text{for } q^{(i+1),\text{pred}}_y \leq 0 \, \wedge \, v^{(i+1)}_y\leq 0 \;, \\
  & \Lambda_U^{(i+1)} = 0 \quad \text{for } q^{(i+1),\text{pred}}_y > 0 \, \vee \, \left(q^{(i+1),\text{pred}}_y\leq 0 \, \wedge \, v^{(i+1)}_y>0\right) \;,
  \label{normal2}
\end{align}
and the tangential frictional impulse $\Lambda_{T}^{(i+1)}$ (if the respective normal impulse is active, i.e. $q^{(i+1),\text{pred}}_y \leq 0$) 
\begin{align}
  \label{tangential1}
  & \Lambda_T^{(i+1)} = -\frac{v^{(i+1)}_x}{\| v^{(i+1)}_x \|}\mu mg \Delta t \quad \text{for } v^{(i+1)}_x \neq 0 \;, \\
  & \|\Lambda_T^{(i+1)}\| \leq \mu mg \Delta t \quad \text{for } v^{(i+1)}_x = 0 \;.
  \label{tangential2}
\end{align}
The prediction is defined by
\begin{align}
  q^{(i+1),\text{pred}}_y = q^{(i)}_y + \gamma v^{(i)}_y \Delta t
\end{align}
with $0\leq\gamma\leq 1$, usually $\gamma=\frac{1}{2}$. Obviously a penetration is allowed with this classic scheme, however it is possible to avoid it by adding the unilateral constraint on position level with an extra Lagrange multiplier~\cite{Sch12d}. For an implementation, the active normal equations \eqref{normal1}-\eqref{normal2} and active tangential equations \eqref{tangential1}-\eqref{tangential2} can be written as
\begin{align}
  f_{U}(\Lambda_U^{(i+1)}, v^{(i+1)}_y) &:= \Lambda_U^{(i+1)} - \text{prox}_{C_U}(\Lambda_U^{(i+1)}-r_U v^{(i+1)}_y) = 0 \; , \\
  f_T(\Lambda_T^{(i+1)},v^{(i+1)}_x) &:= \Lambda_T^{(i+1)} - \text{prox}_{C_T(\mu m g \Delta t)}(\Lambda_T^{(i+1)}-r_T v^{(i+1)}_x) = 0\; ,
\end{align}
with the proximal point
\begin{align}
 & \text{prox}_{C_U}(x)=\arg\min_{x^*\in C_U}\|x-x^*\| \; , \\
 & \text{prox}_{C_T(\mu m g \Delta t)}(x)=\arg\min_{x^*\in C_T(\mu m g \Delta t)}\|x-x^*\| \;,
\end{align}
arbitrary parameters $r_U>0 \, , r_T>0$, the convex cone and the friction disc
\begin{align}
&  C_U = \{x\in\mathbb{R}\;|\; x \geq 0 \} \;, \\
&  C_T(\mu m g \Delta t) = S^0_{\mu m g \Delta t} = \{x\in\mathbb{R}\;|\;\|x\|\leq \mu m g \Delta t\} \;.
\end{align}
The functions $f_U$ and $f_T$ can be both evaluated by distinguishing different cases
\begin{align}
	f_U(\Lambda_U^{(i+1)},v^{(i+1)}_y) &= \begin{cases}r_U v^{(i+1)}_y & \text{if } \Lambda_U^{(i+1)}-r_U v^{(i+1)}_y \geq 0 \\ \Lambda_U^{(i+1)} & \text{if } \Lambda_U^{(i+1)}-r_U v^{(i+1)}_y < 0 \end{cases} \;, \\
        f_T(\Lambda_T^{(i+1)},v^{(i+1)}_x) &= \begin{cases}r_T v^{(i+1)}_x & \text{if }\|\Lambda_T^{(i+1)}-r_T v^{(i+1)}_x\|\leq\mu m g \Delta t\\ \Lambda_T^{(i+1)}-\frac{\Lambda_T^{(i+1)}-r_T v^{(i+1)}_x}{\|\Lambda_T^{(i+1)}-r_T v^{(i+1)}_x\|}\mu m g \Delta t & \text{if }\|\Lambda_T^{(i+1)}-r_T v^{(i+1)}_x\|>\mu m g \Delta t \end{cases} \;.
\end{align}
Typically, one inserts the velocity formulas $v^{(i+1)}_y$, $v^{(i+1)}_x$ into $f_U, f_T$ and gets nonsmooth and nonlinear equations for $\Lambda_U^{(i+1)}, \Lambda_T^{(i+1)}$:
\begin{align}
  &	f_U \left(\Lambda_U^{(i+1)}, v^{(i)}_y+\frac{1}{m}\left[h_y(q^{(i)}_y+\frac{\Delta t}{2}v^{(i)}_y,v^{(i)}_y)\Delta t + \Lambda_{U}^{(i+1)}\right]\right) = 0 \;, \label{normal_nonsmooth} \\
  &	f_T \left(\Lambda_T^{(i+1)}, v^{(i)}_x+\frac{1}{m}\left[h_x(q^{(i)}_x+\frac{\Delta t}{2}v^{(i)}_x,v^{(i)}_x)\Delta t + \Lambda_{T}^{(i+1)}\right]\right) = 0 \;.
\label{tangential_nonsmooth}
\end{align}
These two equations can be solved by fixed-point or root-finding schemes. After that one calculates $v^{(i+1)}_x, v^{(i+1)}_y$ and $q^{(i+1)}_x, q^{(i+1)}_y$ by the discrete equation of motion~\cite{Sch11a}.\par
For the frictional oscillator, the normal impulse is trivial; for the tangential frictional impulse, we could simplify the numerical calculations in a similar way as for the analytical calculations. Therefore, we distinguish different cases:
\begin{itemize}
	\item $v^{(i+1)}_x=0$ yields the necessary condition
	\begin{align}
		& \mu mg \Delta t \geq \|-mv^{(i)}_x - h_x(q^{(i)}_x+\frac{\Delta t}{2}v^{(i)}_x,v^{(i)}_x)\Delta t\| \;. 
	\end{align}
	\item $v^{(i+1)}_x\neq 0$ yields the necessary condition
  \begin{align}
		& \mu mg \Delta t < \|-mv^{(i)}_x - h_x(q^{(i)}_x+\frac{\Delta t}{2}v^{(i)}_x,v^{(i)}_x)\Delta t\| 
	\end{align}
  for the consistent case of $\text{sgn}(v^{(i+1)}_x)$, i.e. left or right sliding.
\end{itemize}
The other way round, we have to try the two possibilities in the second case.
\begin{remark}
Physically, the Prox function determines the criteria of sticking occurrence, then force $V_x^{i+1}=0$. Thus, practically, within the complementary formulation, we can force  $V_x^{i+1}=0$ as well when the sticking occurrence detected, though the criteria is not so strict. 
In the complementary formulation, the $\Lambda ^{i}$ will be evaluated except the sticking case, but there is no difference from the $\Lambda ^{i+1}$ evaluated via the Prox formulation in most cases. Only when the slip-stick and left-right slip cases occur, the $\Lambda ^{i}$ will be one step falls behind.
\end{remark}
\subsection{Algorithm description}
A detailed algorithmic description of the fixed-point iteration scheme for solving the nonsmooth equations \eqref{normal_nonsmooth}-\eqref{tangential_nonsmooth} is presented in Table~\ref{algo_nor} and Table~\ref{algo_tan}.
\begin{table}[hbt]
  \centering
  \caption{\label{algo_nor}Algorithm for $\Lambda_U$}
  \begin{tabular}{ l }
  \hline
    Evaluate active $\Lambda_U$ at every step \\ (with $\Delta t$, and $q^{(i)}_y, v^{(i)}_y, h_y$ the given constants) \\ [2pt] \hline
    Set tolerance $\varepsilon$, parameter $r_U$, and initial guess $\Lambda_{U(0)}^{(i+1)}=mg\Delta t$\\ \hline
    While Loop ($j$th iteration with $j_{max}$) \\ [2pt] \hline
    \quad\quad Calculate $v_{y(j)}^{(i+1)}=v^{(i)}_y+\frac{1}{m}\left[h_y(q^{(i)}_y+\frac{\Delta t}{2}v^{(i)}_y,v^{(i)}_y)\Delta t + \Lambda_{U(j)}^{(i+1)}\right]$ \\ [6pt] \hline
    \quad \quad if $\Lambda_{U(j)}^{(i+1)}-r v^{(i+1)}_{y(j)} \geq 0$ \\
    \quad \quad \quad \quad $\Lambda_{U(j+1)}^{(i+1)}= -m v^{(i)}_y - h_y^{(i)}\Delta t$ \\
    \quad \quad  else if $\Lambda_{U(j)}^{(i+1)}-r v^{(i+1)}_{y(j)} < 0$ \\
    \quad \quad \quad \quad $\Lambda_{U(j+1)}^{i+1}=0$ \\ [2pt] \hline
    \quad \quad Calculate the difference $\sigma=\| \Lambda_{U(j+1)}^{(i+1)} - \Lambda_{U(j)}^{(i+1)} \|$ \\ [2pt] \hline
    \quad \quad if $\sigma<\varepsilon$ \\
    \quad \quad \quad \quad Return $\Lambda_{U(j+1)}^{(i+1)}$ \\
    \quad \quad else if $j>j_{max}$ \\
    \quad \quad \quad \quad Return $\Lambda_{U(j+1)}^{(i+1)}$, ''Desired tolerance cannot be reached'' \\ [2pt] \hline
    \end{tabular}
\end{table}
\begin{table}
\centering
    \caption{\label{algo_tan}Algorithm for $\Lambda_T$}
    \begin{tabular}{ l }
    \hline
    Evaluate $\Lambda_T$ at every step \\ (with $\Delta t$, and $q^{(i)}_x, v^{(i)}_x, h_x$ the given constants) \\ [2pt] \hline
    Set tolerance $\varepsilon$, parameter $r_T$, and initial guess $\Lambda_{T(0)}^{(i+1)}=0$\\ \hline
    While Loop ($j$th iteration with $j_{max}$) \\ [2pt] \hline
    \quad\quad Calculate $v_{x(j)}^{(i+1)}=v^{(i)}_x+\frac{1}{m}\left[h_x(q^{(i)}_x+\frac{\Delta t}{2}v^{(i)}_x,v^{(i)}_x)\Delta t + \Lambda_{T(j)}^{(i+1)}\right]$ \\ [6pt] \hline
    \quad \quad if $\|\Lambda_{T(j)}^{(i+1)} - rv^{(i+1)}_{x(j)}\| \leq \mu m g \Delta t$  \\
    \quad \quad \quad \quad $\Lambda_{T(j+1)}^{i+1}=-m v^{(i)}_x - h_x^{(i)} \Delta t$ \\
    \quad \quad  else $\|\Lambda_{T(j)}^{(i+1)} - rv^{(i+1)}_{x(j)} \| > \mu m g \Delta t$ \\
    \quad \quad \quad \quad $\Lambda_{T(j+1)}^{i+1}=\frac{\Lambda_{T(j)}^{(i+1)}-rv^{(i+1)}_{x(j)}}{\| \Lambda_{T(j)}^{(i+1)} - r v^{(i+1)}_{x(j)}\|}\mu \Lambda_U$ \\ [2pt] \hline
    \quad \quad Calculate the difference $\sigma=\| \Lambda_{T(j+1)}^{(i+1)} - \Lambda_{T(j)}^{(i+1)} \|$ \\ [2pt] \hline
    \quad \quad if $\sigma<\varepsilon$ \\
    \quad \quad \quad \quad Return $\Lambda_{T(j+1)}^{(i+1)}$ \\
    \quad \quad else if $j>j_{max}$ \\
    \quad \quad \quad \quad Return $\Lambda_{T(j+1)}^{(i+1)}$, ''Desired tolerance cannot be reached'' \\ [2pt] \hline
    \end{tabular}
\end{table}
    
\subsection{Simulation results}
\begin{figure}
  \centering
  \includegraphics[width=0.85\linewidth]{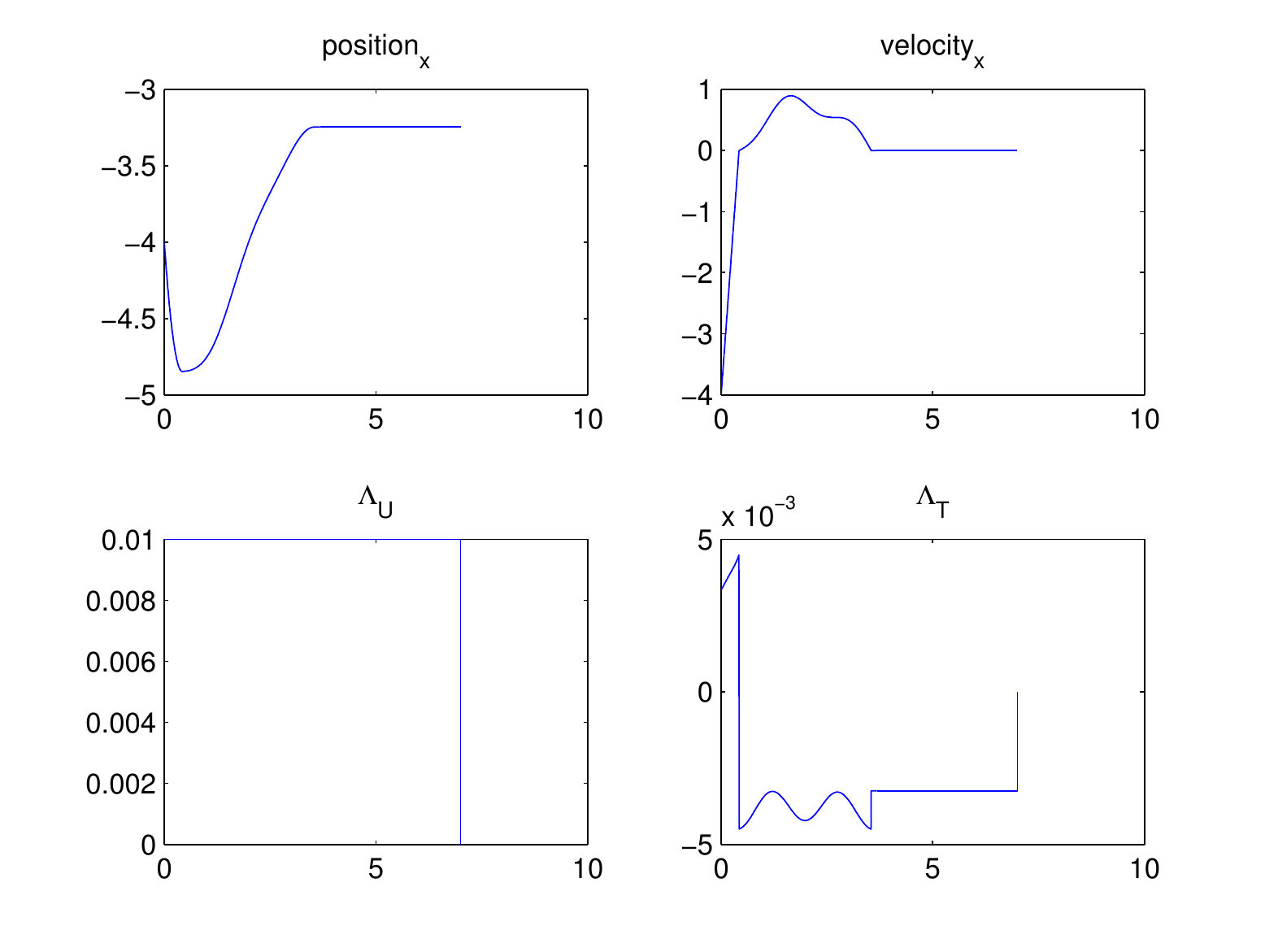}
  \caption{Simulation of oscillator free motion via prox function from \cite{zhang14_1}.}
  \label{fig:simulate motion}
\end{figure}

\begin{figure}
  \centering
  \includegraphics[width=0.85\linewidth]{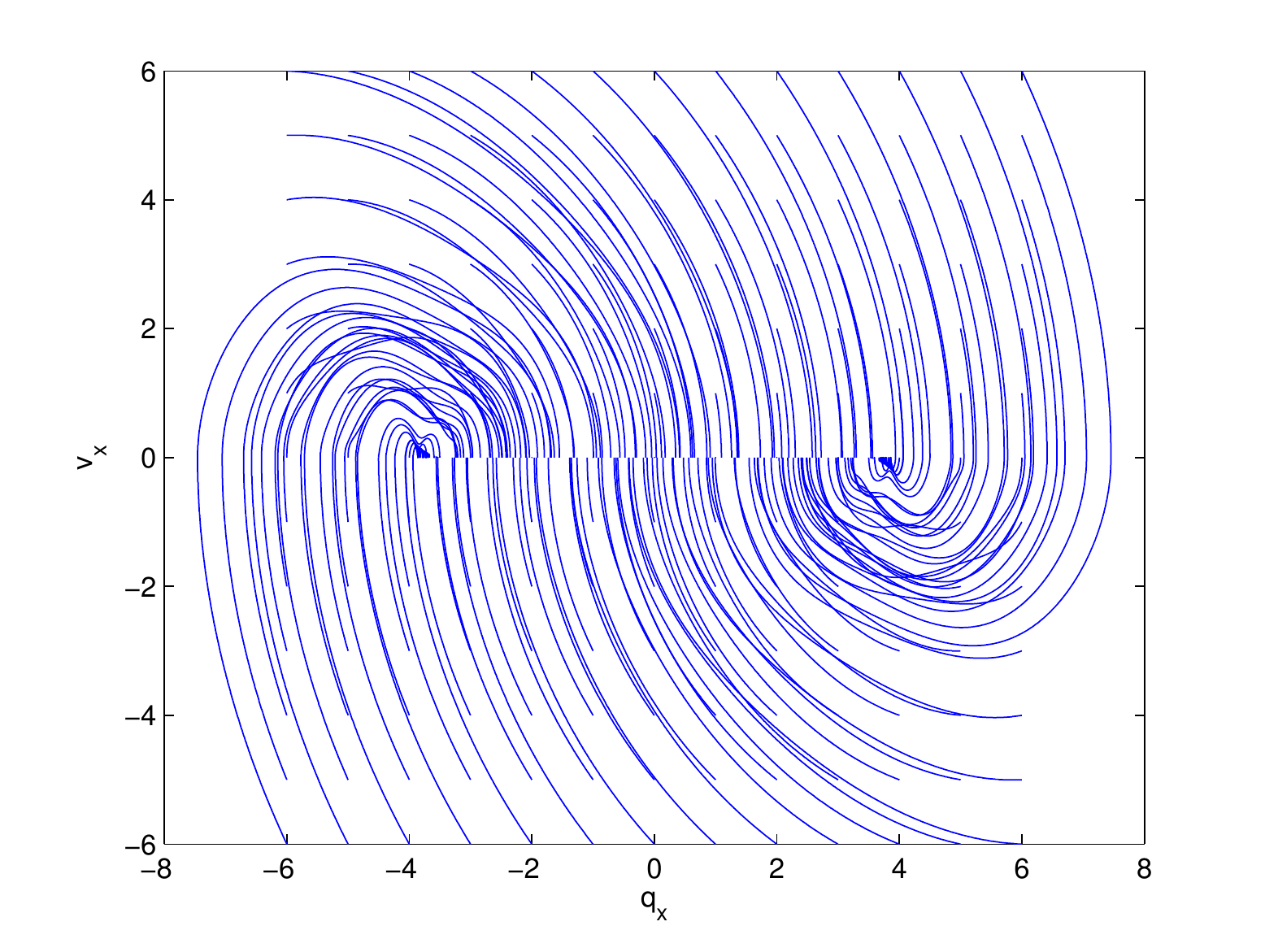}
  \caption{Phase diagram of oscillator behavior via prox function from \cite{zhang14_1}.}
  \label{fig:orbit diagram of motion}
\end{figure}

The example from Wouw and Leine \cite{Wou12} is reproduced for validating our numerical algorithms. For the dynamic equations \eqref{eq:motion1}-\eqref{eq:motion2} of the frictional oscillator, we have the mass $m=\unit[1]{kg}$, the gravitational acceleration $g=\unitfrac[10]{m}{s^2}$, the spring parameter $k_1=\unitfrac[1]{N}{m}$ and the damping parameter $k_2=\unitfrac[0.5]{Ns}{m}$. The friction coefficient is of the form
\[
\mu (q_x,v_x,t) = \frac{\mu_1 - \mu_2}{1+\unitfrac[0.5]{s}{m} \| v_x \|} + \mu_2 + \mu_3 \sin (\Omega t) ,
\]
where $\mu_1=0.4 , \mu_2=0.3 , \mu_3 = 0.05 , \Omega=\unitfrac[4]{1}{s}$. The initial conditions are given by
\[
  \mathbf q = \begin{pmatrix}  q_x \\  q_y \end{pmatrix} = \begin{pmatrix} -4 \\ 0 \end{pmatrix}\unit{m} \;, \quad \mathbf v = \begin{pmatrix}  v_x \\  v_y \end{pmatrix} = \begin{pmatrix} -4 \\ 0 \end{pmatrix}\unitfrac{m}{s} \;.
\]
The simulation within 7 seconds via solving the prox function is shown in Fig. \ref{fig:simulate motion}.
In order to analyze the phase diagram of this system on tangential direction, the initial condition space is determined by $q_x \in [\unit[-6]{m},\unit[6]{m}], v_x \in [\unitfrac[-6]{m}{s},\unitfrac[6]{m}{s}]$. As the simulated result in Fig.~\ref{fig:orbit diagram of motion} shows, the attraction line is defined by $v_x=\unitfrac[0]{m}{s}, q_x \in [\unit[-4]{m},\unit[4]{m}]$. It is noticed that some trajectories cross each other, which is due to the time-dependent friction coefficient.
\newpage
\section{Impulse Feedback Control Law on Impact Level}
Following the formulation of impulse-based control from Wouw and Leine \cite{Wou12}, we assume the controller consists of a classical state-feedback control force $u$ and an impulse control force $U$. The impulse control force is restricted to the impact equation that
\[U= m(v_x^+(t_i) - v_x^-(t_i) ) \; , \]
where the pre-impact velocity $v_x^-(t_i)=0$ at time $t_i$, and $v_x^+(t_i)$ is the post-impact velocity at time $t_i$.
\par
\begin{figure}[hbt] 
	\centering
	\def\svgwidth{0.7\textwidth}
	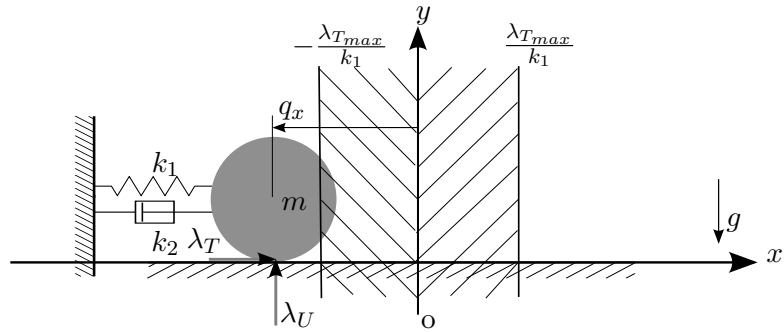
  \caption{Depiction of impulsive control area of oscillator}
  \label{fig:oscillator_control_inkscape}
\end{figure}

Regarding the motion of tangential direction, the equations (\ref{tan_motion1}) - (\ref{tan_motion2}) adding impulse can be rewritten into the form of differential inclusion that
\begin{align*}
  \mathrm{d}{q}_x &= v_x \mathrm{d}t \;, \\
  m\mathrm{d}{v}_x &\in  \Lambda_T + u \mathrm{d} t + U \mathrm{d} \eta \; , 
\end{align*}
where $\mathrm{d} t$ is the Lebesgue measure, $\mathrm{d} \eta$ is a differential atomic measure consisting of a sum of Dirac point measures, and the damping and spring effects are regarded as the control forces. The control forces are given by
\begin{equation}
u(q_x, v_x)=-k_1 q_x -k_2 v_x \; , \quad k_1, k_2>0 \; , 
\end{equation}
\begin{equation}
 U(q_x, v_x^-)=\left\{ \begin{split}\begin{aligned}
&U_T (q_x) \quad  \text{if}   \;\; v_x^-=0 \wedge \| q_x \| \leq \frac{\lambda_{T_{max}}}{k_1}  \\ 
&0 \quad \quad \quad \; \;  \text{else} 
\end{aligned}\end{split} \right. \; .
\end{equation}
The active area of impulsive controller is the shaded part depicted in Fig. \ref{fig:oscillator_control_inkscape}.

This implies that when the oscillator is outside of the interval \[q_x \in [-\frac{\lambda_{T_{max}}}{k_1}, \frac{\lambda_{T_{max}}}{k_1}] \; ,\]
it experiences the slip-slip transition, while only inside of that boundary the slip-stick transition will occur. Thus once the slip-stick transition occurs not at the equilibrium point, the impulse force will be applied on the system leading to the state jump at that time instant $t_i$. The state can be reseted as the followings
\begin{align}
&q_x^+ (t_i)=q_x^- (t_i) \; , \\
&v_x^+ (t_i)=v_x^- (t_i) + \frac{U_T ( q_x^- )}{m}
\end{align}
Marking the first sticking time instance as $t_1$, after this time instance, the feedback control law for $u$ has to be modified by changing the value of $k_2$
\begin{equation}
 k_2(t)=\left\{ \begin{split}\begin{aligned}
&k_{2pre} \quad \;  \text{for}   \;\; t_0\leq t < t_1  \\ 
&k_{2post} \quad  \text{for} \; \;  t\geq t_1 
\end{aligned}\end{split} \right. \; .
\end{equation}
\section{Impulse Estimating}
When the sticking occurs, the impulse is required to be estimated such that the controller could stabilize the oscillator to the equilibrium point at the origin. The impulse on tangential direction at the sticking instant is formulated by
\begin{align}
\label{impulse}
U_T=m(v_{x_{S}}^+ - v_{x_{S}}^-),
\end{align}
where $v_{x_{S}}^-=0$ at the sticking point. Here we rewrite the impulsive control law from \cite{Wou12} in detail for a completed consideration on the both cases of $\underline{q}_x(t_0^*)<0$ and $\underline{q}_x(t_0^*)>0$.

In order to obtain a robust impulse-based control, the lower bound of the friction coefficient $\underline{\mu}$ is employed to approximately estimate the impulse, where the friction coefficient is of the form $\underline{\mu} \leq \mu(q_x,v_x,t) \leq \overline{\mu}$.

For the post-impulse period, the control is added into this system to reinforce the damping effect by replacing $k_{2pre}$ with $k_{2post}$. Thus the reduced differential equation in tangential direction, corresponding to $\underline{\mu}$, can be rewritten as
\begin{align}
\label{2order_reduced1}
& \underline{\dot q}_x = \underline{v}_x , \\
& \underline{\dot v}_x = -\omega_n^2 \underline{q}_x - 2 \vartheta \omega_n \underline{v}_x - \text{sgn}({\underline{v}}_x) g \underline{\mu}\; ,
\label{2order_reduced2}
\end{align}
where \[\omega_n=\sqrt{\frac{k_1}{m}}\; , \quad \vartheta=\frac{k_{2post}}{2\sqrt{m k_1}}>1\; .\]
The characteristic roots of this equation are given by
\begin{align}
\lambda_1=-\omega_n \vartheta + \omega_n \sqrt{\vartheta^2-1}\\
\lambda_2=-\omega_n \vartheta - \omega_n \sqrt{\vartheta^2-1}
\end{align}
The boundary conditions of this systems are given by 
\begin{align*}
& q_x(t^*_{0})=q_{xS}^*, \; v_x(t^*_{0})=v_{xS}^+,\\ 
& q_x(t^*_{end})=0, \; v_x(t^*_{end})=0,
\end{align*}
where $v_{xS}^+$ and  $t^*_{end}$ are unknowns. Therefore, the analytical solution of this reduced system can de obtained by taking $t^*_{end}$ as reference time, and separating into two cases with $\underline{c}=g\underline{\mu}/\omega_n^2$; particularly, the underlined $\underline{q}_x, \underline{v}_x$ indicates the solution of the reduced differential equations (\ref{2order_reduced1})-(\ref{2order_reduced2}) in the post impulse period, but not that of the real system:
\begin{itemize}
\item $q_{xS}^*<0 \; , v_{xS}^+>0$, in the case of $\vartheta>1$, the post impulse velocity $\underline{v}_x$ is always positive, that is $\text{sgn}(\underline{v}_x)=1$.
\begin{align}
\label{BVPequation_minus_q_x_focus_q}
& \underline{q}_x(t) = \underline{c}\left( \frac{\lambda_2}{\lambda_2 - \lambda_1} e^{\lambda_1 (t -t^*_{end})} - \frac{\lambda_1}{\lambda_2 - \lambda_1} e^{\lambda_2 (t - t^*_{end})} -1 \right) \;, & t^*_0 \leq t \leq t^*_{end} \;, \\
& \underline{v}_x(t) = \underline{c} \left( \frac{\lambda_1 \lambda_2}{\lambda_2 - \lambda_1}(e^{\lambda_1 (t - t^*_{end})} - e^{\lambda_2 (t - t^*_{end})}) \right) \;, & t^*_0 \leq t \leq t^*_{end} \;.
\label{BVPequation-q}
\end{align}
\item $q_{xS}^*>0\; , v_{xS}^+<0$, in the case of $\vartheta>1$, the post impulse velocity $\underline{v}_x$ is always negative, that is $\text{sgn}(\underline{v}_x)=-1$.

\begin{align}
& \underline{q}_x(t) = -\underline{c}\left( \frac{\lambda_2}{\lambda_2 - \lambda_1} e^{\lambda_1 (t -t^*_{end})} - \frac{\lambda_1}{\lambda_2 - \lambda_1} e^{\lambda_2 (t - t^*_{end})} -1 \right) \;, & t^*_0 \leq t \leq t^*_{end} \;, \\
& \underline{v}_x(t) = -\underline{c} \left( \frac{\lambda_1 \lambda_2}{\lambda_2 - \lambda_1}(e^{\lambda_1 (t - t^*_{end})} - e^{\lambda_2 (t - t^*_{end})}) \right) \;, & t^*_0 \leq t \leq t^*_{end} \;.
\label{BVPequation+q}
\end{align}
\end{itemize}
In oder to obtain the unknown of $t^*_{end}$, the following nonlinear equation $f(t')=0$ is required to be solved at the instance of $t=t^*_{0}$ by taking $t^*_{end}=t'$ as the variable:
\begin{align}
f(t') =\left\{ \begin{array}{cc} \underline{c}\left( \frac{\lambda_2}{\lambda_2 - \lambda_1} e^{\lambda_1 (t^*_0 -t')} - \frac{\lambda_1}{\lambda_2 - \lambda_1} e^{\lambda_2 (t^*_0 - t')} -1 \right) - \underline{q}_x(t^*_0) \;, & \underline{q}_x(t^*_0)<0 \\
-\underline{c}\left( \frac{\lambda_2}{\lambda_2 - \lambda_1} e^{\lambda_1 (t^*_0 -t')} - \frac{\lambda_1}{\lambda_2 - \lambda_1} e^{\lambda_2 (t^*_0 - t')} -1 \right) - \underline{q}_x(t^*_0) \;, & \underline{q}_x(t^*_0)>0 \end{array} \right. .
\label{nonlinear_equation_for_t_end}
\end{align}
The Newton iteration scheme is employed to solve this equation for $t^*_{end}$, the $i$th iteration of $t'$ can be computed numerically by
\begin{align}
t'_{(i+1)}=t'_{(i)}-\frac{f(t'_{(i)})}{f'(t'_{(i)})},
\end{align}
where
\begin{align}
f'(t')= \left\{ \begin{array}{cc} \underline{c} \left( \frac{\lambda_1 \lambda_2}{\lambda_2 - \lambda_1}( -e^{\lambda_1 (t^*_0 - t')} + e^{\lambda_2 (t^*_0 - t')})\right) \; , & \underline{q}_x(t^*_0)<0 \\
 -\underline{c} \left( \frac{\lambda_1 \lambda_2}{\lambda_2 - \lambda_1}( -e^{\lambda_1 (t^*_0 - t')} + e^{\lambda_2 (t^*_0 - t')}) \right) \; , & \underline{q}_x(t^*_0)>0
\end{array}\right. .
\label{nonlinear_equation_derivative_f}
\end{align}
\label{}
 Then substitute $t^*_{end}$ back to the equation (\ref{BVPequation-q}) or (\ref{BVPequation+q}) and take $t=t^*_0$, the velocity after the impulse can be obtained by
 \begin{align}
 v_{xS}^+=\underline{v}_x(t^*_0).
 \end{align}
Thus, the impulse can be estimated via \eqref{impulse}.
\subsection{Approximate impulsive-based control law}
After the estimated impulse, the following impulses are guaranteed by the approximated impulsive control law \cite{Wou12},
\begin{align}
\label{approx_control}
\Lambda_{approx}(q_x)=- \text{sgn}(q_x)\sqrt{2 \underline{c} \, \omega_n^2 m^2 \| q_x \|} = m v^+_{xS},
\end{align}
which is valid when $\| q_x \|$ is small. We rewrite the proof with minor corrections compared to the original one.
\begin{proof}
Firstly, consider the $q_{xS}^*<0 \; , v_{xS}^+>0$ case, via equations (\ref{BVPequation_minus_q_x_focus_q}),  (\ref{nonlinear_equation_for_t_end}) and $f(t^*_{end};\underline{q}_x(t^*_0))=0$, it yields that 
\begin{align}
\frac{d\underline{q}_x(t^*_0)}{dt^*_{end}}=f'(t^*_{end}).
\end{align}
According to equations (\ref{BVPequation-q}) and (\ref{nonlinear_equation_derivative_f}), it also holds that
\[
\underline{v}_x(t^*_0)=-f'(t^*_{end})= \underline{c} \left( \frac{\lambda_1 \lambda_2}{\lambda_2 - \lambda_1}(e^{\lambda_1 (t^*_0 - t^*_{end})} - e^{\lambda_2 (t^*_0 - t^*_{end})}) \right) \; .
\]
Thus, we have
\begin{align}
\frac{d\underline{v}_x(t^*_0)}{dt^*_{end}}=-f''(t^*_{end}) \; .
\end{align}
Therefore, the slope of the impulsive control law $\Lambda(\underline{q}_x(t^*_0))$ is given by
\[
\frac{d\Lambda(\underline{q}_x(t^*_0))}{d\underline{q}_x(t^*_0)}=m \frac{d\underline{v}_x(t^*_0)}{d\underline{q}_x(t^*_0)} = m \frac{d\underline{v}_x(t^*_0)}{dt^*_{end}} \frac{dt^*_{end}}{d\underline{q}_x(t^*_0)} = -m \frac{f''(t^*_{end})}{f'(t^*_{end})}\; .
\]
By using
\[
f'(t^*_{end})=-\underline{v}_x(t^*_0)=-\frac{\Lambda(\underline{q}_x(t^*_0))}{m} \; ,
\]
the slope of $\Lambda(\underline{q}_x(t^*_0))$ can be further written by
\begin{align}
\Lambda'(\underline{q}_x(t^*_{end}))=m^2 \frac{f''(t^*_{end})}{\Lambda(\underline{q}_x(t^*_{end}))}.
\end{align}
Moreover, construct the inequality by
\begin{align*}
 f''(t^*_{end})+\underline{c}\omega_n^2+\lambda_1 f'(t^*_{end})
= \underline{c}\omega_n^2 \left( 1- e^{ (-\omega_n \vartheta - \omega_n \sqrt{\vartheta^2-1})(t^*_0 - t^*_{end}) } \right) \leq 0,
\end{align*}
due to the values of $\vartheta>1$, $\lambda_1=-\omega_n \vartheta - \omega_n \sqrt{\vartheta^2-1}<0$, and $t^*_0 - t^*_{end}\leq 0$,
where
\begin{align*}
& f''(t^*_{end})=\underline{c} \left( \frac{\lambda_1 \lambda_2}{\lambda_2 - \lambda_1}( \lambda_1 e^{\lambda_1 (t^*_0 - t^*_{end})} + \lambda_2 e^{\lambda_2 (t^*_0 - t^*_{end})})\right) \; , \\
& f'(t^*_{end})=\underline{c} \left( \frac{\lambda_1 \lambda_2}{\lambda_2 - \lambda_1}( -e^{\lambda_1 (t^*_0 - t^*_{end})} + e^{\lambda_2 (t^*_0 - t^*_{end})})\right) \; , \\
& \lambda_1 \lambda_2 = \omega_n^2 \; , \\
& \lambda_2 - \lambda_1 = -2\omega_n \sqrt{\vartheta^2-1} \; .
\end{align*}
Thus, the inequality is in the form of
\begin{align}
 f''(t^*_{end})\leq-\underline{c}\omega_n^2-\lambda_1 f'(t^*_{end}).
\end{align}
Then the differential inequality  can be obtained by
\begin{align}
\Lambda'(x) \leq -m\lambda_1 - \frac{\underline{c}\omega_n^2m^2}{\Lambda(x)}
\label{differential_inequality}
\end{align}
with the domain 
\[
x=\underline{q}_x(t^*_0)\;.
\]
In the domain $x\leq0$, with the boundary condition $\Lambda(0)=0$, the solution of equation (\ref{differential_inequality}), the impulsive control law, is bounded by
\begin{align}
& \Lambda(x) \geq \sqrt{-2\underline{c}\,\omega_n^2 x} \; , \quad x\leq0 \; , \\
& \Lambda(x) \geq  - m \lambda_1 x \; ,\quad x\leq0 \; .
\label{linear_bnd1}
\end{align}

Secondly, consider the $q_{xS}^*>0 \; , v_{xS}^+<0$ case, due to the symmetry of the system, the $\Lambda(x)$ is therefore unevenness that $\Lambda(x)=-\Lambda(-x)$. Thus, in the domain $x\geq0$, the impulsive control law is symmetrically bounded by
\begin{align}
& \Lambda(x) \leq - \sqrt{2\underline{c}\,\omega_n^2 x} \; , \quad x\geq0 \; , \\
& \Lambda(x) \leq  m \lambda_1 x \; , \quad x\geq0 \; .
\label{linear_bnd2}
\end{align}

Finally, in the range of small $x$, regardless of the linear boundaries (\ref{linear_bnd1}) and (\ref{linear_bnd2}), the impulsive control law can be approximated by
\begin{align}
\Lambda_{approx}(x)=- \text{sign}(x)\sqrt{2 \underline{c} \, \omega_n^2 m^2 \| x \|},
\end{align}
\end{proof}

\subsection{Shooting method}
According to the dynamics equation \eqref{eq:motion1}-\eqref{eq:motion2}, the boundary conditions are given by
\[
q_x(t_{end})=0, \quad v_x(t_{end})=0, \quad q_x(t_0)=q_x^*.
\]
For the given initial guess $v_x^+(t_0)=s$ and the time sequence
\[
\mathbf t=[0,t_1,t_2,\dots, t_n]^T,
\]
the sequences of velocity and position can be numerically computed
\begin{align*}
& \mathbf v_x=[s, v_x(t_1;s), v_x(t_2;s),\dots, v_x(t_n;s)]^T,\\
& \mathbf q_x=[q_x^*, q_x(t_1;s), q_x(t_2;s), \dots, q_x(t_n;s)]^T.
\end{align*}
The boundary constraints are in the form of
\begin{align}
\label{bnd_q}
F_q(s)=q_x(t_n;s)=0,\\
F_v(s)=v_x(t_n;s)=0.
\label{bnd_v}
\end{align}
Since $\dot q=v$, the position-level constraint (\ref{bnd_q}) shall be satisfied before the velocity-level constraint (\ref{bnd_v}). Thus, the zero root of (\ref{bnd_q}) can be iterated by
\[
s_q^{(i+1)}=s_q^{(i)}-\frac{F_q(s_q^{(i)})}{\Delta F_q(s_q^{(i)})}
\]
where 
\[
\Delta F_q(s_q^{(i)})=\frac{F_q(s_q^{(i)}+\Delta s)-F_q(s_q^{(i)})}{\Delta s}
\]
with $\Delta s$ a given value.

Once the convergent solution of $s_q^*$ is found, set $s_v^{(1)}=s_q^*$, then the zero root of (\ref{bnd_v}) near $s_q^*$ can be iterated by
\[
s_v^{(i+1)}=s_v^{(i)}-\frac{F_v(s_v^{(i)})}{\Delta F_v(s_v^{(i)})}
\]
where 
\[
\Delta F_v(s_v^{(i)})=\frac{F_v(s_v^{(i)}+\Delta s)-F_v(s_v^{(i)})}{\Delta s}
\]
Practically, finding the zero root of equation (\ref{bnd_q}) is enough for the impulse estimating, while the velocity level constraint (\ref{bnd_v}) is not so important to be restrictedly satisfied. 
\subsection{Simulation comparison}
\begin{figure}[hbt]
  \centering
  \includegraphics[width=0.65\linewidth]{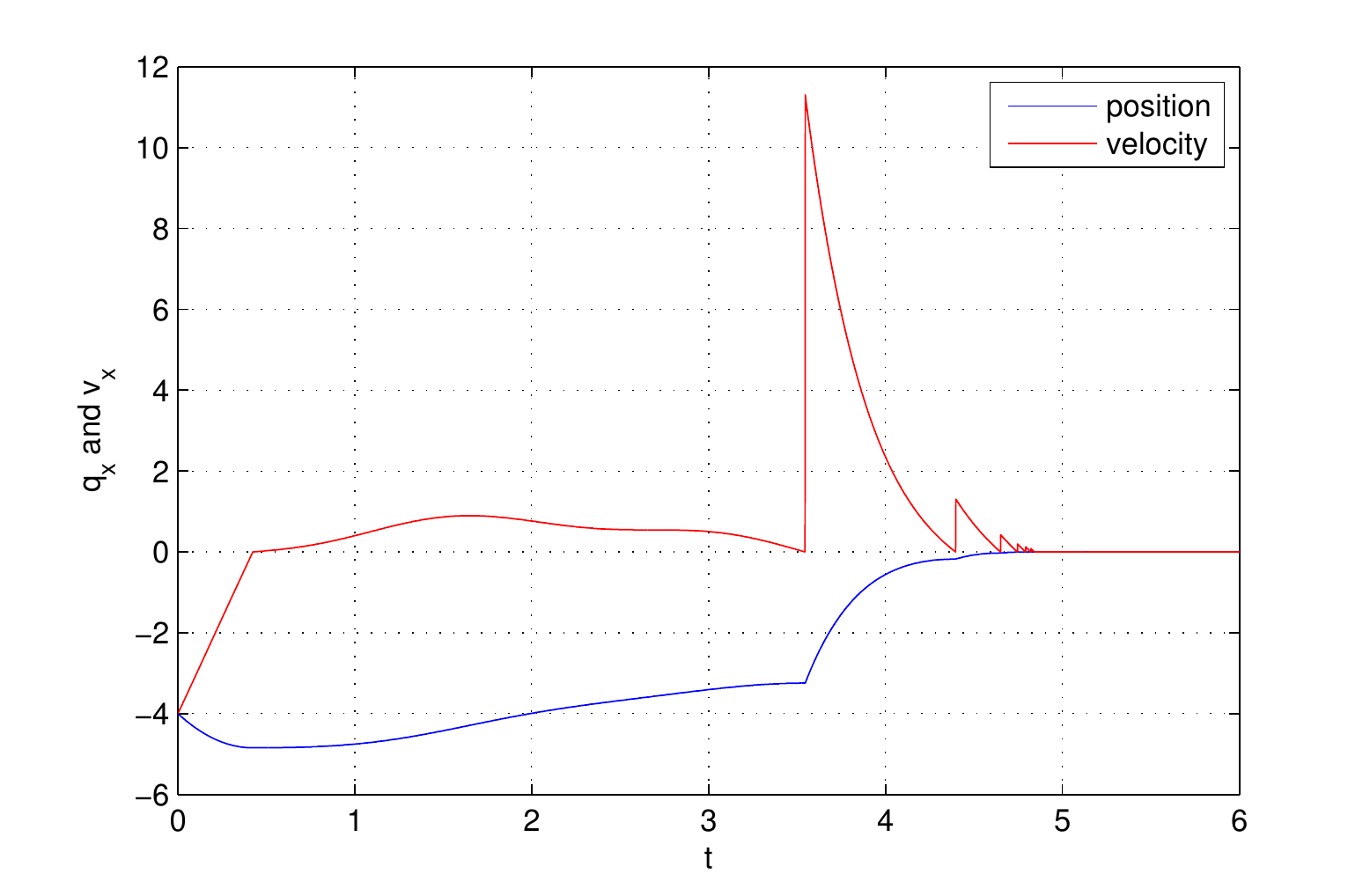}
  \caption{Wouw and Leine's method via Prox function from \cite{zhang14_1}}
  \label{fig:Leine_k3_prox}
\end{figure}

\begin{figure}
  \centering
  \includegraphics[width=0.65\linewidth]{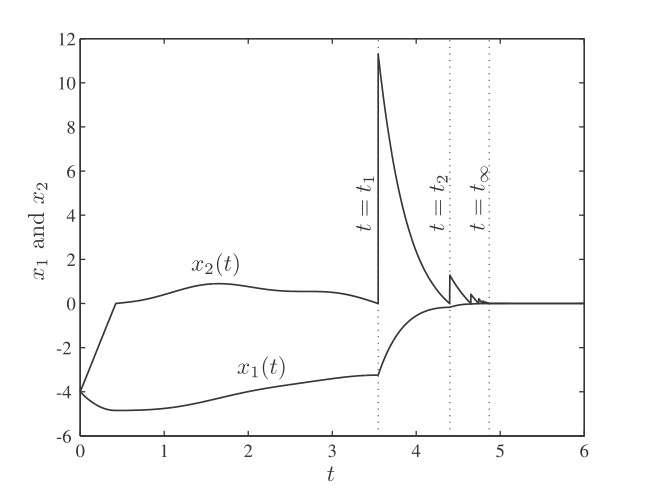}
  \caption{Wouw's simulation result}
  \label{fig:wouw}
\end{figure}

\begin{figure}
  \centering
  \includegraphics[width=0.65\linewidth]{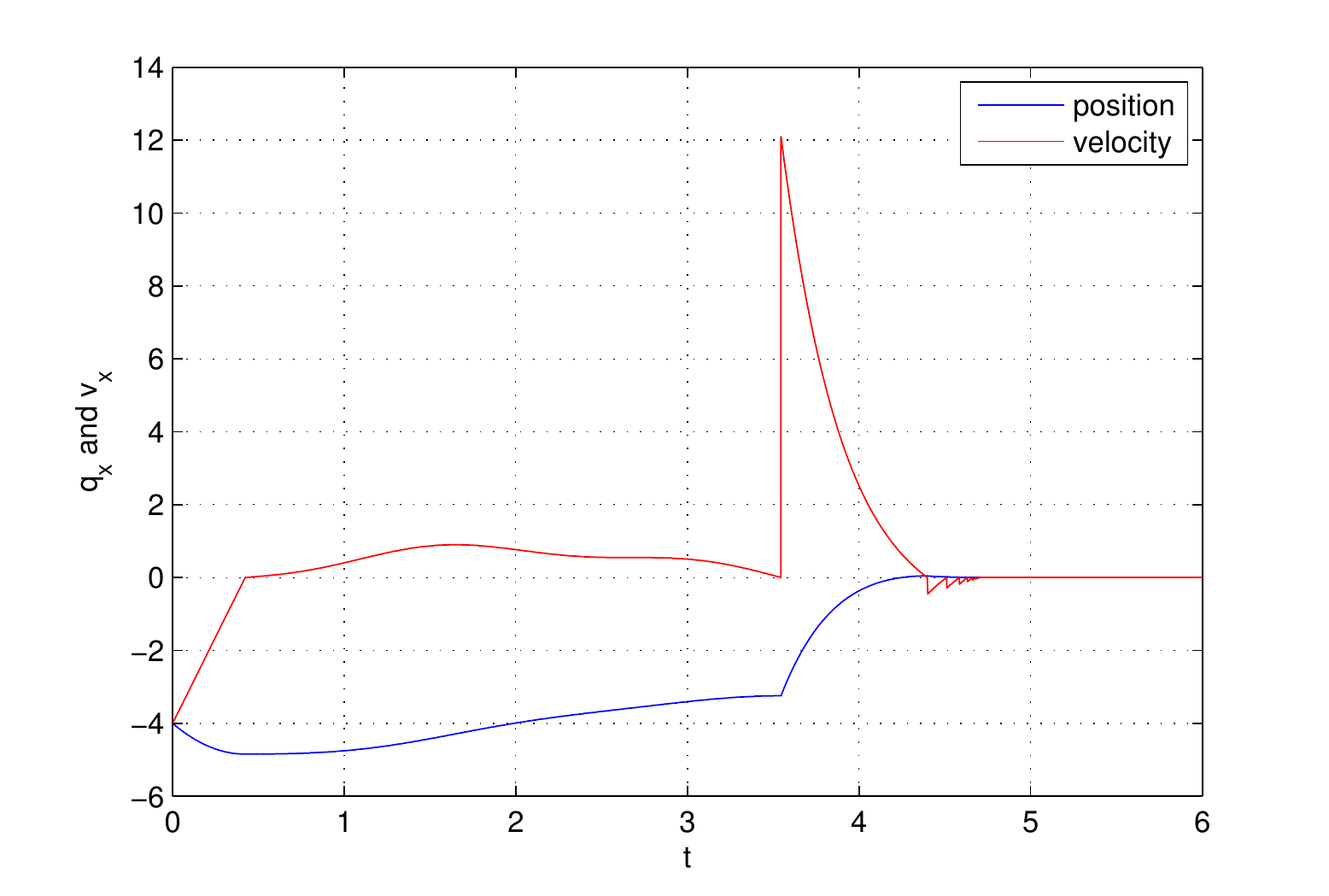}
  \caption{Shooting method via Prox function from \cite{zhang14_1}}
  \label{fig:shooting_k3_prox}
\end{figure}

With the methods and example indicated above, setting the control law with $k_1=1, k_{2pre}=0.5, k_{2post}=3$, the simulation result is presented in Fig. \ref{fig:Leine_k3_prox} to compare with the result from \cite{Wou12} shown in Fig. \ref{fig:wouw}. All the data are the same with those presented in Section 2.2. Additionally, the simulation result via shooting method for impulse estimating is also presented in Fig. \ref{fig:shooting_k3_prox}.\par
Fig. \ref{fig:Leine_k3_prox} and Fig. \ref{fig:wouw} are almost exactly the same. However, in this case, the estimating via solving the boundary value problem is activated for three times. The core idea for estimating shall be finding the best impulse to force the oscillator into the equilibrium point directly. While the three times estimating is away from its core idea, in own opinion, losing its value in some sense. There could be some alternative methods, which are more general and also robust, to achieve this goal instead of solving the BVP.\par
\begin{figure}
\centering
 \begin{subfigmatrix}{2}
  \subfigure[Normal frictional impulse$\Lambda_{U}$]{\includegraphics{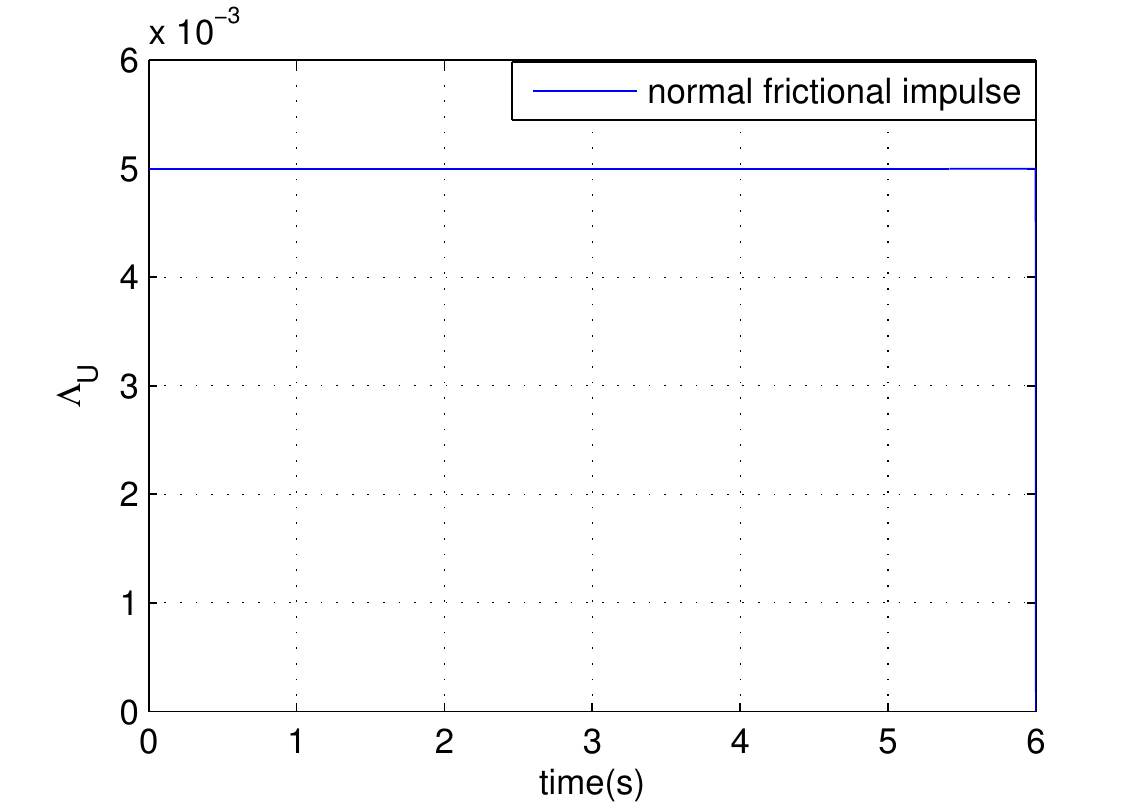}}
  \subfigure[Tangential frictional impulse $\Lambda_{T}$]{\includegraphics{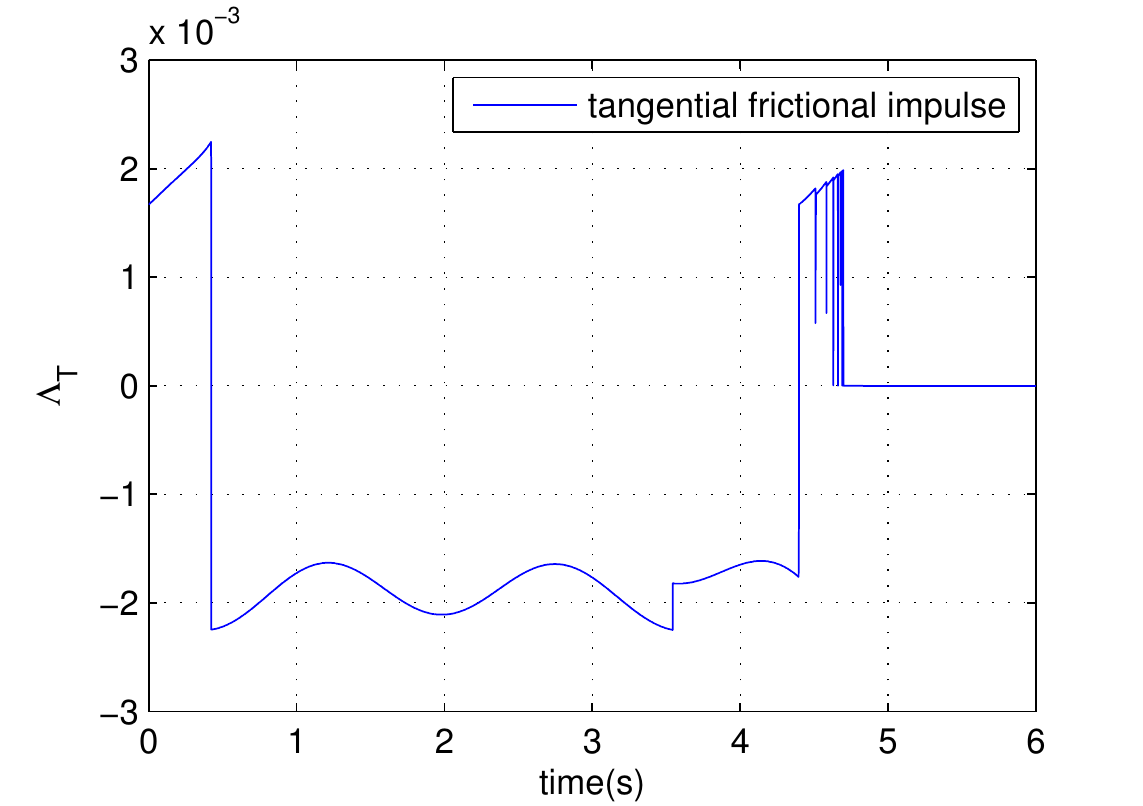}}
  \subfigure[Feedback control $u$]{\includegraphics{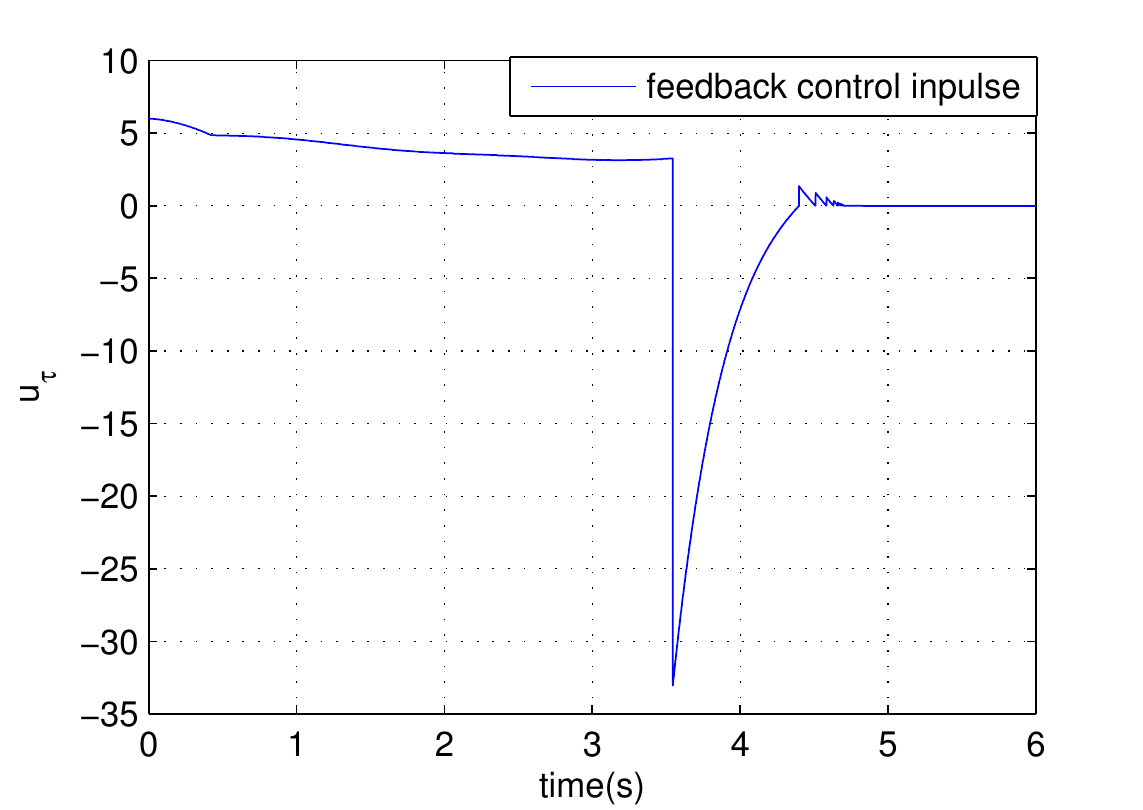}}
  \subfigure[Impulse control $U$ ]{\includegraphics{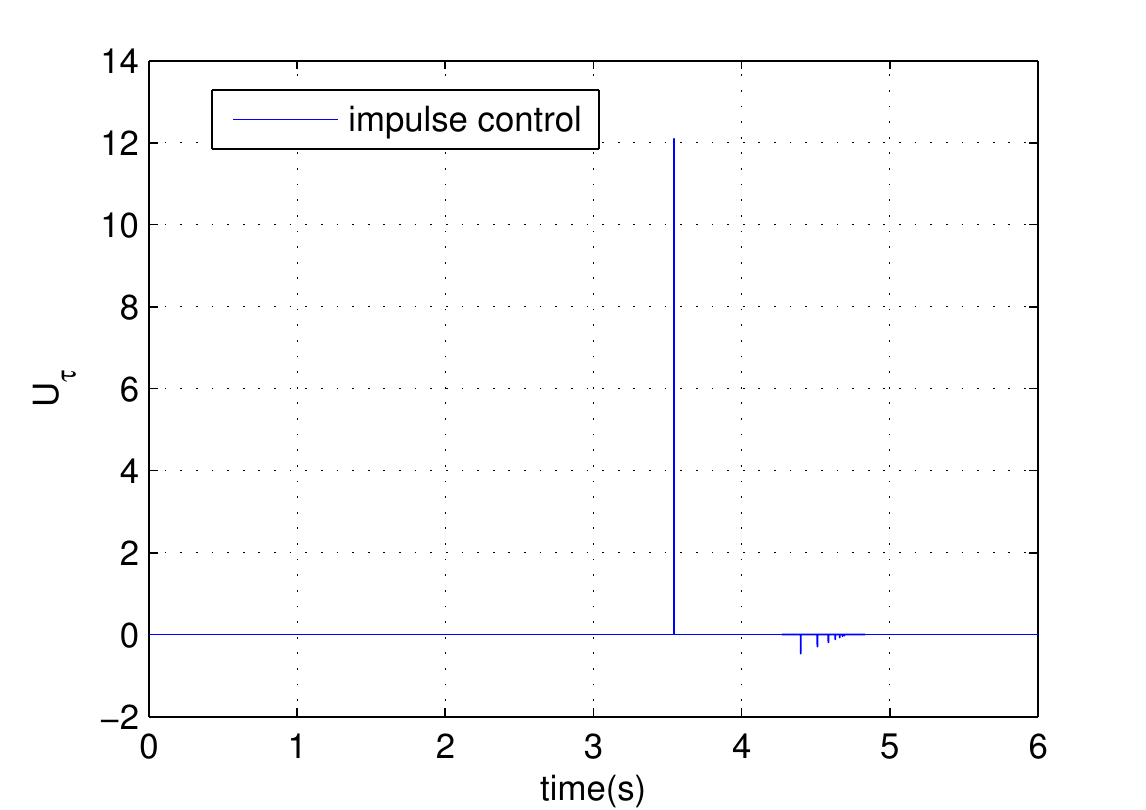}}
 \end{subfigmatrix}
 \caption{Time history of frictional impulse, feedback and impulsive control of controlled oscillator with shooting method}
 \label{fig:oscillator_control}
\end{figure}

Compare Fig.~\ref{fig:Leine_k3_prox} and Fig.~\ref{fig:wouw} to Fig. \ref{fig:shooting_k3_prox}, the shooting method shows the advantage in impulse estimating. Therefore, in the case of shooting method is applied, the time history of normal and tangential frictional impulse, feedback and the impulsive controls are plotted in Fig. \ref{fig:oscillator_control}
\par
\begin{figure}
  \centering
  \includegraphics[width=0.8\linewidth]{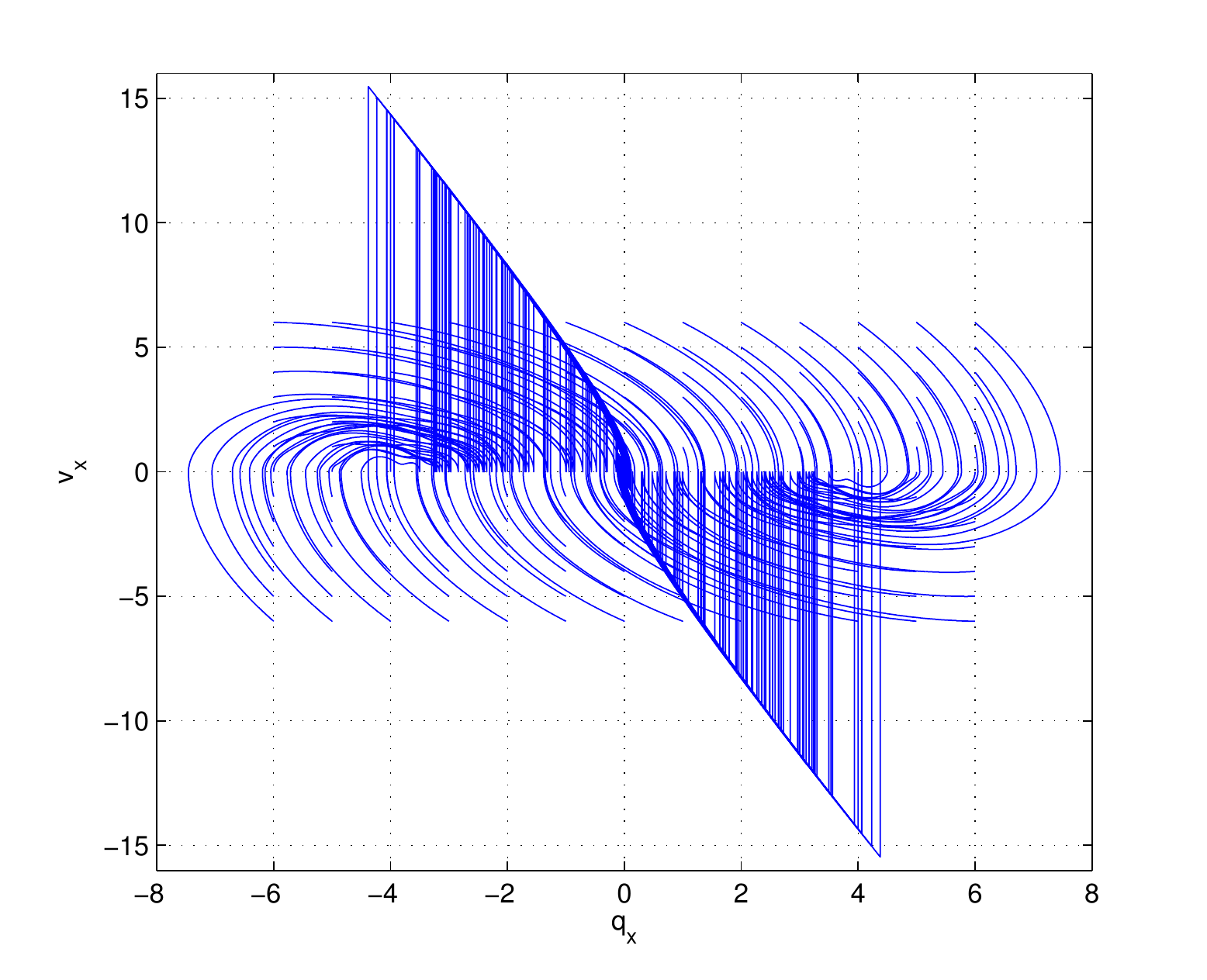}
  \caption{Phase diagram of oscillator controlled behavior with shooting method}
  \label{fig:oscillator_control_diagram}
\end{figure}

In order to analyze the phase diagram of this impulse-based controlled system on tangential direction, the initial condition space is determined as the previous example by $q_x \in [\unit[-6]{m},\unit[6]{m}], v_x \in [\unitfrac[-6]{m}{s},\unitfrac[6]{m}{s}]$. The simulation result is shown in Fig. \ref{fig:oscillator_control_diagram}, the attraction point is defined by $v_x=\unitfrac[0]{m}{s}, q_x =\unit[0]{m}$.



\chapter{Dynamics of Furuta Pendulum} 

\label{Chapter4} 

\lhead{Chapter 4. \emph{Dynamics of Furuta Pendulum}} 

\section{Motion of a Frictional Furuta Pendulum}
\begin{figure}[hbt] 
	\centering
	\def\svgwidth{0.5\textwidth}
	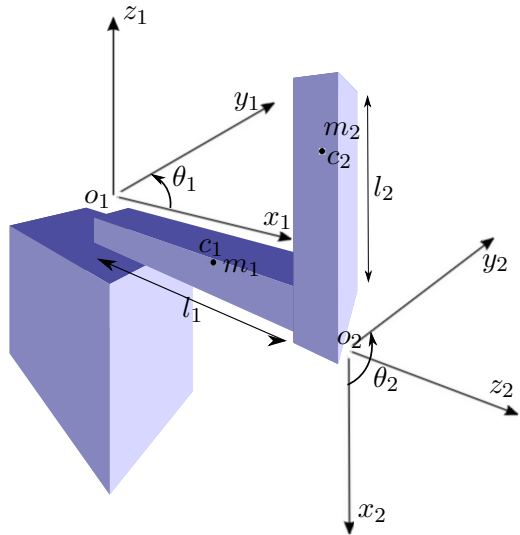
  \caption{Pendulum geometry from \cite{zhang14_2}.}
  \label{fig:pendulum}
\end{figure}

We followed the formulation of dynamics of furuta pendulum by Cazzolato and Prime \cite{caz2011pendulum}, while the frictions were not taken into account. The geometry of the Furuta pendulum is illustrated in Fig. \ref{fig:pendulum}.

The actuated arm has its body-fixed frame $O_1X_1Y_1Z_1$, the pendulum has its body-fixed frame $O_2X_2Y_2Z_2$, and the inertial frame $OXYZ$ is not illustrated in the above figure. The length of the arm and pendulum are $l_1,l_2$, and total mass are $m_1,m_2$ located at the centers of the mass, whose length  in their own frames are $c_1,c_2$, respectively. Assume that the body-fixed coordinates are the principal axes of the arm and pendulum, then the principal inertial tensor are given by
\begin{align}
\mathbf J_1= \begin{pmatrix} J_{1xx} & 0 & 0 \\ 0 & J_{1yy} & 0  \\ 0 & 0 & J_{1zz} \end{pmatrix},\\
\mathbf J_2= \begin{pmatrix} J_{2xx} & 0 & 0 \\ 0 & J_{2yy} & 0  \\ 0 & 0 & J_{2zz} \end{pmatrix}.
\end{align}
The rotated angle of the arm and pendulum are $\theta_1, \theta_2$ along the direction of axes $z_1,z_2$ (counterclockwise). The rotation matrix from the inertial frame to arm frame is given by
\begin{align}
\mathbf R_1=\begin{pmatrix} \cos \theta_1 & \sin \theta_1 & 0 \\ -\sin \theta_1 & \cos \theta_1 & 0 \\ 0 & 0 & 1 \end{pmatrix},
\end{align}
and the rotation matrix from arm frame to pendulum frame is given by
\begin{align}
\mathbf R_2 = \begin{pmatrix} \cos \theta_2 & \sin \theta_2 & 0 \\ -\sin \theta_2 & \cos \theta_2 & 0 \\ 0 & 0 & 1 \end{pmatrix} \begin{pmatrix} \cos \frac{\pi}{2} & 0 & -\sin \frac{\pi}{2} \\ 0 & 1 & 0 \\ \sin \frac{\pi}{2} & 0 & \cos \frac{\pi}{2} \end{pmatrix} = \begin{pmatrix} 0 & \sin \theta_2 & -\cos \theta_2 \\ 0 & \cos \theta_2 & \sin \theta_2 \\ 1 & 0 & 0 \end{pmatrix}.
\end{align}
The angular velocity of the arm in the arm frame is given by
\[
\omega_1 = {\begin{pmatrix}0 & 0 & \dot \theta_1\end{pmatrix}}^T.
\]
The joint between arm and the base is fixed such that its velocity is
\[
\mathbf v_{J1}={\begin{pmatrix}0 & 0 & 0\end{pmatrix}}^T.
\]
Thus, the velocity of the mass center in the arm frame is given by
\[
\mathbf v_{1c}=\mathbf v_{J1} + \omega_1 \times {\begin{pmatrix} c_1 & 0 & 0 \end{pmatrix}}^T={\begin{pmatrix} 0 & \dot \theta_1 c_1 & 0 \end{pmatrix}}^T.
\]
Then, the angular velocity of the pendulum in the pendulum frame is given by
\[
\omega_2=\mathbf R_2 \omega_1 + {\begin{pmatrix} 0&0&\dot \theta_2 \end{pmatrix}}^T={\begin{pmatrix} -\dot\theta_1\cos \theta_2 & \dot \theta_1 \sin \theta_2 &\dot \theta_2 \end{pmatrix}}^T.
\]
The velocity of the joint between the arm and pendulum in the pendulum frame is given by
\[
\mathbf v_{J2}=\mathbf R_2 \left( \omega_1 \times {\begin{pmatrix} l_1&0&0 \end{pmatrix}}^T \right) ={\begin{pmatrix} \dot \theta_1 l_1 \sin \theta_2 & \dot \theta_1 l_1 \cos \theta_2 & 0 \end{pmatrix}}^T
\]
Thus, the velocity of the mass center of the pendulum in the pendulum frame is given by
\[
\mathbf v_{2c}=\mathbf v_{J2}+ \omega_2 \times {\begin{pmatrix} l_2&0&0 \end{pmatrix}}^T={\begin{pmatrix} \dot \theta_1 l_1 \sin \theta_2 & \dot \theta_1 l_1 \cos \theta_2 + \dot \theta_2 l_2 & -\dot \theta_1 l_2 \sin \theta_2 \end{pmatrix}}^T
\]
The potential and kinetic energy of the arm and pendulum are given by
\begin{align*}
E_{p1}= & 0, \\
E_{p2}= & g m_2 c_2 (1-\cos \theta_2), \\
E_{k1}= & \frac{1}{2}(\mathbf v_{1c}^T m_1 \mathbf v_{1c} + \omega_1^T \mathbf J_1 \omega_1)= \frac{1}{2}\dot \theta_1^2(m_1 c_1^2+J_{1zz}),\\
E_{k2}= & \frac{1}{2}(\mathbf v_{2c}^T m_2 \mathbf v_{2c} + \omega_2^T \mathbf J_2 \omega_2)= \frac{1}{2} \dot \theta_1^2 \left[ m_2 l_2^2+(m_2 c_2^2 +J_{2yy}) \sin^2 \theta_2 + J_{2xx}\cos^2 \theta_2 \right] \\
& + \frac{1}{2} \dot\theta_2^2 ( J_{2zz}+m_2 c_2^2) + m_2 l_1 c_2 \cos(\theta_2) \dot\theta_1 \dot\theta_2.
\end{align*}
Therefore, the Lagrangian of the system can be obtained by
\begin{align*}
L=E_k-E_p \; ,
\end{align*}
where $E_k=E_{k1}+K_{k2}$, $E_p=E_{p1}+E_{p2}$.
Substitute this into the Euler-Lagrange equation that
\begin{align*}
\frac{d}{dt}\left( \frac{\partial L}{\partial \dot{q_i}} \right) - \frac{\partial L}{\partial q_i}=Q_i \; ,
\end{align*}
where $q_i=[\theta_1, \theta_2]^T$ is the generalised coordinate, and $Q_i=[\tau,0]$ is the generalised external torque, then the equation of motion without friction can be obtained by
\begin{align}
\mathbf{M} \begin{pmatrix}
\ddot{\theta_1} \\ \ddot{\theta_2}
\end{pmatrix}
= \mathbf{h} +  \begin{pmatrix}
\tau \\ 0
\end{pmatrix} \; ,
\end{align}
where
\[
\mathbf{M}=\begin{pmatrix} J_{1zz}+m_1 c_1^2+m_2 l_1^2 + (J_{2yy}+m_2 c_2^2) \sin^2 \theta_2 + J_{2xx} \cos^2 \theta_2 & m_2 l_1 c_2 \cos \theta_2 \\
m_2 l_1 c_2 \cos \theta_2 & m_2 c_2^2 + J_{2zz} 
\end{pmatrix} \; ,
\]
\[
\mathbf{h}=\begin{pmatrix}
\dot{\theta}_2^2 m_2 l_1 c_2 \sin \theta_2 - \dot{\theta}_1 \dot{\theta}_2 \sin 2\theta_2 (m_2 c_2^2 + J_{2yy} - J_{2xx})  \\
\frac{1}{2}\dot{\theta}_1^2 \sin 2\theta_2 (m_2 c_2^2 + J_{2yy} - J_{2xx}) - g m_2 c_2 \sin \theta_2
\end{pmatrix} \; .
\]
Finally, adding the frictional moments $M_{f1}, M_{f2}$ into the equation, we have the equation that
\begin{align}
\label{eom}
\mathbf{M} \begin{pmatrix}
\ddot{\theta}_1 \\ \ddot{\theta}_2
\end{pmatrix}
= \mathbf{h}  + \begin{pmatrix}
M_{f1} \\ M_{f_2}
\end{pmatrix} +  \begin{pmatrix}
\tau \\ 0
\end{pmatrix} \; ,
\end{align}
where the frictional moments are represented by
\begin{align*}
M_{fi}\left\{ \begin{array}{lll}
= -M_{ci} & \text{if} \; \dot \theta_i > 0 \\ 
\in [-M_{si}, M_{si}] & \text{if} \; \dot \theta_i = 0 \\
=M_{ci} & \text{if} \; \dot \theta_i < 0
\end{array} \right.
\end{align*}
\section{Frictional Moment Formulation}
\begin{figure}[hbt] 
	\centering
	\def\svgwidth{0.5\textwidth}
	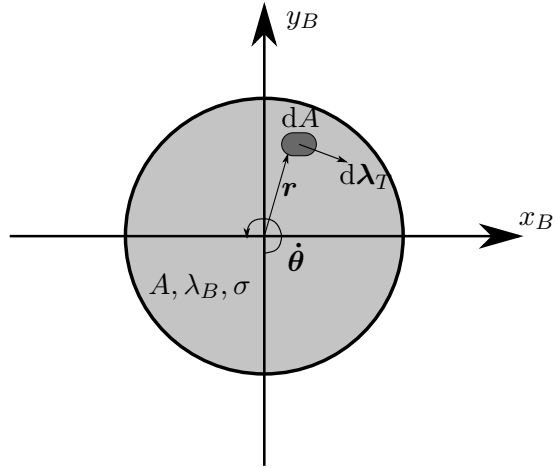
  \caption{Tangential plane of contact surface with uniform pressure}
  \label{fig:frictional_moment}
\end{figure}

The frictional moment applied around the pendulum joint should be calculated via the surface integration on the whole contacted frictional plate. The frictional plate on the joint provides a uniform pressure because the contact is between locally flat bodies. The friction of the kind of uniform pressure distribution has been discussed in detail by Leine and Glocker \cite{Lei03}. Following their ideas, we derive the formulation of frictional moment under the bilateral uniform contact. The depiction of the forces on the tangential plane of contact surface is shown in Fig. \ref{fig:frictional_moment}.
\par
First we use the notation that $\lambda_{B}$ is the normal force on the contact surface between the joint, and $A$ is the area of contact surface, such that $\sigma=\lambda_{B}/{A}$ is the uniform pressure. In the case of $\dot{\bm\theta} \neq 0$, the tangential frictional force on $\text{d}A$ is given by
\begin{align}
\text{d}\bm\lambda_T=-\mu\sigma \text{d}A \frac{\dot{\bm \theta}}{\| \dot{\bm \theta} \|} \times \frac{\bm r}{\| \bm r \|} \; ,
\end{align}
where $\bm r$ is the position vector pointing from the center of the joint to $\text{d}A$. Then the frictional moment on $\text{d}A$ can be obtained by
\begin{align}
\text{d} \bm M_{fric} = \bm r \times \text{d}\bm \lambda_T =-\mu\sigma r \text{d}A \frac{\bm r}{\| \bm r \|} \times\left( \frac{\dot{\bm \theta}}{\| \dot{\bm \theta} \|} \times \frac{\bm r}{\| \bm r \|} \right) \; .
\end{align}
Using the property of
\[
\frac{\bm r}{\| \bm r \|} \times\left( \frac{\dot{\bm \theta}}{\| \dot{\bm \theta} \|} \times \frac{\bm r}{\| \bm r \|} \right) = \frac{\dot{\bm \theta}}{\| \dot{\bm \theta} \|} \; ,
\]
we finally have the integration formula that
\begin{align}
\bm M_{fric}= - \iint_{A}\mu \sigma r \text{d}A \frac{\dot{\bm \theta}}{\| \dot{\bm \theta} \|} \; .
\end{align}
\par
Particularly, we consider the case that the frictional plate is of the form with the maximum radius $R_1$, minimum radius $R_2$, and the area $A=\pi (R_1^2 - R_2^2)$, then the integration formula can be written as
\begin{align}
\begin{split} \bm M_{fric} & = - \int^{2\pi}_{0} \int^{R_1}_{R_2}\mu \sigma r^2 \text{d}r \text{d}\theta \frac{\dot{\bm \theta}}{\| \dot{\bm \theta} \|} \\ 
& = - \frac{2}{3} \pi \mu \sigma ( R_1^3 - R_2^3 )\frac{\dot{\bm \theta}}{\| \dot{\bm \theta} \|} \\
& = - \frac{2}{3} \mu \sigma A \frac{R_1^3 - R_2^3}{R_1^2 - R_2^2}\frac{\dot{\bm \theta}}{\| \dot{\bm \theta} \|} \\
& = - \mu \lambda_{B} R_E \frac{\dot{\bm \theta}}{\| \dot{\bm \theta} \|} \; ,
\end{split}
\end{align}
where the equivalent moment arm is
\[
R_E=\frac{2(R_1^3 - R_2^3)}{3(R_1^2 - R_2^2)} \; .
\]
Therefore, the frictional moment can be formulated by
\begin{align}
\bm M_{fric} \left\{ \begin{array}{ll}
=  - \mu \lambda_{B} R_E \frac{\dot{\bm \theta}}{\| \dot{\bm \theta} \|}  & \text{if} \; \dot{\bm \theta}_i \neq 0  \\ 
\in [- \mu \lambda_{B} R_E, \; \mu \lambda_{B} R_E] & \text{if} \; \dot{\bm \theta_i }= 0
\end{array}\right. \; .
\end{align}
\subsection{Normal contact force formulation}
The joints provide static contact normal forces $\lambda_{B1, static}, \lambda_{B2, static}$, while the real contact normal forces become dynamic once their values exceed the static normal forces. Because of the orthogonal arms, the dynamics contact normal force can be formulated separately by
\begin{align}
& N_1=(m_1+m_2)g+m_2 c_2 \dot{\theta}_2^2 \cos{\theta_2} \; , \\
& N_2=m_2 l_1 \dot{\theta}_1^2 \; .
\end{align}
The real normal contact force can be obtain by
\begin{align}
\lambda_{B1}=\left\{ \begin{array}{ll} \lambda_{B1, static} & \text{if} \; \| N_1 \| \leq \lambda_{B1, static} \\
N_1 & \text{if} \;  \lambda_{B1, static} \leq \| N_1 \|
\end{array} \right. \; , \\
\lambda_{B2}=\left\{ \begin{array}{ll} \lambda_{B2, static} & \text{if} \; \| N_1 \| \leq \lambda_{B2, static} \\
N_2 & \text{if} \;  \lambda_{B2, static} \leq \| N_2 \|
\end{array} \right. \; .
\end{align}
Therefore, the motion of the Furuta pendulum will certainly influence the tangential frictional moments

\section{Simplification of EoM}
Pendulums usually have slender rotational symmetric arms, which indicates that the moment of inertia along the axis of arms is negligible, and the moments of inertial in two of the principal axes are equal. Thus, the inertial tensors of driving and pendulum arms can be simplified into
\begin{align*}
\mathbf J_1= \begin{pmatrix} 0 & 0 & 0 \\ 0 & J_{1} & 0  \\ 0 & 0 & J_{1} \end{pmatrix}, \quad \mathbf J_2= \begin{pmatrix} 0 & 0 & 0 \\ 0 & J_{2} & 0  \\ 0 & 0 & J_{2} \end{pmatrix}.
\end{align*}
Then the matrices of the equation (\ref{eom}) can be written as
\[
\mathbf{M}=\begin{pmatrix} J_{1}+m_1 c_1^2+m_2 l_1^2 + (J_{2}+m_2 c_2^2) \sin^2 \theta_2 & m_2 l_1 c_2 \cos \theta_2 \\
m_2 l_1 c_2 \cos \theta_2 & m_2 c_2^2 + J_{2} 
\end{pmatrix} \; ,
\]
\[
\mathbf{h}=\begin{pmatrix}
\dot{\theta}_2^2 m_2 l_1 c_2 \sin \theta_2 - \dot{\theta}_1 \dot{\theta}_2 \sin 2\theta_2 (m_2 c_2^2 + J_{2})  \\
\frac{1}{2}\dot{\theta}_1^2 \sin 2\theta_2 (m_2 c_2^2 + J_{2}) - g m_2 c_2 \sin \theta_2
\end{pmatrix} \; .
\]
Then the second-order equation can be represented by
\begin{align}
\label{motion_pendulum1}
& \dot{\mathbf{q}}_\theta = \begin{pmatrix} \dot{\theta}_1 \\ \dot{\theta}_2 \end{pmatrix} = \mathbf{v}_\theta \; ,
\\
& \dot{\mathbf{v}}_\theta=
\begin{pmatrix}
\ddot{\theta}_1 \\ \ddot{\theta}_2
\end{pmatrix}
= \mathbf{M}^{-1} \left( \mathbf{h}  + \begin{pmatrix}
M_{f1} \\ M_{f_2}
\end{pmatrix} +  \begin{pmatrix}
\tau \\ 0
\end{pmatrix}\right) \; .
\label{motion_pendulum2}
\end{align}

\section{Numerical Implementation with Coupled Frictions}
Employing Moreau's midpoint rule again, the impact equation of motion on velocity level can be obtained by
\begin{align}
  & \mathbf v^{(i+1)}_\theta = \begin{pmatrix}  \dot{\theta}_1^{(i)} \\ \dot{\theta}_2^{(i)} \end{pmatrix}+\mathbf{M}^{-1}_M \left[ \left( \mathbf{h}_M(\mathbf{q}_\theta^{(i)}+\frac{\Delta t}{2}\mathbf{v}_\theta^{(i)},\mathbf{v}_\theta^{(i)})  + \begin{pmatrix} \tau \\  0 \end{pmatrix} \right) \Delta t + \begin{pmatrix} \Lambda_{M1}^{(i+1)} \\ \Lambda_{M2}^{(i+1)} \end{pmatrix} \right] \;, \\
  & \mathbf q_\theta^{(i+1)} = \mathbf q_\theta^{(i)}+\frac{\mathbf v_\theta^{(i+1)}+ \mathbf v_\theta^{(i)}}{2}{\Delta t},
\end{align}
with the frictional moment impulse on the joint of driving arm $\Lambda_{M1}^{(i+1)}$ 
\begin{align}
\label{first_moment1}
  & \Lambda_{M1}^{(i+1)} = -\frac{\dot{\theta}_1^{(i+1)}}{\| \dot{\theta}^{(i+1)}_1 \|}\mu \lambda_{B1} R_{E1} \Delta t \quad \text{for } \dot{\theta}^{(i+1)}_1 \neq 0 \;, \\
  & \|\Lambda_{M1}^{(i+1)}\| \leq \mu  \lambda_{B1} R_{E1} \Delta t \quad \text{for } \dot{\theta}^{(i+1)}_1 = 0 \; ,
\label{first_moment2}
\end{align}
and that on the joint of pendulum $\Lambda_{M2}^{(i+1)}$ 
\begin{align}
\label{second_moment1}
  & \Lambda_{M2}^{(i+1)} = -\frac{\dot{\theta}_2^{(i+1)}}{\| \dot{\theta}^{(i+1)}_2 \|}\mu \lambda_{B2} R_{E2} \Delta t \quad \text{for } \dot{\theta}^{(i+1)}_2 \neq 0 \;, \\
  & \|\Lambda_{M2}^{(i+1)}\| \leq \mu  \lambda_{B2} R_{E2} \Delta t \quad \text{for } \dot{\theta}^{(i+1)}_2 = 0 \; ,
\label{second_moment2}
\end{align}
For an implementation, the first active frictional moment equations \eqref{first_moment1}-\eqref{first_moment2} and the second active frictional equations \eqref{second_moment1}-\eqref{second_moment2} can be written as
\begin{align}
  f_1(\Lambda_{M1}^{(i+1)},\dot{\theta}^{(i+1)}_1) &:= \Lambda_{M1}^{(i+1)} - \text{prox}_{C_1(\mu \lambda_{B1} R_{E1} \Delta t)}(\Lambda_{M1}^{(i+1)}-r_1 \dot{\theta}^{(i+1)}_1) = 0\; , \\
  f_2(\Lambda_{M2}^{(i+1)},\dot{\theta}^{(i+1)}_2) &:= \Lambda_{M2}^{(i+1)} - \text{prox}_{C_2(\mu \lambda_{B2} R_{E2} \Delta t)}(\Lambda_{M2}^{(i+1)}-r_2 \dot{\theta}^{(i+1)}_2) = 0\; ,
\end{align}
with the proximal point
\begin{align}
 & \text{prox}_{C_1(\mu \lambda_{B1} R_{E1} \Delta t)}(x)=\arg\min_{x^*\in C_1(\mu \lambda_{B1} R_{E1} \Delta t)}\|x-x^*\| \; , \\
 & \text{prox}_{C_2(\mu \lambda_{B2} R_{E2} \Delta t)}(x)=\arg\min_{x^*\in C_2(\mu \lambda_{B2} R_{E2} \Delta t)}\|x-x^*\| \;,
\end{align}
arbitrary parameters $r_1>0 \, , r_2>0$, the friction discs
\begin{align}
&  C_1(\mu \lambda_{B1} R_{E1} \Delta t) = S^0_{\mu \lambda_{B1} R_{E1} \Delta t} = \{x\in\mathbb{R}\;|\;\|x\|\leq \mu \lambda_{B1} R_{E1} \Delta t\} \;, \\
&  C_2(\mu \lambda_{B2} R_{E2} \Delta t) = S^0_{\mu \lambda_{B2} R_{E2} \Delta t} = \{x\in\mathbb{R}\;|\;\|x\|\leq \mu \lambda_{B2} R_{E2} \Delta t\} \;.
\end{align}
The functions $f_1$ and $f_2$ can be both evaluated by distinguishing different cases
\begin{align}
	 f_1(\Lambda_{M1}^{(i+1)},\dot{\theta}^{(i+1)}_1) &= \begin{cases}r_1 \dot{\theta}^{(i+1)}_1 & \text{if }\|\Lambda_{M1}^{(i+1)}-r_1 \dot{\theta}^{(i+1)}_1\|\leq\mu \lambda_{B1} R_{E1} \Delta t\\ \Lambda_{M1}^{(i+1)}-\frac{\Lambda_{M1}^{(i+1)}-r_1 \dot{\theta}^{(i+1)}_1}{\|\Lambda_{M1}^{(i+1)}-r_1 \dot{\theta}^{(i+1)}_x\|}\mu \lambda_{B1} R_{E1} \Delta t & \text{if }\|\Lambda_{M1}^{(i+1)}-r_1 \dot{\theta}^{(i+1)}_1\|>\mu \lambda_{B1} R_{E1} \Delta t \end{cases} \;, \\
        f_2(\Lambda_{M2}^{(i+1)},\dot{\theta}^{(i+1)}_2) &= \begin{cases}r_2 \dot{\theta}^{(i+1)}_2 & \text{if }\|\Lambda_{M2}^{(i+1)}-r_2 \dot{\theta}^{(i+1)}_2\|\leq\mu \lambda_{B2} R_{E2} \Delta t\\ \Lambda_{M2}^{(i+1)}-\frac{\Lambda_{M2}^{(i+1)}-r_2 \dot{\theta}^{(i+1)}_2}{\|\Lambda_{M2}^{(i+1)}-r_2 \dot{\theta}^{(i+1)}_x\|}\mu \lambda_{B2} R_{E2} \Delta t & \text{if }\|\Lambda_{M2}^{(i+1)}-r_2 \dot{\theta}^{(i+1)}_2\|>\mu \lambda_{B2} R_{E2} \Delta t \end{cases} \;.
\end{align}
Typically, one inserts the velocity formulas $\dot{\theta}^{(i+1)}_1$, $\dot{\theta}^{(i+1)}_2$ into $f_1, f_2$ and gets nonsmooth and nonlinear equations for $\Lambda_{M1}^{(i+1)}, \Lambda_{M2}^{(i+1)}$:
\begin{align}
  &	f_{1} \left(\Lambda_{M1}^{(i+1)}, \dot{\theta}^{(i)}_1+\left\{ \mathbf{M}^{-1}_M \left[ \left( \mathbf{h}_M(\mathbf{q}_\theta^{(i)}+\frac{\Delta t}{2}\mathbf{v}_\theta^{(i)},\mathbf{v}_\theta^{(i)})  + \begin{pmatrix} \tau \\  0 \end{pmatrix} \right) \Delta t + \begin{pmatrix} \Lambda_{M1}^{(i+1)} \\ \Lambda_{M2}^{(i+1)} \end{pmatrix} \right] \right\}_{1\bullet} \right) = 0 \;, \label{first_nonsmooth} \\
  &	f_{2} \left(\Lambda_{M2}^{(i+1)}, \dot{\theta}^{(i)}_2+\left\{ \mathbf{M}^{-1}_M \left[ \left( \mathbf{h}_M(\mathbf{q}_\theta^{(i)}+\frac{\Delta t}{2}\mathbf{v}_\theta^{(i)},\mathbf{v}_\theta^{(i)})  + \begin{pmatrix} \tau \\  0 \end{pmatrix} \right) \Delta t + \begin{pmatrix} \Lambda_{M1}^{(i+1)} \\ \Lambda_{M2}^{(i+1)} \end{pmatrix} \right] \right\}_{2\bullet} \right) = 0 \; ,
\label{second_nonsmooth}
\end{align}
where the index $1\bullet, 2\bullet$ indicate the first line and the second line of the matrix.
These two equations can be solved by fixed-point or root-finding schemes. After that one calculates $\dot{\theta}^{(i+1)}_1, \dot{\theta}^{(i+1)}_2$ and $\theta^{(i+1)}_1, \theta^{(i+1)}_2$ by the discrete equation of motion.\par
\newpage
\subsection{Algorithm description of coupled frictions system}
A detailed algorithmic description of the fixed-point iteration scheme for solving the nonsmooth equations \eqref{first_nonsmooth}-\eqref{second_nonsmooth} with coupled frictional moment impulses is presented in Table~\ref{algo_coupled}.

\begin{table}[hbt]
\centering
    \caption{\label{algo_coupled}Algorithm for $\Lambda_{M1}, \Lambda_{M2}$}
    \begin{tabular}{ l }
    \hline
    Evaluate $\Lambda_{M1}, \Lambda_{M2}$ at every step \\ [3pt]
    (with $\Delta t$, and $\theta^{(i)}_1, \dot{\theta}^{(i)}_1, \theta^{(i)}_2, \dot{\theta}^{(i)}_2, \mathbf{h}_M, \mathbf M_M, \tau$ the given constants) \\ [4pt] \hline
    Set tolerance $\varepsilon$, parameter $r$, and initial guess $\Lambda_{M1(0)}^{(i+1)}=0, \Lambda_{M2(0)}^{(i+1)}=0$\\ [5pt] \hline
    While Loop ($j$th iteration with $j_{max}$) \\ [3pt] \hline
    \quad\quad Calculate $\mathbf v_{\theta (j)}^{(i+1)}=\mathbf v_{\theta}^{(i)}+ \mathbf{M}^{-1}_M \left[ \left( \mathbf{h}_M(\mathbf{q}_\theta^{(i)}+\frac{\Delta t}{2}\mathbf{v}_\theta^{(i)},\mathbf{v}_\theta^{(i)})  + \begin{pmatrix} \tau \\  0 \end{pmatrix} \right) \Delta t + \begin{pmatrix} \Lambda_{M1}^{(i+1)} \\ \Lambda_{M2}^{(i+1)} \end{pmatrix} \right] $ \\ [13pt] \hline

\quad \quad 1, for $\Lambda_{M1(j)}^{(i+1)}$: \\ [5pt]
\quad \quad \quad \quad if $\|\Lambda_{M1(j)}^{(i+1)} - r_1 \dot{\theta}^{(i+1)}_{1(j)}\| \leq \mu \lambda_{B1} R_{E1} \Delta t$  \\ [5pt]
\quad \quad \quad \quad \quad \quad $\Lambda_{M1(j+1)}^{i+1}=- \left(\mathbf{M}_M \mathbf{v}_\theta^{(i)} \right)_{1\bullet} - \left[ \left(\mathbf{h}_M^{(i)} \right)_{1\bullet} + \tau \right] \Delta t$ \\ [5pt]
\quad \quad \quad \quad  else $\|\Lambda_{M1(j)}^{(i+1)} - r_1 \dot{\theta}^{(i+1)}_{1(j)} \| > \mu \lambda_{B1} R_{E1} \Delta t$ \\ [5pt]
\quad \quad \quad \quad \quad \quad $\Lambda_{M1(j+1)}^{i+1}=\frac{\Lambda_{M1(j)}^{(i+1)}-r_1 \dot{\theta}^{(i+1)}_{1(j)}}{\| \Lambda_{M1(j)}^{(i+1)} - r_1 \dot{\theta}^{(i+1)}_{1(j)}\|}\mu \lambda_{B1} R_{E1} \Delta t$ \\ [10pt]

\quad \quad 2, for $\Lambda_{M2(j)}^{(i+1)}$: \\ [5pt]
\quad \quad    \quad \quad if $\|\Lambda_{M2(j)}^{(i+1)} - r_2 \dot{\theta}^{(i+1)}_{2(j)}\| \leq \mu \lambda_{B2} R_{E2} \Delta t$  \\ [5pt]
\quad \quad    \quad \quad \quad \quad $\Lambda_{M2(j+1)}^{i+1}=- \left(\mathbf{M}_M \mathbf{v}_\theta^{(i)} \right)_{2\bullet} - \left(\mathbf{h}_M^{(i)} \right)_{2\bullet} \Delta t$ \\ [5pt]
\quad \quad    \quad \quad  else $\|\Lambda_{M2(j)}^{(i+1)} - r_2 \dot{\theta}^{(i+1)}_{2(j)} \| > \mu \lambda_{B2} R_{E2} \Delta t$ \\ [5pt]
\quad \quad    \quad \quad \quad \quad $\Lambda_{M2(j+1)}^{i+1}=\frac{\Lambda_{M2(j)}^{(i+1)}-r_2 \dot{\theta}^{(i+1)}_{2(j)}}{\| \Lambda_{M2(j)}^{(i+1)} - r_2 \dot{\theta}^{(i+1)}_{2(j)}\|}\mu \lambda_{B2} R_{E2} \Delta t$ \\ [12pt] \hline
    \quad \quad Calculate the differences $\sigma_1=\| \Lambda_{M1(j+1)}^{(i+1)} - \Lambda_{M1(j)}^{(i+1)} \|$ and $\sigma_2=\| \Lambda_{M2(j+1)}^{(i+1)} - \Lambda_{M2(j)}^{(i+1)} \|$ \\ [5pt] \hline
    \quad \quad if $\sigma_1<\varepsilon$ and $\sigma_2<\varepsilon$ \\ [3pt]
    \quad \quad \quad \quad Return $\Lambda_{M1(j+1)}^{(i+1)}$ and $\Lambda_{M2(j+1)}^{(i+1)}$ \\ [3pt]
    \quad \quad else if $j>j_{max}$ \\ [3pt]
     \quad \quad \quad \quad Change the value of $r$, return to the iteration loop \\ [3pt]
      \quad \quad else if $j>2 j_{max}$ \\ [3pt]
    \quad \quad \quad \quad Return $\Lambda_{M1(j+1)}^{(i+1)}$ and $\Lambda_{M2(j+1)}^{(i+1)}$, ''Desired tolerance cannot be reached'' \\ [5pt] \hline
    \end{tabular}
\end{table}

\newpage
\subsection{Simulation results}
\begin{figure}[hbt]
\centering
 \begin{subfigmatrix}{2}
  \subfigure[$\theta_1$]{\includegraphics{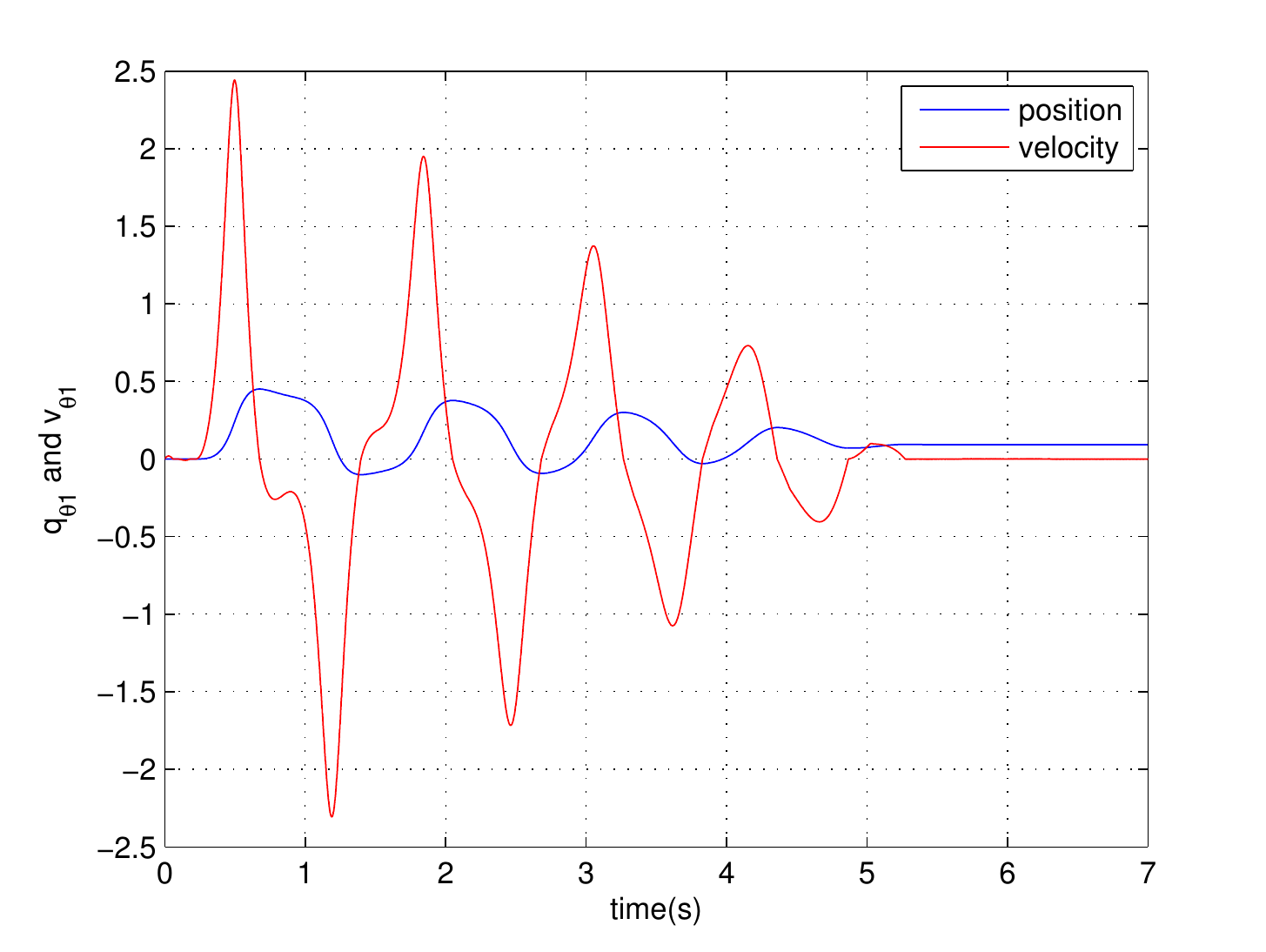}}
  \subfigure[$\theta_2$]{\includegraphics{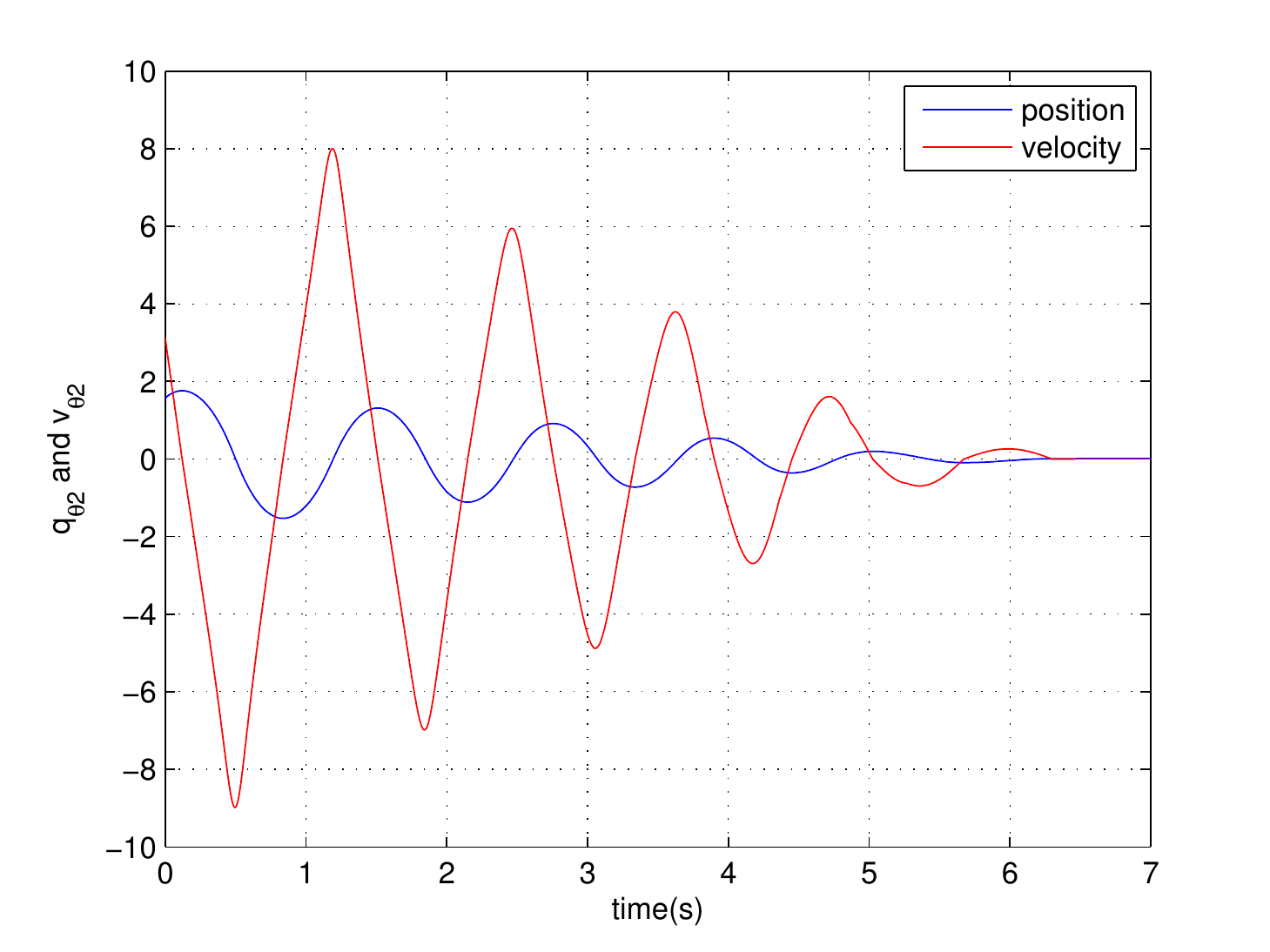}}
 \end{subfigmatrix}
 \caption{Pendulum motion without control torque}
 \label{fig:pendulum_motion}
\end{figure}

\begin{figure}
  \centering
  \includegraphics[width=0.85\linewidth]{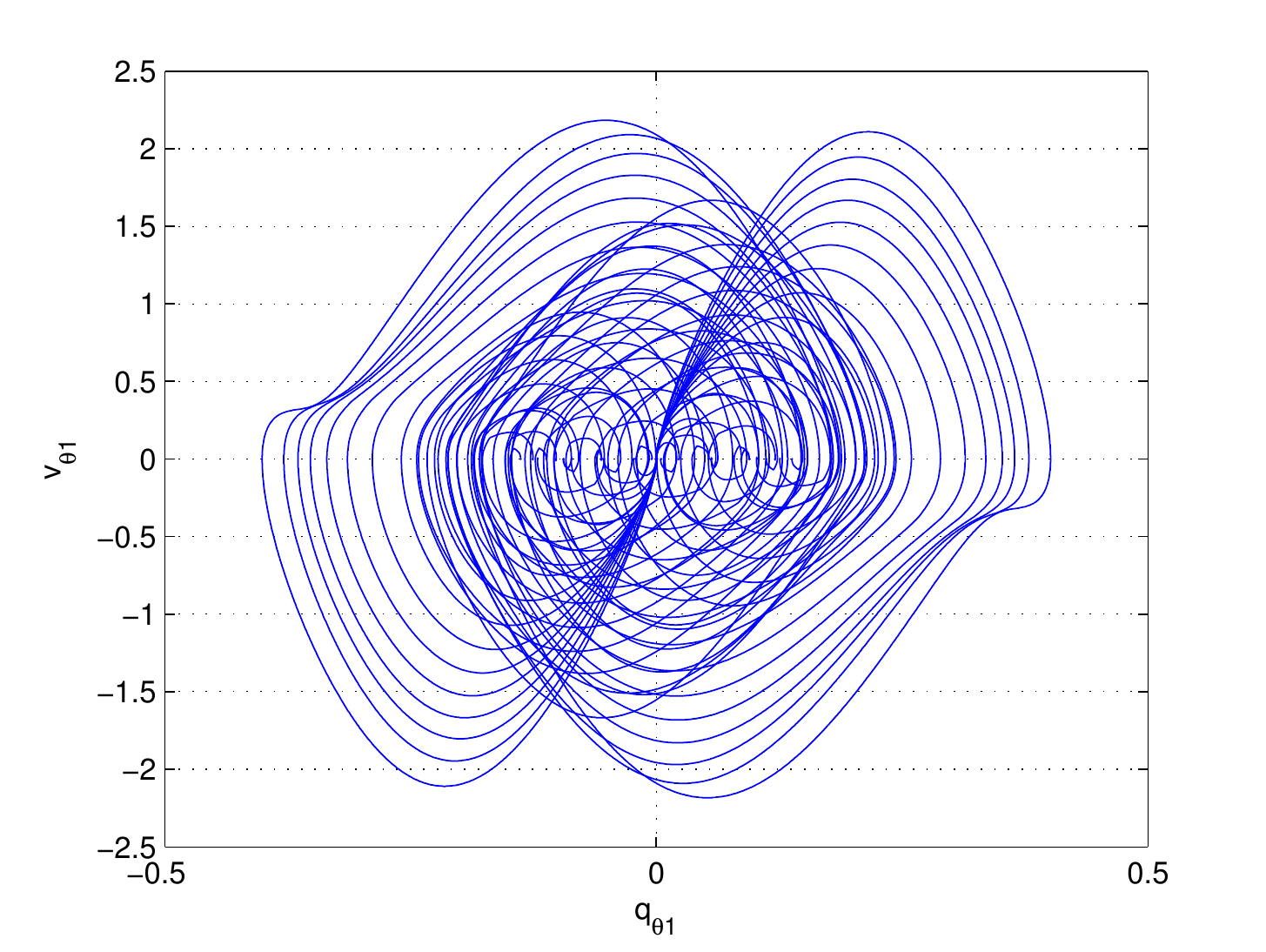}
  \caption{Phase diagram of pendulum free behavior on $\theta_1$}
  \label{fig:phase_theta1}
\end{figure}

\begin{figure}
  \centering
  \includegraphics[width=0.85\linewidth]{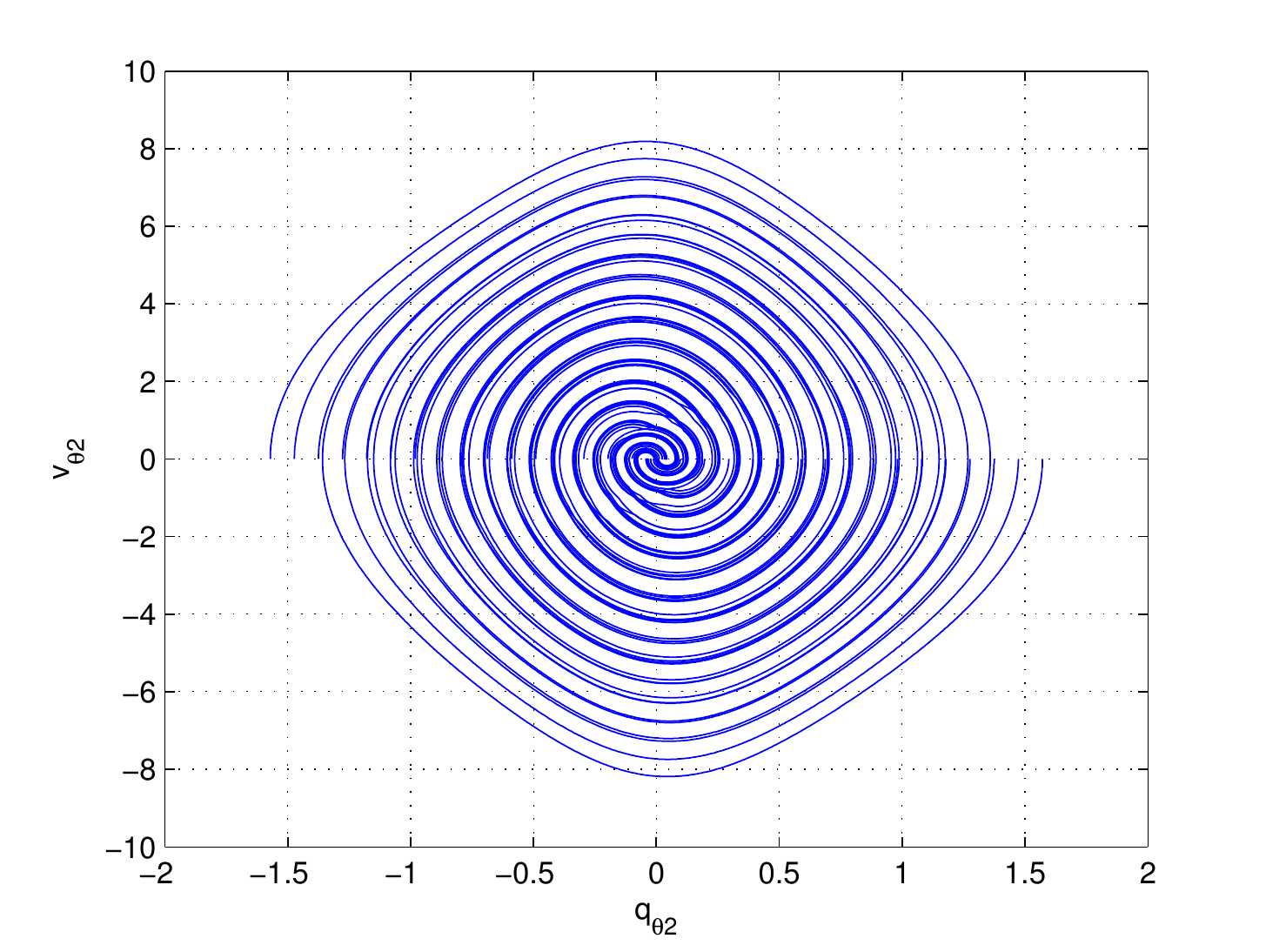}
  \caption{Phase diagram of pendulum free behavior on $\theta_2$}
  \label{fig:phase_theta2}
\end{figure}

The parameters of the pendulum of this example are presented in the table \ref{parameter} in Appendix \ref{AppendixA}. The initial conditions are given by
\[
  \mathbf q_\theta = \begin{pmatrix}  \theta_1 \\  \theta_2 \end{pmatrix} = \begin{pmatrix} 0 \\ \pi/2 \end{pmatrix}\unit{m} \;, \quad \mathbf v_\theta = \begin{pmatrix}  \dot{\theta}_1 \\  \dot{\theta}_2 \end{pmatrix} = \begin{pmatrix} 0 \\ \pi \end{pmatrix}\unitfrac{m}{s} \;,
\]
and the external torque is set to zero
\[
\tau=0 \; .
\]
The simulation result of pendulum motion without control are shown in Fig. \ref{fig:pendulum_motion}.
It is noticed that this coupled frictional case has been solved successfully. 
In order to obtain an insight of the pendulum behavior, we determine the initial condition space by 
\begin{align*}
& \theta_1=0, \quad \theta_2 \in [\frac{-\pi}{2}, \frac{\pi}{2}] \;, \\
& \dot{\theta}_1=0, \quad \dot{\theta}_2=0 \;.
\end{align*}
The phase diagrams of this system on $\theta_1, \theta_2$ are presented in Fig. \ref{fig:phase_theta1} and Fig. \ref{fig:phase_theta2}. As the simulated results shows, the attraction lines lie on $\dot{\theta}_1=\unitfrac[0]{m}{s}, \dot{\theta}_2=\unitfrac[0]{m}{s}$.



\chapter{Impulse-based Control of Frictional Furuta Pendulum} 

\label{Chapter5} 

\lhead{Chapter 5. \emph{Impulse-based Control of Frictional Furuta Pendulum}} 

\section{Impulse-based Control Law}
Similar to the control method of the frictional oscillator in Chapter \ref{Chapter3}, the controller of the Furuta pendulum consists of a classical state-feedback control torque $u_\tau$ and an impulse control torque $U_\tau$. The impulse control torque applied on the driving arm is restricted to the impact equation that
\begin{align}
\begin{pmatrix}U_\tau(\bm q_\theta, \bm v_\theta) \\ 0 \end{pmatrix} =  \mathbf{M}(\bm{v}_{\theta}^+(t_i) - \bm{v}_{\theta}^-(t_i) )  \; ,
\end{align}
which implies that
\begin{align}
\label{impulse_torque_pre}
U_\tau(\bm q_\theta, \bm v_\theta) & =
\begin{pmatrix}J_{1}+m_1 c_1^2+m_2 l_1^2 + (J_{2}+m_2 c_2^2) \sin^2 \theta_2 \\ m_2 l_1 c_2 \cos \theta_2 \end{pmatrix}^{T}
\begin{pmatrix}
\dot{\theta}_1^+(t_i) - \dot{\theta}_1^-(t_i) \\
\dot{\theta}_2^+(t_i) - \dot{\theta}_2^-(t_i)
\end{pmatrix} \; , \\
 0 & = 
\begin{pmatrix}
m_2 l_1 c_2 \cos \theta_2 & m_2 c_2^2 + J_{2} 
\end{pmatrix}
\begin{pmatrix}
\dot{\theta}_1^+(t_i) - \dot{\theta}_1^-(t_i) \\
\dot{\theta}_2^+(t_i) - \dot{\theta}_2^-(t_i)
\end{pmatrix} \; .
\end{align}
After the rearrangement, we have
\begin{align}
\label{arrange1}
\dot{\theta}_2^+(t_i) - \dot{\theta}_2^-(t_i) = -\frac{m_2 l_1 c_2 \cos \theta_2}{m_2 c_2^2 + J_{2}} \left( \dot{\theta}_1^+(t_i) - \dot{\theta}_1^-(t_i) \right) \; , \\
\dot{\theta}_1^+(t_i) - \dot{\theta}_1^-(t_i) = -\frac{m_2 c_2^2 + J_{2}}{m_2 l_1 c_2 \cos \theta_2} \left( \dot{\theta}_2^+(t_i) - \dot{\theta}_2^-(t_i) \right) \; .
\label{arrange2}
\end{align}
Substitute equations \eqref{arrange1} - \eqref{arrange2} into \eqref{impulse_torque_pre}, we have two representations of the impulsive torque
\begin{align}
U_\tau(\theta_2, \dot{\theta}_1)=
\begin{pmatrix}J_{1}+m_1 c_1^2+m_2 l_1^2 + (J_{2}+m_2 c_2^2) \sin^2 \theta_2 \\ m_2 l_1 c_2 \cos \theta_2 \end{pmatrix}^{T}
\begin{pmatrix}
1 \\ -\frac{m_2 l_1 c_2 \cos \theta_2}{m_2 c_2^2 + J_{2}}
\end{pmatrix}
\left( \dot{\theta}_1^+(t_i) - \dot{\theta}_1^-(t_i) \right) \; , \\
U_\tau(\theta_2, \dot{\theta}_2)=
\begin{pmatrix}J_{1}+m_1 c_1^2+m_2 l_1^2 + (J_{2}+m_2 c_2^2) \sin^2 \theta_2 \\ m_2 l_1 c_2 \cos \theta_2 \end{pmatrix}^{T}
\begin{pmatrix}
 -\frac{m_2 c_2^2 + J_{2}}{m_2 l_1 c_2 \cos \theta_2}  \\ 1
\end{pmatrix}
\left( \dot{\theta}_2^+(t_i) - \dot{\theta}_2^-(t_i) \right) \; .
\end{align}
The impulsive control torque $U_\tau$ is applied on the driving arm when $\dot{\theta}_1^-(t_i)=0$ or $\dot{\theta}_2^-(t_i)=0$, except the Furuta pendulum has been stabilized on the inverted position that $\theta_2(t_i)=\pi/2+ i \, 2\pi, i \in \mathbb N$.
\par
Regarding the motion of Furuta pendulum, the equations \eqref{motion_pendulum1} - \eqref{motion_pendulum2} adding impulse can be rewritten into the form of differential inclusion that
\begin{align}
\mathrm{d}{\mathbf{q}}_\theta & = \mathbf{v}_\theta \mathrm{d}t \; ,
\\
\mathbf{M} \, \mathrm{d}{\mathbf{v}}_\theta & \in
\left( \mathbf{h}  + \begin{pmatrix}
M_{f1} \\ M_{f_2}
\end{pmatrix} \right) \mathrm{d}t +  
\begin{pmatrix}
u_\tau \\ 0
\end{pmatrix}\mathrm{d}t +  
\begin{pmatrix}
U_\tau \\ 0
\end{pmatrix}\mathrm{d} \eta\; .
\end{align}
where $\mathrm{d} t$ is the Lebesgue measure, $\mathrm{d} \eta$ is a differential atomic measure consisting of a sum of Dirac point measures. The state feedback control law is given by
\begin{equation}
u_\tau(\bm q_\theta, \bm v_\theta)=-\begin{pmatrix}
k_1 & k_2 & k_3 & k_4 
\end{pmatrix}
\begin{pmatrix}
\theta_1-\theta_{ref}\\ \dot{\theta}_1 \\ \theta_2 - (\pi + i \, 2\pi) \\ \dot{\theta}_2
\end{pmatrix}
\; , \quad i= \left\lfloor \frac{\theta_2}{2\pi} \right\rfloor , \quad k_1, k_2,k_3,k_4>0 \; , 
\end{equation}
where $\lfloor \bullet \rfloor$ indicates the operation of rounding down. The impulse control law is given by
\begin{equation}
\label{impulse_control_law}
 U_\tau(\bm q_\theta, \bm v_\theta^-)=\left\{ \begin{array}{ll}
U_{\tau 1} (\theta_2, \dot{\theta}_1^-) & \text{if}   \;\; \dot{\theta}_1^-=0  \wedge \dot{\theta}_2 \neq 0 \wedge \theta_2 \neq \pi/2+ i \, 2\pi  \\ 
U_{\tau 2} (\theta_2, \dot{\theta}_2^-) &  \text{if}   \;\; \dot{\theta}_2^-=0  \wedge \dot{\theta}_1 \neq 0 \wedge \theta_2 \neq \pi/2+ i \, 2\pi \\ 
U_{\tau 3} (\theta_2, \dot{\theta}_1^-, \dot{\theta}_2^-) & \text{if}   \;\; \dot{\theta}_2^-=0  \wedge \dot{\theta}_1^- = 0 \wedge \theta_2 \neq \pi/2+ i \, 2\pi \\ 
0 & \text{else} 
\end{array} \right. \; ,
\end{equation}
which indicates four distinguishing cases corresponding to sticking driving arm and moving pendulum arm, moving driving arm and sticking pendulum arm, sticking driving and pendulum arm outside the inverted position, and the sticking arms at the unstable equilibrium (inverted) position, respectively.
\par
Thus, once the slip-stick transition occurs outside at the inverted position, the impulse control torque will be applied on the system leading to the state jump at that time instant $t_i$. The state can be reseted as the followings
\begin{align}
&\bm q_\theta^+ (t_i)=\bm q_\theta^- (t_i) \; , \\
&\bm v_\theta^+ (t_i)=\bm v_\theta^- (t_i) + \mathbf M^{-1} \begin{pmatrix}U_\tau(\bm q_\theta, \bm v_\theta^-) \\ 0 \end{pmatrix} \; .
\end{align}
Marking the first sticking time instance of $\theta_1$ as $t_{\theta 1}$ and of $\theta_2$ as $t_{\theta 2}$, after this time instance, the feedback control law for $u_\tau$ is modified by
\begin{equation}
 k_2(t)=\left\{ \begin{split}\begin{aligned}
&k_{2pre} \quad \;  \text{for}   \;\; t_0\leq t < t_{\theta 1}  \\ 
&k_{2post} \quad  \text{for} \; \;  t\geq t_{\theta 1} 
\end{aligned}\end{split} \right. \; ,
\end{equation}
\begin{equation}
 k_4(t)=\left\{ \begin{split}\begin{aligned}
&k_{4pre} \quad \;  \text{for}   \;\; t_0\leq t < t_{\theta 2}  \\ 
&k_{4post} \quad  \text{for} \; \;  t\geq t_{\theta 2} 
\end{aligned}\end{split} \right. \; .
\end{equation}

\subsection{Estimating method and reduced control law}
According to the related impact equations \eqref{arrange1}-\eqref{arrange2}, it is noticed that the angular velocity jumps of $(\dot{\theta}_1^+(t_i) - \dot{\theta}_1^-(t_i))$ and $(\dot{\theta}_2^+(t_i) - \dot{\theta}_2^-(t_i))$ caused by the impulse $U_\tau$ are coupled. Thus, the three cases of $U_{\tau 1}, U_{\tau 2}, U_{\tau 3}$ in the impulse control law \eqref{impulse_control_law} are not really distinguishing, because the impulse control torque applied on the driving arm will certainly lead the angular velocity jump on both driving and pendulum arms. \par
Therefore, we rewrite the impulse control law into a reduced form of
\begin{equation}
\label{impulse_control_law_reduced}
 U_\tau(\bm q_\theta, \bm v_\theta^-)=\left\{ \begin{array}{ll}
U_{\tau}  & \text{if} \;  (\dot{\theta}_2^-=0 \vee \dot{\theta}_1^-=0) \wedge \theta_2 \neq \pi/2+ i \, 2\pi  \\ 
0  & \text{else} 
\end{array} \right. \; ,
\end{equation}
which indicates that once the sticking occurs outside the unstable equilibrium (inverted) position, the impulse torque are enforced. Because of the dissipation caused by the friction, and the spring and damping effects provided by the full-state feedback control law, it is sufficient to estimate the impulse control torque on the position level to push the pendulum arm to the inverted position. Thus, we employ the shooting method again for the estimation.
\par
Based on the dynamic equations \eqref{motion_pendulum1}-\eqref{motion_pendulum2}, we consider the case of moving driving arm and sticking pendulum arm as a demonstration, where the boundary conditions are given by
\[
\theta_2(t_{end})=\pi+2\pi i, i=\left\lfloor \frac{\theta_2}{2\pi}\right\rfloor, \quad \dot{\theta}_2(t_{end})=0, \quad \theta_2(t_0)=\theta_2^*.
\]
For the given initial guess $\theta_2^+(t_0)=s$ and the time sequence
\[
\mathbf t=[0,t_1,t_2,\dots, t_n]^T,
\]
the sequences of velocity and position can be numerically computed
\begin{align*}
& \mathbf v_\theta=[s, \dot{\theta}_2(t_1;s), \dot{\theta}_2(t_2;s),\dots, \dot{\theta}_2(t_n;s)]^T,\\
& \mathbf q_\theta=[\theta_2^*, \theta_2(t_1;s), \theta_2(t_2;s), \dots, \theta_2(t_n;s)]^T.
\end{align*}
The boundary constraints are in the form of
\begin{align}
\label{bnd_theta2_q}
F_q(s)=\theta_2(t_n;s)-(\pi + 2\pi i)=0 \; .
\end{align}
The zero root of (\ref{bnd_theta2_q}) can be iterated by
\begin{align}
s_q^{(i+1)}=s_q^{(i)}-\frac{F_q(s_q^{(i)})}{\Delta F_q(s_q^{(i)})}
\end{align}
where 
\[
\Delta F_q(s_q^{(i)})=\frac{F_q(s_q^{(i)}+\Delta s)-F_q(s_q^{(i)})}{\Delta s}
\]
with $\Delta s$ a given value. The algorithm for the shooting method is presented in Table \ref{algo_shooting}.
\begin{table}[hbt]
  \centering
  \caption{\label{algo_shooting}Algorithm for shooting method}
  \begin{tabular}{ l }
  \hline
    Evaluate $F_q(s)$, with $\Delta t$, and $\mathbf q_\theta(t_0), \mathbf v_\theta(t_0), \Delta s, i$ the given constants \\ [2pt] \hline
    Set tolerance $\varepsilon$, and initial guess $s^{(0)}=-\text{sign}(\theta_2-(\pi + 2\pi i)) \sqrt{10 \| \theta_2-(\pi + 2\pi i) \|}$\\ [3pt] \hline
    While Loop ($j$th iteration with $j_{max}$) \\ [2pt] \hline
    \quad\quad Calculate $\theta_2(t_n;s^{(i)})$ and $\theta_2(t_n;s^{(i)}+\Delta s)$ via numerical integration \\ [2pt] \hline
    \quad \quad if $\theta_2(t_n;s^{(i)}+\Delta s) - \theta_2(t_n;s^{(i)}) \geq \varepsilon$ \\ [2pt]
    \quad \quad \quad \quad $s_q^{(i+1)}=s_q^{(i)}-\frac{F_q(s_q^{(i)})}{\Delta F_q(s_q^{(i)})}$ \\[2pt]
    \quad \quad  else if $j \geq j_{max}$ \\ [2pt]
    \quad \quad \quad \quad  Return $s^{(i+1)}$,''Desired tolerance cannot be reached'' \\ [2pt]
    \quad \quad else  Return $s^{(i+1)}$ \\ [2pt] \hline
    \end{tabular}
\end{table}
\par
After the iteration, the velocity jumps can be calculated by
\begin{align}
& \dot{\theta}_2^+(t_i) - \dot{\theta}_2^-(t_i) = s^{(end)} \; , \\
& \dot{\theta}_1^+(t_i) - \dot{\theta}_1^-(t_i) = -\frac{m_2 c_2^2 + J_{2}}{m_2 l_1 c_2 \cos \theta_2} s^{(end)} \; ,
\end{align}
and the impulsive control torque can be obtained by
\begin{align}
U_\tau=
\begin{pmatrix}J_{1}+m_1 c_1^2+m_2 l_1^2 + (J_{2}+m_2 c_2^2) \sin^2 \theta_2 \\ m_2 l_1 c_2 \cos \theta_2 \end{pmatrix}^{T}
\begin{pmatrix}
 -\frac{m_2 c_2^2 + J_{2}}{m_2 l_1 c_2 \cos \theta_2}  \\ 1
\end{pmatrix}
s^{(end)}\; .
\end{align}

\section{Simulation Results}
\begin{figure}[hbt]
\centering
 \begin{subfigmatrix}{2}
  \subfigure[$(\theta_1,\dot{\theta}_1)$ time history]{\includegraphics{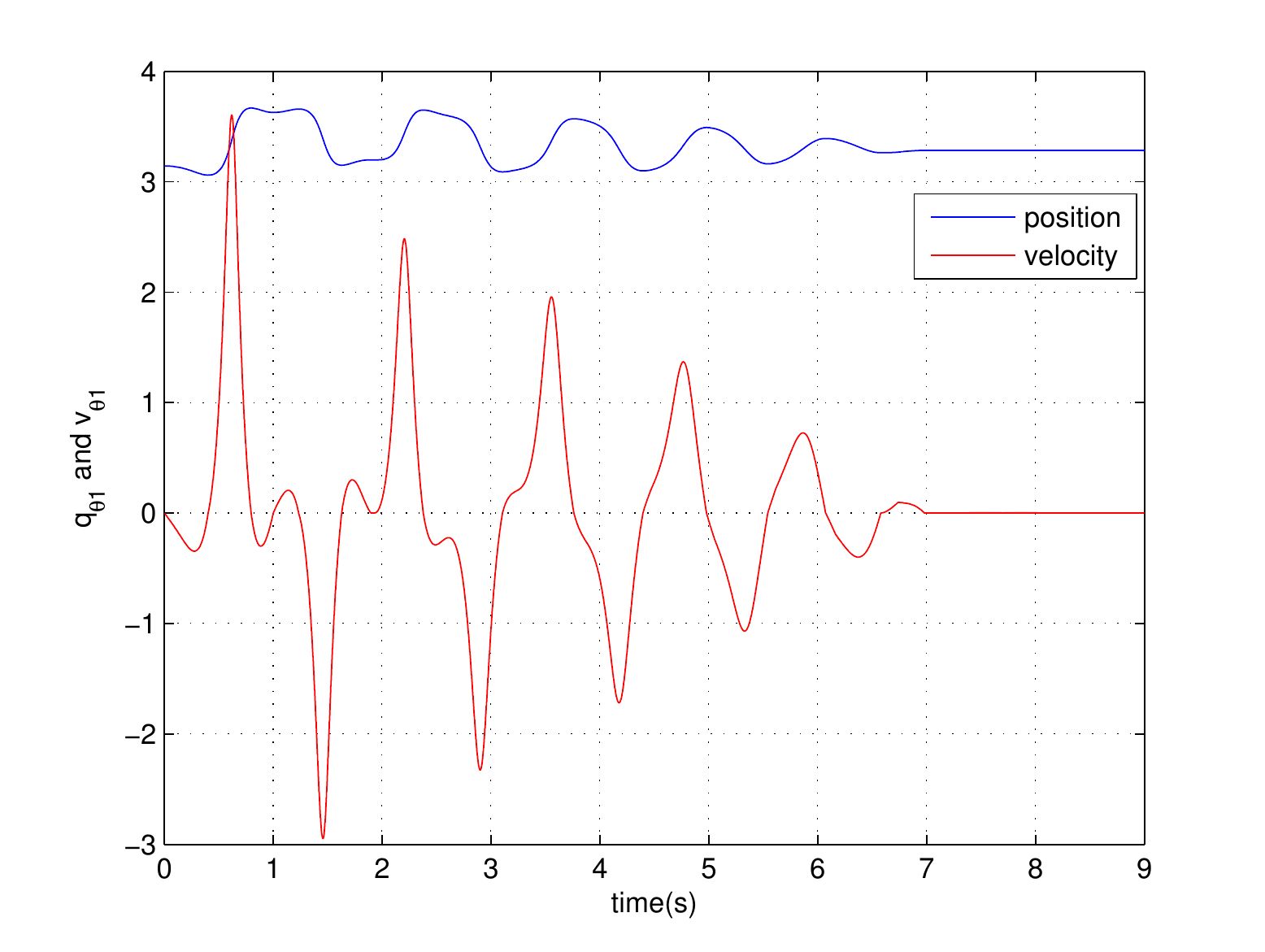}}
  \subfigure[$(\theta_2, \dot{\theta}_2)$ time history]{\includegraphics{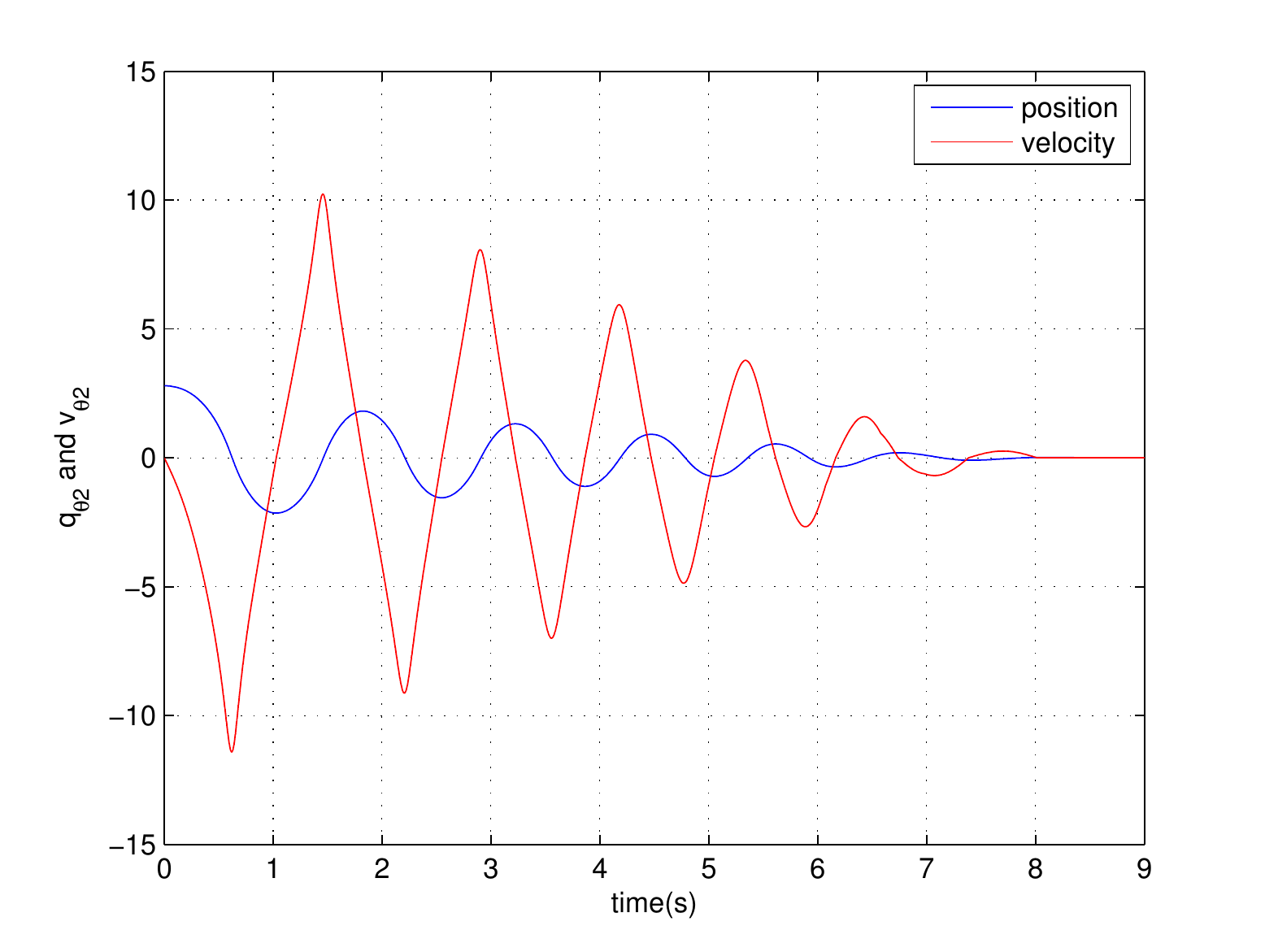}}
 \end{subfigmatrix}
 \caption{Pendulum motion without control}
 \label{fig:pendulum_no_control}
\end{figure}

\begin{figure}
\centering
 \begin{subfigmatrix}{1}
  \subfigure[$(\theta_1,\dot{\theta}_1)$ time history]{\includegraphics[width=0.8\linewidth]{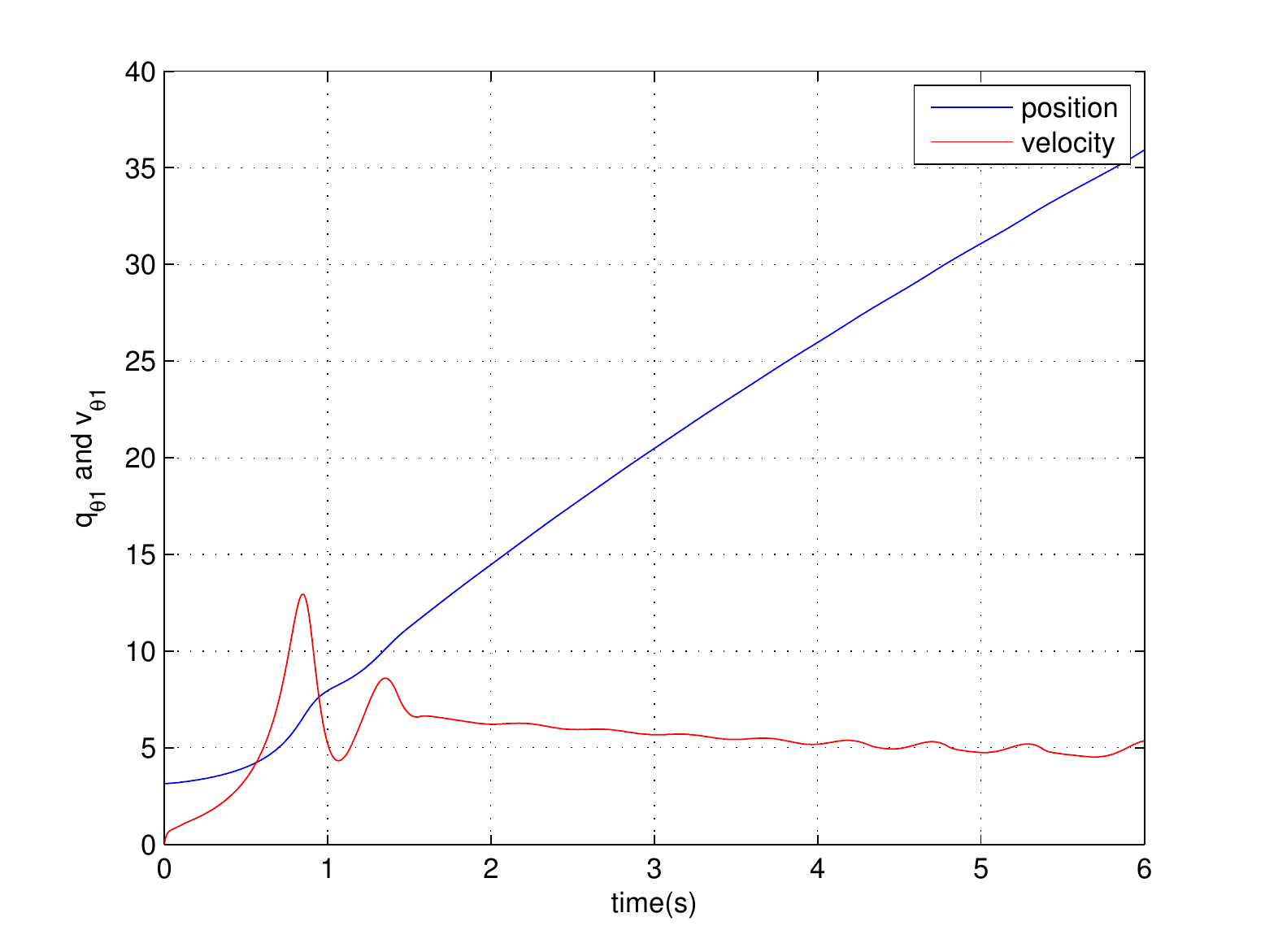}}
  \subfigure[$(\theta_2,\dot{\theta}_2)$ time history]{\includegraphics[width=0.8\linewidth]{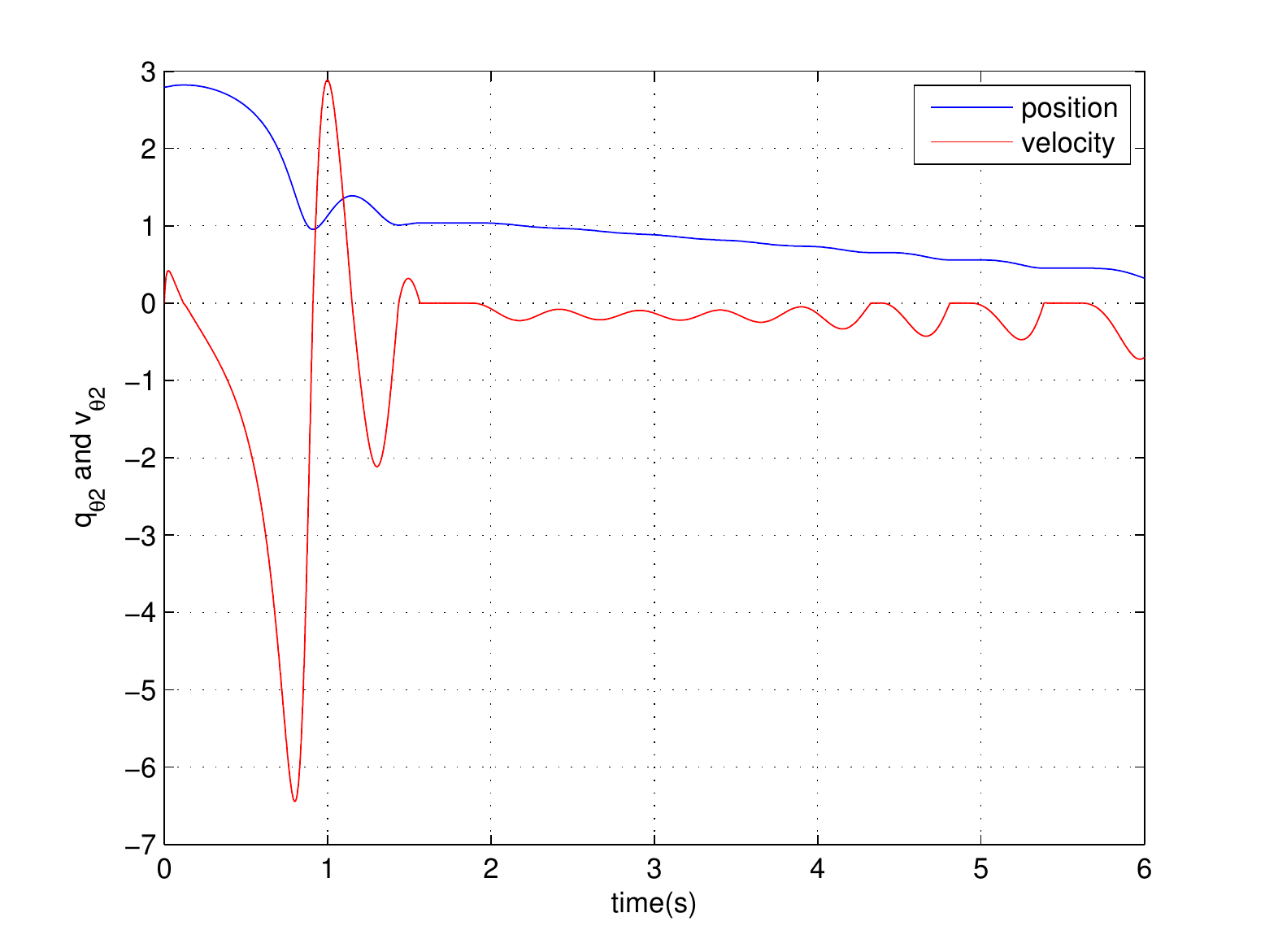}}
 \end{subfigmatrix}
 \caption{Pendulum motion with feedback control switch on}
 \label{fig:pendulum_feedback}
\end{figure}

\begin{figure}
\centering
 \begin{subfigmatrix}{1}
  \subfigure[$(\theta_1,\dot{\theta}_1)$ time history]{\includegraphics[width=0.8\linewidth]{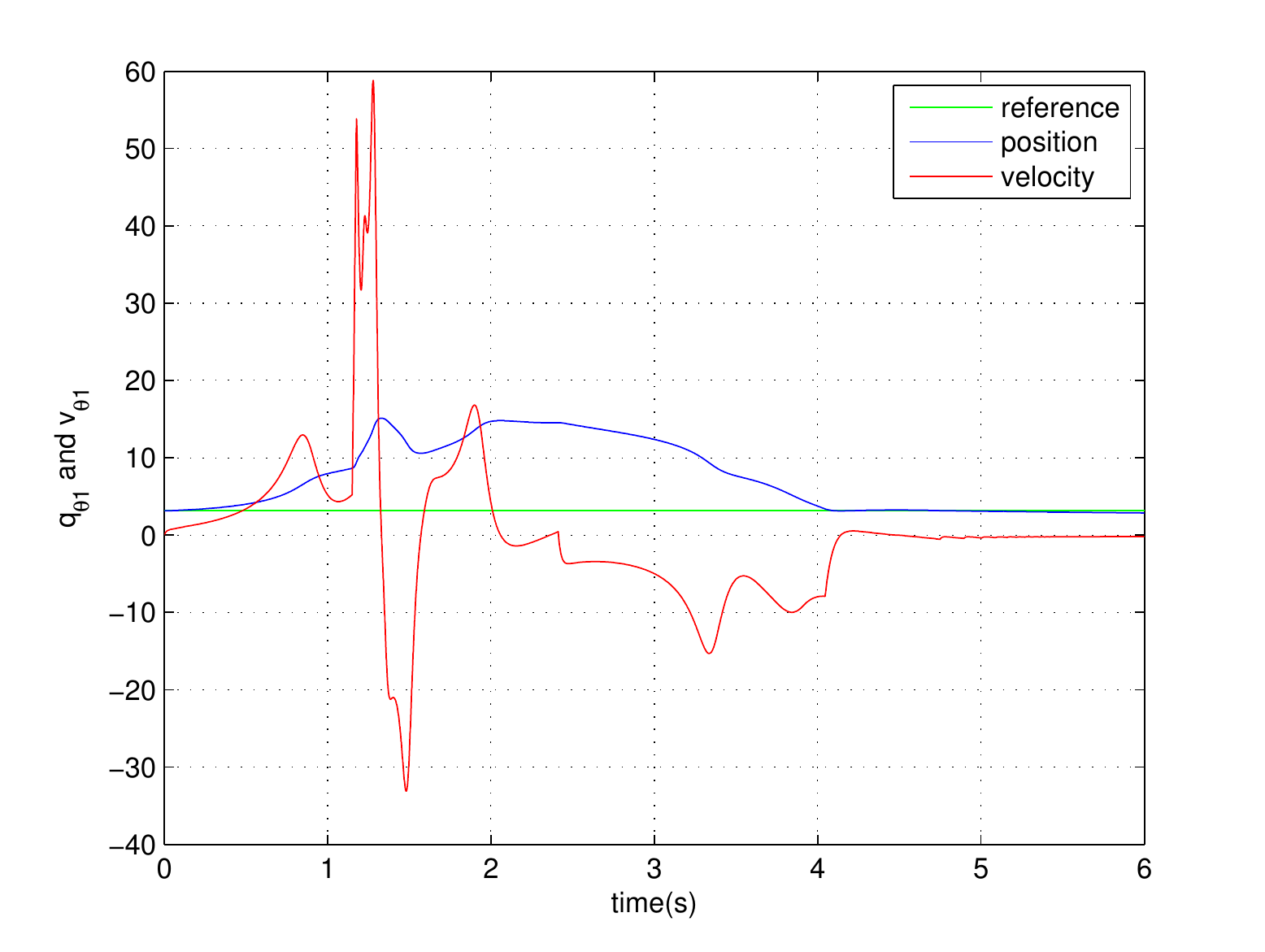}}
  \subfigure[$(\theta_2,\dot{\theta}_2)$ time history]{\includegraphics[width=0.8\linewidth]{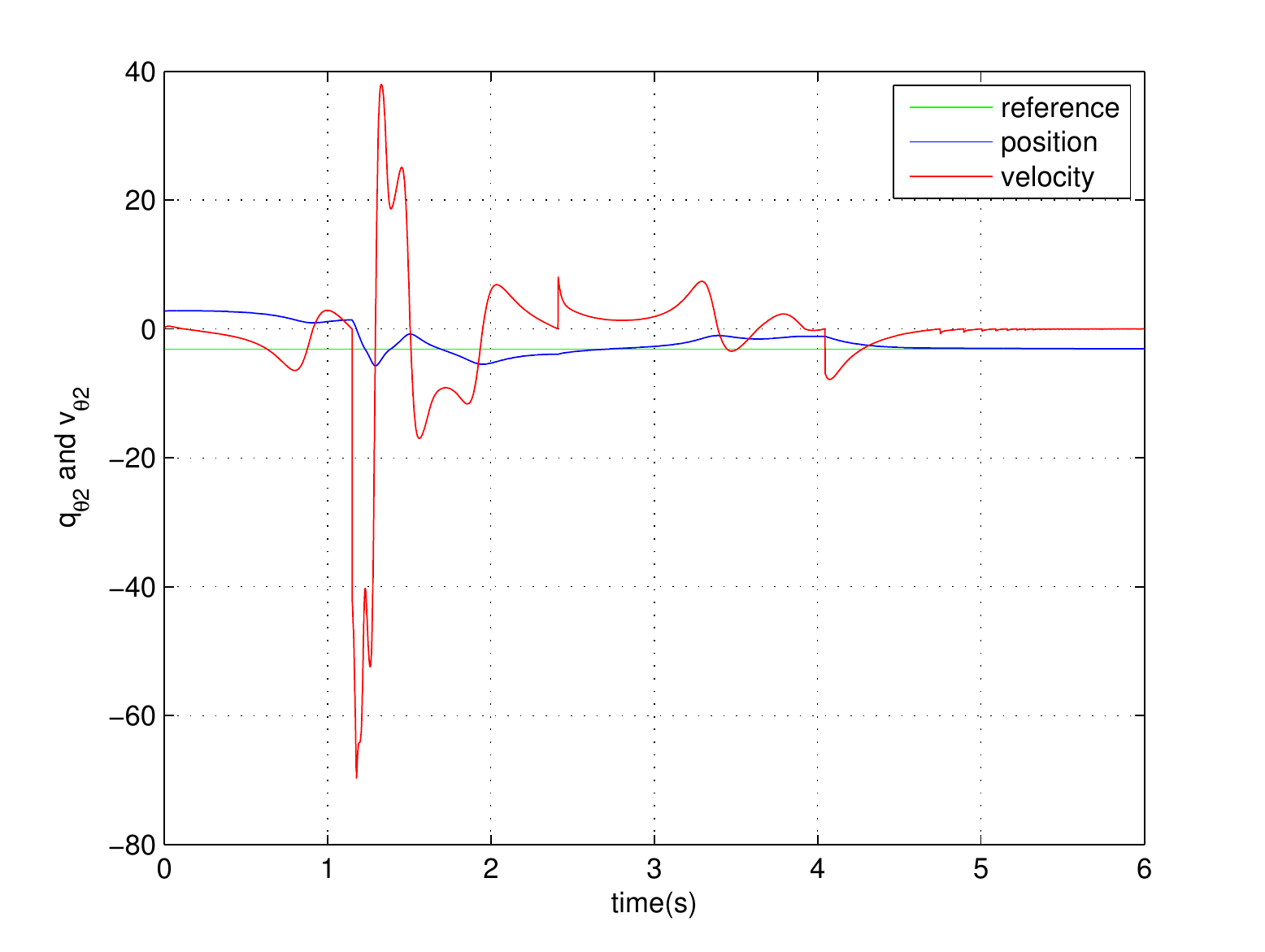}}
\end{subfigmatrix}
 \caption{Pendulum motion with impulse and feedback control switch on}
 \label{fig:pendulum_impulse}
\end{figure}

\begin{figure}
\centering
 \begin{subfigmatrix}{2}
  \subfigure[$\Lambda_{M1}$]{\includegraphics{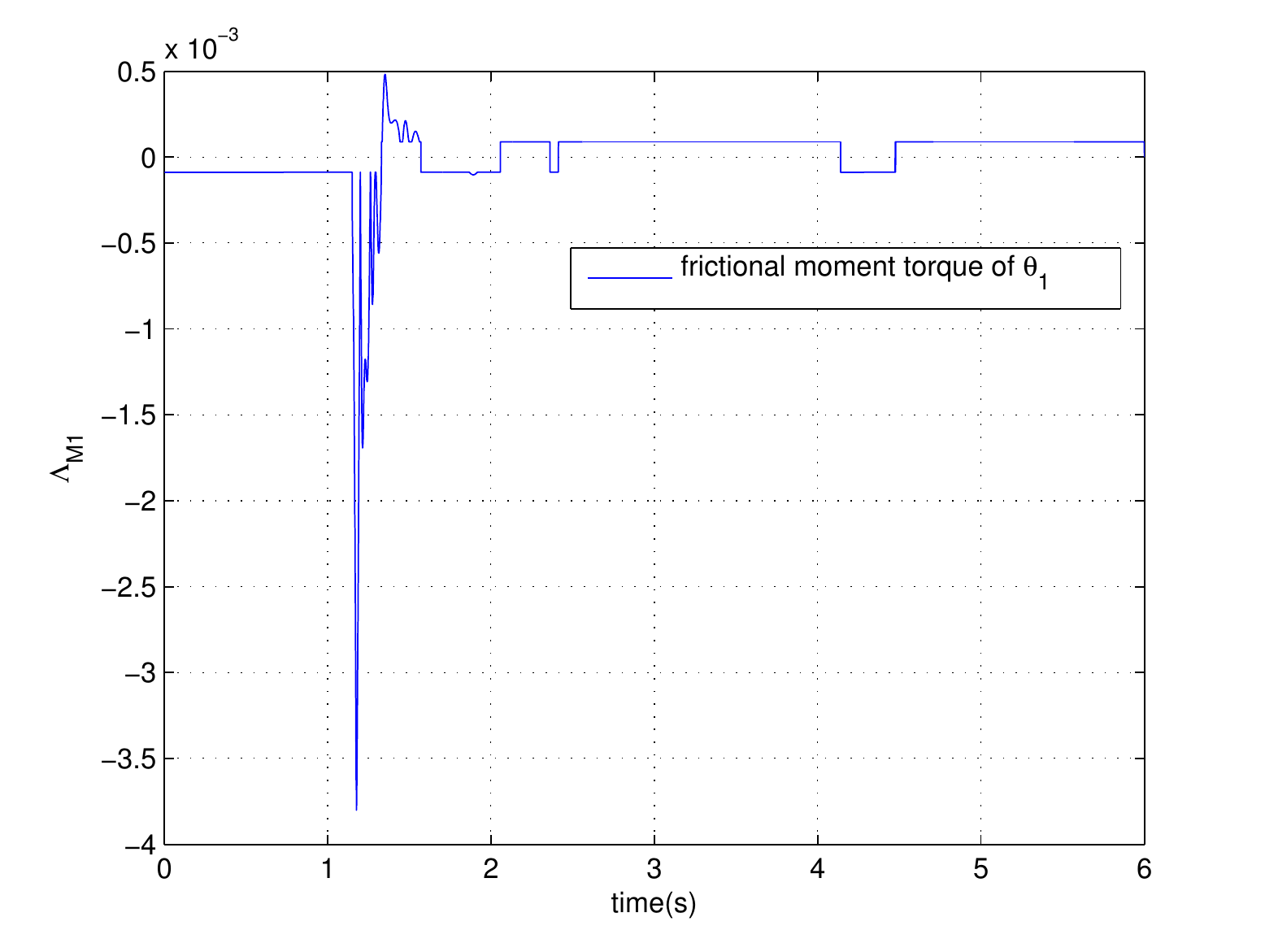}}
  \subfigure[$\Lambda_{M2}$]{\includegraphics{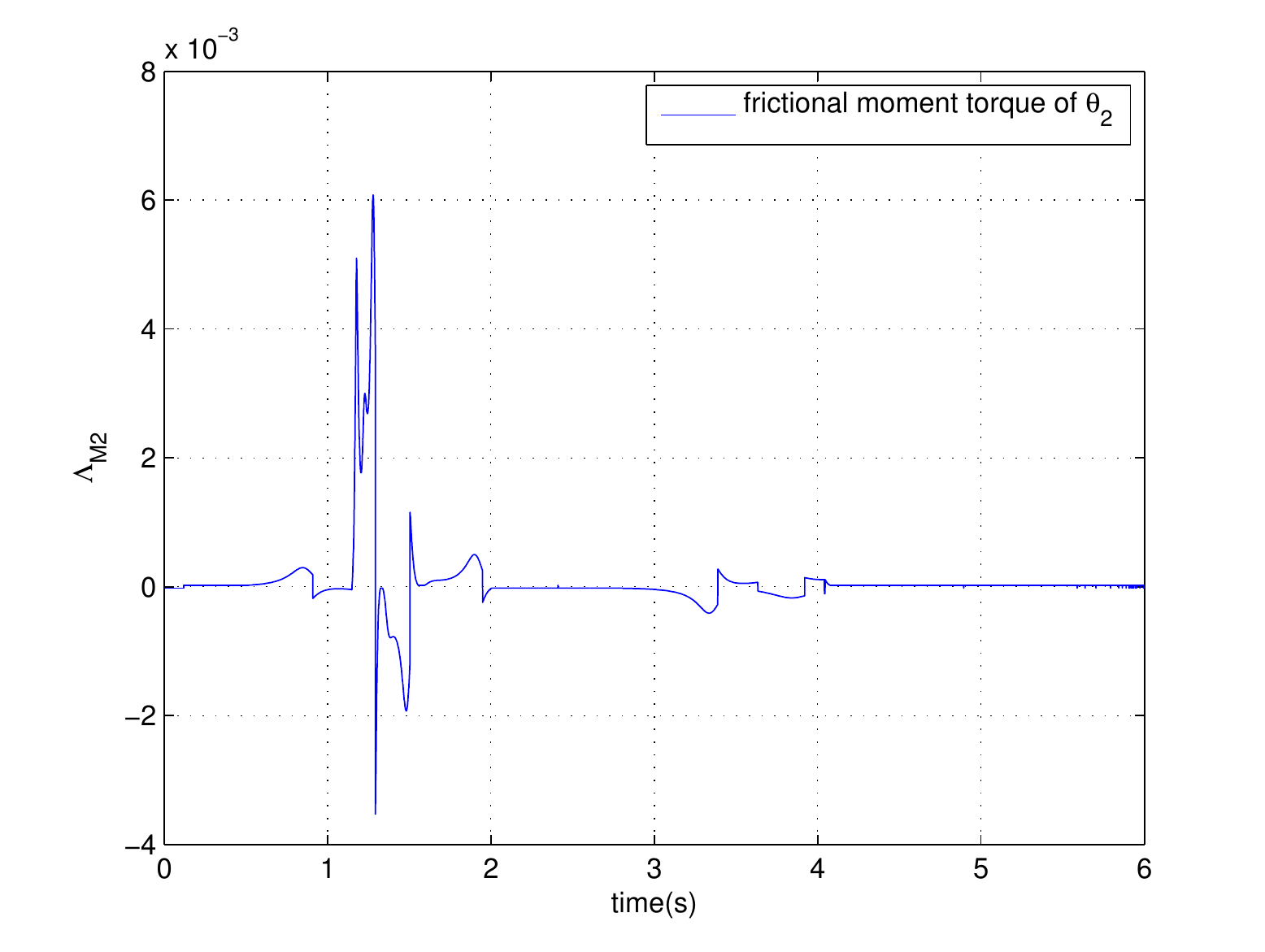}}
  \subfigure[$u_\tau$]{\includegraphics{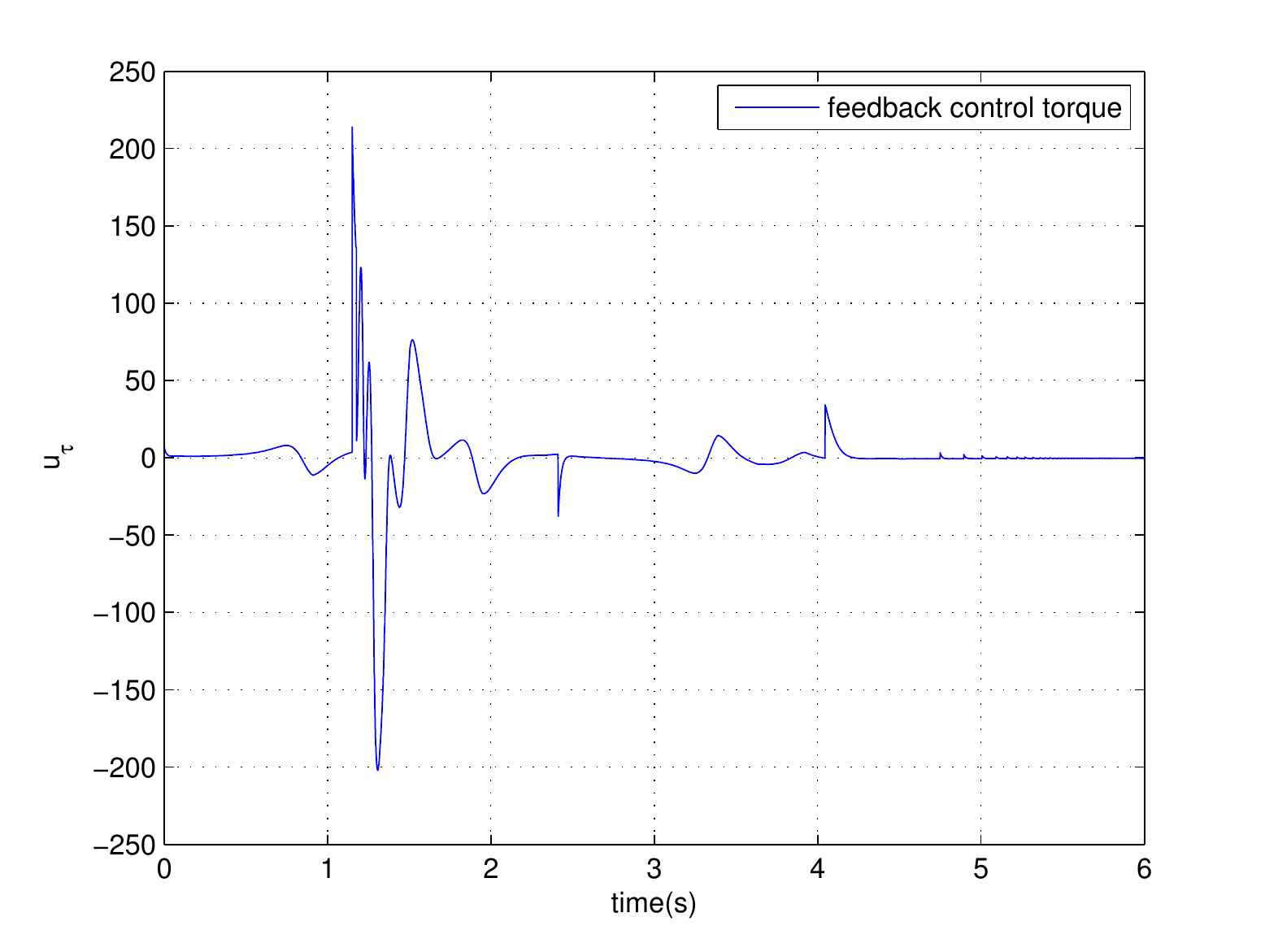}}
  \subfigure[$U_\tau$]{\includegraphics{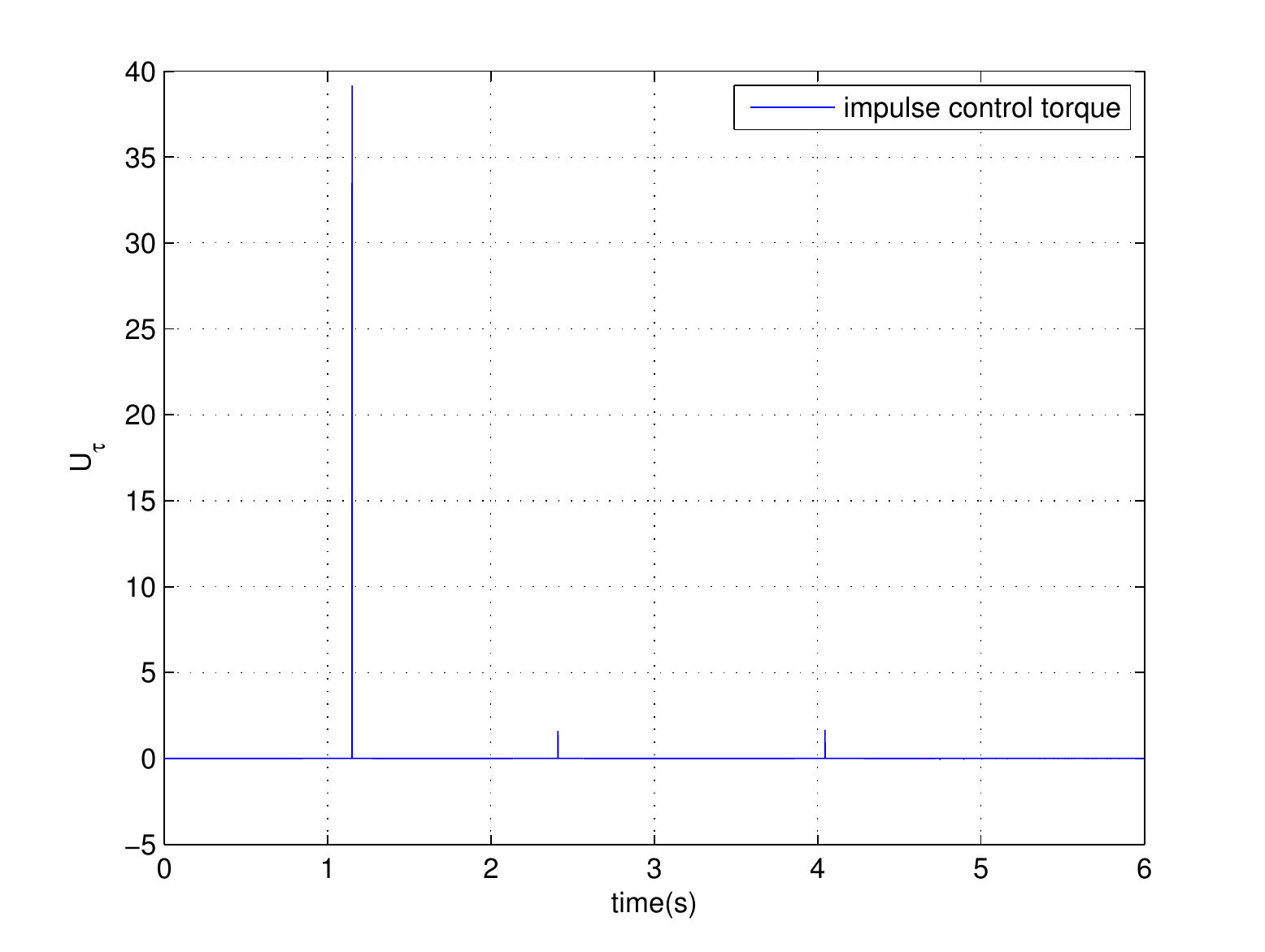}}
 \end{subfigmatrix}
 \caption{Time history of frictional moment impulse, feedback control torque, and impulsive control torque of controlled pendulum}
 \label{fig:pendulum_impulse_lmd_tau}
\end{figure}

The parameters in Table \ref{parameter} in Appendix \ref{AppendixA} are employed for simulation. The feedback control parameters are set as
\[k_1=1, \quad k_2=5, \quad k_3=20, \quad k_4=5 \; ,\]
and there is no difference between $k_{pre}$ and $k_{post}$. The initial conditions are given by
\[
  \mathbf q_\theta = \begin{pmatrix}  \theta_1 \\  \theta_2 \end{pmatrix} = \begin{pmatrix} \pi \\ \frac{8}{9}\pi \end{pmatrix}\unit{m} \;, \quad \mathbf v_\theta = \begin{pmatrix}  \dot{\theta}_1 \\  \dot{\theta}_2 \end{pmatrix} = \begin{pmatrix} 0 \\ 0 \end{pmatrix}\unitfrac{m}{s} \;.
\]
The simulation result of free motion of the pendulum is shown in Fig. \ref{fig:pendulum_no_control}, the result with feedback control switching on is shown in Fig. \ref{fig:pendulum_feedback}, and the result with both feedback and impulse control switching on is presented in Fig. \ref{fig:pendulum_impulse}. Addtionally, in the case of feedback and impulse control switching on, the time history of frictional moment impulse, feedback and impulsive control torques are plotted in Fig. \ref{fig:pendulum_impulse_lmd_tau}.

\par
Particularly, in Fig. \ref{fig:pendulum_impulse}, the green solid lines indicates the reference positions for the pendulum that
\[
\theta_1=\pi, \quad, \theta_2=-\frac{\pi}{2}\;.
\]
It is noticed that the state $(\theta_2, \dot{\theta}_2)$ of pendulum arm finally converges the desired state that
\[
\theta_{2, converge}=-\frac{\pi}{2}, \quad \dot{\theta}_{2, converge}=0\;,
\]
and the driving arm also experience a stable motion with very limited angular velocity around its reference position. Due to the dissipation effect resulted from friction, switching off the control, the pendulum will stick on the desired unstable equilibrium (inverted) position. 
\par
\begin{figure}[hbt]
  \centering
  \includegraphics[width=0.8\linewidth]{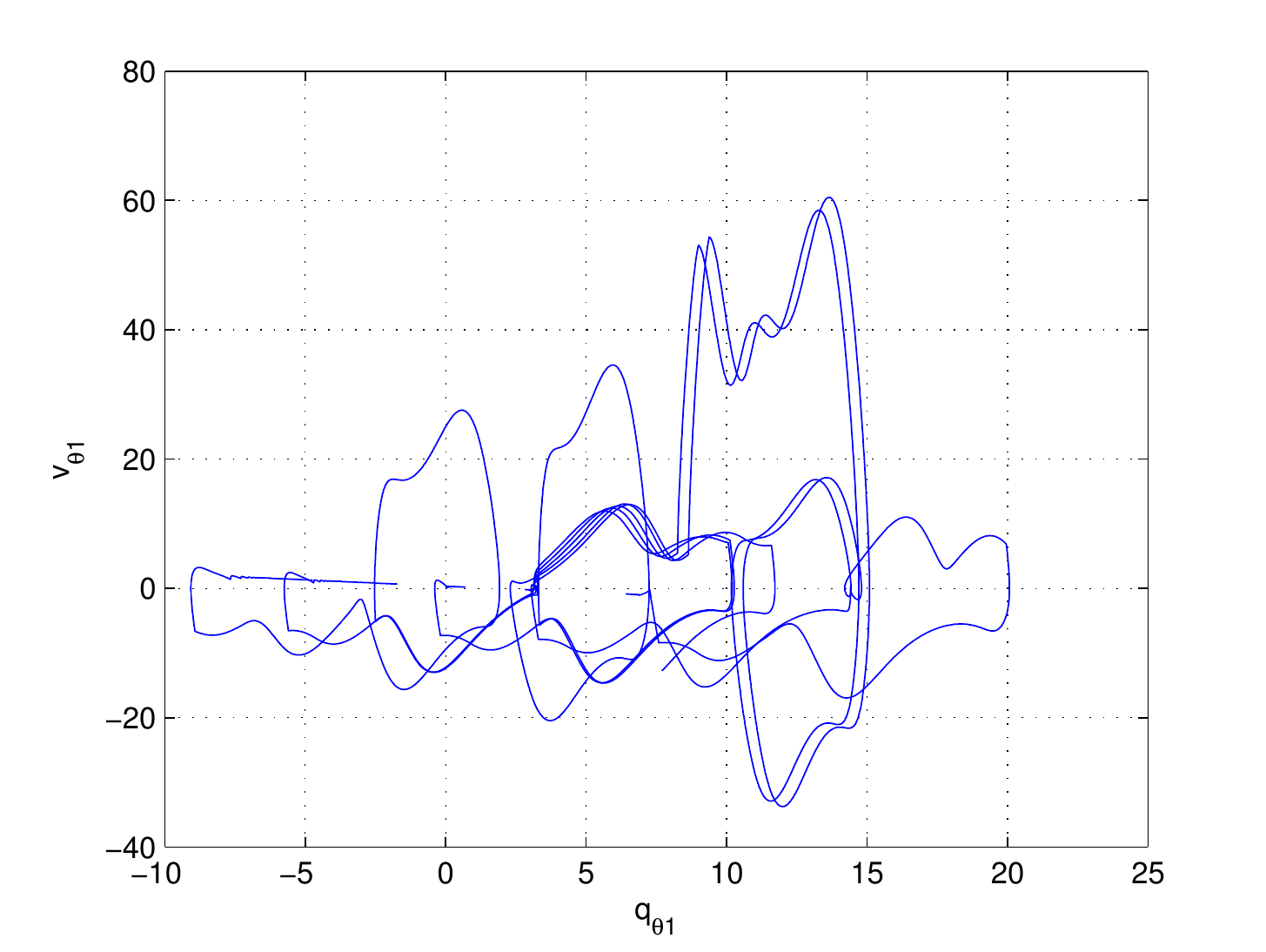}
  \caption{Phase diagram of pendulum controlled behavior on $\theta_1$}
  \label{fig:phase_controlled_theta1}
\end{figure}
\begin{figure}
  \centering
  \includegraphics[width=0.8\linewidth]{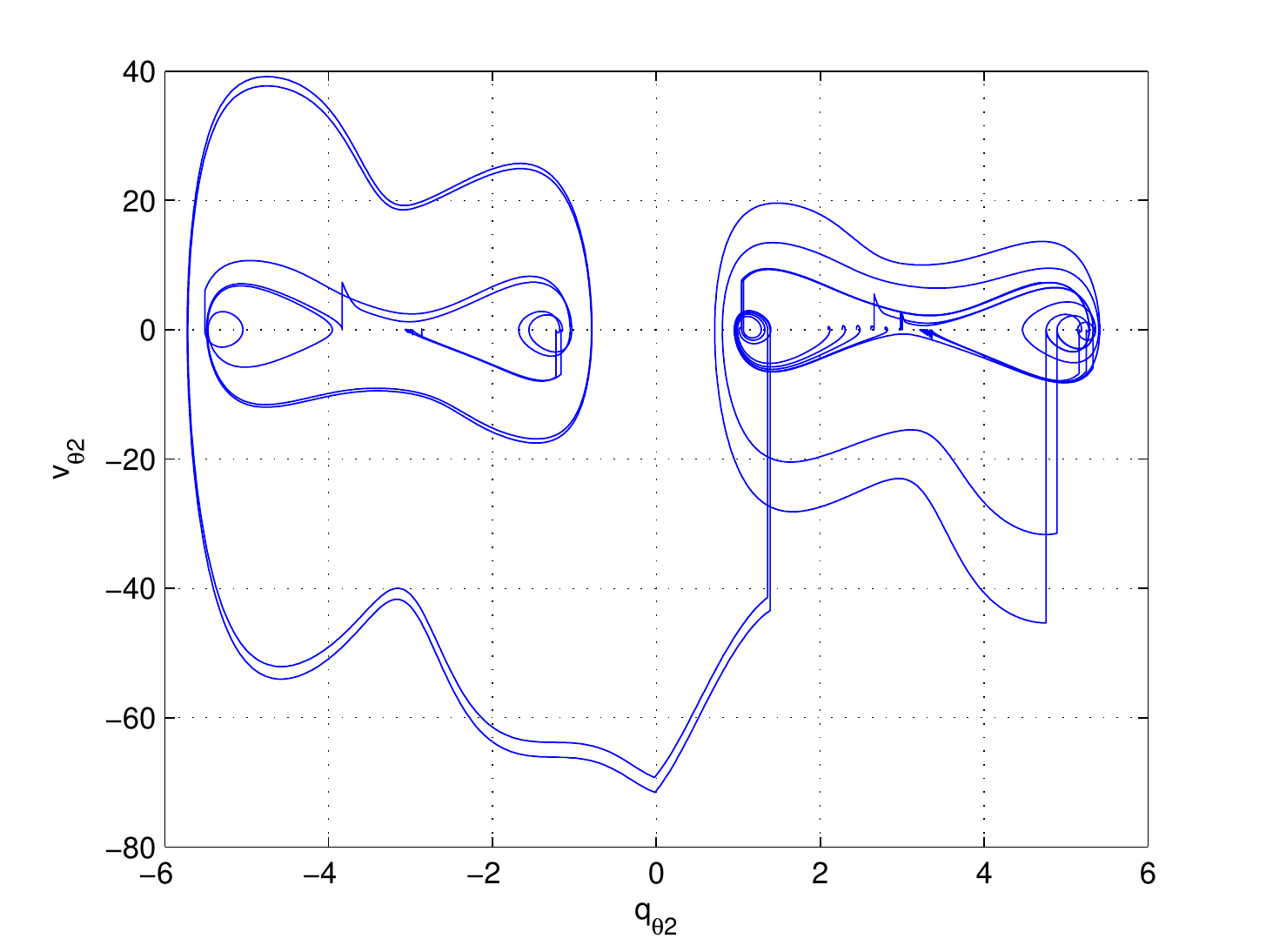}
  \caption{Phase diagram of pendulum controlled behavior on $\theta_2$}
  \label{fig:phase_controlled_theta2}
\end{figure}

In order to analyze the controlled pendulum behavior and the robust of our method, we test out method with the initial condition space by 
\begin{align*}
\theta_1=0, \quad \theta_2 \in [\frac{2\pi}{3}, \frac{\pi}{2}), \quad \dot{\theta}_1=0, \quad \dot{\theta}_2=0 \;.
\end{align*}
The phase diagrams of this system on $\theta_1, \theta_2$ are presented in Fig. \ref{fig:phase_controlled_theta1} and Fig. \ref{fig:phase_controlled_theta2}. As the simulated results shows, the driving arm is most likely to experience a stable motion with limited constant angular velocities, while only an exceptional case has a changing angular velocity within 6 seconds of simulation time; the pendulum arm has been successfully stabilized to the inverted position with $(\theta_2, \dot{\theta}_2)$ converges to the points defined by $\dot{\theta}_2=\unitfrac[0]{rad}{s}, \theta_2=\pm \, {\pi}/{2}\,$.




\chapter{Conclusion} 

\label{Chapter6} 

\lhead{Chapter 6. \emph{Conclusion}} 


In this paper, a non-penetration and physically consistent non-smooth numerical approach has been proposed by employing Prox formulation, Moreau's mid-point time-stepping schemes for the contact dynamics with unilateral decoupled constraints. Under this circumstance, the impulse-based control has been successfully implemented and validated on the motion system of controlled frictional oscillator. Further development has been achieved by utilizing shooting method in the impulse estimating process instead of robust estimation.
\par
This non-smooth numerical method has been applied in the fields of under-actuated friction-coupled mulit-body system, by means of an implementation on the frictional Furuta pendulum. The feedback controller supplemented with reduced impulse-based control law and shooting method has successfully solved the problem of stabilization of the inverted frictional Furuta pendulum, which is suffered from the stiction effect of friction with non-zero steady-state errors.
\par
The applications and validating of this numerical method has shown the importance and advantages of a physically consistent scheme and prox function, which provides the convergent solutions resulted from the iteration of computing the coupled frictions.
\par
For further step on the impulse-based control of Furuta pendulum should be constructing and implementing more advanced, effective and robust control methods and mathematical tools for both feedback controller and impulse estimating process, such as energy shaping method, model predicative method, adaptive control, and sophisticated boundary value problem or optimization solver. Moreover, the speed of computation and the possibility of real-time control should be taken into account.


\addtocontents{toc}{\vspace{2em}} 

\appendix 




\chapter{Parameters of Furuta Pendulum} 

\label{AppendixA} 

\lhead{Appendix A. \emph{Parameters of Pendulum}} 

\begin{table}[hbt]
\centering
\caption{\label{parameter}Parameters of Pendulum}
\begin{tabular}{l p{1.5cm} l}
    \hline
    Geometrical Characteristics & & $l_1=\unit[0.435]{m}$ \\
    &&$c_1=\unit[0.217]{m}$ \\
    &&$l_2=\unit[0.2]{m}$ \\
    &&$c_2=\unit[0.19]{m}$ \\
    &&$R_1=\unit[0.08]{m}$ \\
    &&$R_2=\unit[0.03]{m}$ \\
    &&$R_E=\unit[0.059]{m}$ \\
    \\
    Inertial Properties &&$m_1=\unit[0.4]{kg}$ \\
    &&$m_2=\unit[0.55]{kg}$ \\
    &&$J_1=\unit[0.027]{kg\,m^2}$ \\
    &&$J_2=\unit[0.021]{kg\,m^2}$ \\
    \\
    Force and Friction Elements&& $g=\unitfrac[9.81]{m}{s^2}$ \\
    && $ \mu = 0.25$\\
    && $ \lambda_{B1, {static}} = \unit[12]{N}$\\
    && $ \lambda_{B2, static} = \unit[3]{N}$\\
    \hline
\end{tabular}
\end{table}

\addtocontents{toc}{\vspace{2em}} 

\backmatter


\label{Bibliography}

\lhead{\emph{Bibliography}} 

\bibliographystyle{unsrt} 


\end{document}